\documentclass[superscriptaddress,nofootinbib,notitlepage,pra,aps,10pt]{revtex4-1}
\usepackage[utf8]{inputenc}
\usepackage[colorlinks]{hyperref}
\hypersetup{
	pdfstartview={FitH},
	pdfnewwindow=true,
	colorlinks=true,
	linkcolor=blue,
	citecolor=blue,
	filecolor=blue,
	urlcolor=blue}
\usepackage{braket}
\usepackage{amsmath}
\usepackage{amssymb}
\usepackage{amsthm}
\usepackage{graphicx}
\usepackage{algorithmic}
\usepackage{mathrsfs}
\usepackage{qcircuit}
\usepackage{float}
\usepackage[version=4]{mhchem}
\usepackage{enumitem}
\usepackage{siunitx}
\usepackage{soul}
\usepackage{xcolor}

\newcommand{\Or}{\mathcal{O}}
\newcommand{\RR}{\mathbb{R}}

\newcommand{\wt}{\widetilde}

\newcommand{\subsys}{\mathcal{N}}
\newcommand{\bonda}{\chi}
\newcommand{\bondb}{M}
\newcommand{\Nunit}{N_{\rm un}}
\newcommand{\Nphas}{N}

\newcommand{\nn}{\nonumber \\}

\allowdisplaybreaks

\newcommand{\MQ}{\affiliation{%
School of Mathematical and Physical Sciences,
Macquarie University, Sydney, NSW 2109, Australia} }
\newcommand{\Google}{\affiliation{
Google Quantum AI, Venice, CA 90291, United States}}

\newcommand{\SeoulNational}{\affiliation{
Department of Chemistry, Seoul National University, Seoul 08826, Republic of Korea
}}
\newcommand{\Caltech}{\affiliation{
Division of Chemistry and Chemical Engineering, California Institute of Technology, Pasadena, CA 91125, United States}}
\newcommand{\Cal}{\affiliation{
Department of Mathematics, University of California, Berkeley, CA 94720, United States}}

\begin{document}

\title{Rapid initial state preparation for the quantum simulation of\\ strongly correlated molecules}

\author{Dominic W. Berry}
 \email{dominic.berry@mq.edu.au}\MQ
  
\author{Yu Tong}
\affiliation{
Institute of Quantum Information and Matter, California Institute of Technology, Pasadena, CA 91125, United States}
\affiliation{
Department of Mathematics, Duke University, Durham, NC 27708, United States}
\affiliation{Department of Electrical and Computer Engineering, Duke University, Durham, NC 27708, United States\looseness=-1}

	\author{Tanuj Khattar}
 	\Google
  
	\author{Alec White}
 	\Google
  
        \author{Tae In Kim}
        \SeoulNational       

        \author{Sergio Boixo}
        \Google

        \author{Lin Lin}
        \Cal

        \author{Seunghoon Lee}
        \SeoulNational
        \Caltech

        \author{Garnet Kin-Lic Chan}
        \Caltech
       
	\author{Ryan Babbush}
 	\Google
  
	\author{Nicholas C. Rubin}
        \email{nickrubin@google.com}
 	\Google

\date{\today}
\begin{abstract}
Studies on quantum algorithms for ground state energy estimation often assume perfect ground state preparation; however, in reality the initial state will have imperfect overlap with the true ground state.
Here we address that problem in two ways: by faster preparation of matrix product state (MPS) approximations, and more efficient filtering of the prepared state to find the ground state energy.
We show how to achieve unitary synthesis with a Toffoli complexity about $7 \times$ lower than that in prior work, and use that to derive a more efficient MPS preparation method. For filtering we present two different approaches: sampling and binary search. For both we use the theory of window functions to avoid large phase errors and minimise the complexity. We find that the binary search approach provides better scaling with the overlap at the cost of a larger constant factor, such that it will be preferred for overlaps less than about $0.003$. Finally, we estimate the total resources to perform ground state energy estimation of Fe-S cluster systems, including the FeMo cofactor by estimating the overlap of different MPS initial states with potential ground-states of the FeMo cofactor using an extrapolation procedure. {With a modest MPS bond dimension of 4000, our procedure produces an estimate of $\sim 0.9$ overlap squared with a candidate ground-state of the FeMo cofactor, 
producing a total resource estimate of $7.3 \times 10^{10}$ Toffoli gates; neglecting the search over candidates and assuming the accuracy of the extrapolation, this validates prior estimates that used perfect ground state overlap. This presents an example of a practical path to prepare states of high overlap in a challenging-to-compute chemical system.}
\end{abstract}

\maketitle

\tableofcontents

\section{Introduction}
The complexity of estimating ground state energies of chemical and material systems using quantum phase estimation (QPE) is frequently analysed in the ideal case where the ground state has already been prepared accurately. In this restrictive setting, the main error in QPE originates from the eigenphase differing from the approximate output from QPE--a phenomenon known as ``spectral leakage''~\cite{xiong2022dual} or ``bit discretization error''~\cite{greenaway2024case}. In the case where the ground state (or more generally an eigenstate) is not prepared exactly, the QPE protocol outputs an estimate of the ground state energy with a probability proportional to the square of the overlap of the ground state and the input initial state. In order to estimate the total amount of quantum resources (logical qubits and gates) for the most important simulation problems and determine total runtimes for high confidence eigenenergies, both sources of error in QPE must be quantified and accounted for, along with the cost of performing phase estimation.

The problem of preparing an initial state with high overlap with the ground state has been of recent interest. This is in part because a generic state in the Hilbert space has exponentially small overlap with the ground state, and a non-trivial state preparation might need to be performed for a reasonable success probability in QPE, which could add substantially to the total QPE cost. This has spurred a substantial body of 
work examining the cost of preparing various approximate wavefunctions and the associated success probability of QPE. For example, analysis of product state wavefunctions in the fermionic setting~\cite{tubman2018postponing} along with informed orbital optimized improvements~\cite{ollitrault2024enhancing}, state preparations analyzed in the context of embedding theories~\cite{erakovic2024high}, truncated configuration interaction~\cite{fomichev2024initial}, and the use of matrix-product states (MPS) as input QPE states~\cite{RezaNetworks}. In terms of circuit compilations, direct synthesis costs for wavefunctions are known~\cite{low2018trading} along with a variety of MPS state preparation techniques, including layers of two-qubit operations \cite{RanMPS}, low-depth circuits \cite{,PhysRevLett.132.040404}, and sequences of operations with an ancilla register \cite{Sequential}.
Reference \cite{Sequential} in particular has been analysed in detail for minimising the Toffoli count \cite{fomichev2024initial}. A related consideration in the context of evaluating quantum advantage is that classical heuristic algorithms can also be viewed as having a state preparation step, from which a classical estimate of the ground-state energy is then estimated and, in some cases, efficiently refined~\cite{Lee2023,chan2024quantum}. Hence it is relevant to ask for specific systems, the quantitative cost of preparing a good initial state for QPE.

Given that an appropriate initial state has been prepared, there remains the problem of how to measure the ground state energy in a way that properly distinguishes it from excited states.
If the overlap is not too small, then it may be expected that a simple sampling approach will suffice, and it will be necessary to sample enough times that there will be a high probability of sampling the ground state energy at least once.
The sample that corresponds to the ground state energy can be identified by taking the minimum among all samples.
There are two difficulties with this approach.
First, the large number of samples may mean that there is exceptionally large underestimation error in at least one of the samples, resulting in an erroneously small estimate of the ground state energy.
Second, the number of samples will scale with the inverse of the squared overlap.

Amplitude amplification would suggest that the complexity should scale with the inverse overlap instead of the inverse overlap squared.
Naively, in order to obtain the quadratic improvement, the range of energies for the ground state must be known. Binary search can be employed to go beyond this requirement and search for an unknown ground state. The binary search approach was used previously in \cite{LinTong2022,ZhangWangJohnson2022computing,WanBertaCampbell2022randomized,DongLinTong2022ground,WangStilkFrancaEtAl2022quantum}, but these works, together with \cite{DingLin2023simultaneous,NiLiYing2023low,LiNiYing2023low}, typically aim to optimize for the circuit depth, while this work focuses on the number of non-Clifford gates needed for the whole algorithm, a more relevant metric for fault-tolerant quantum computers.

In this work we provide improved results for both MPS preparation and filtering to determine the ground state energy.
For MPS preparation we start by developing a method for synthesising unitaries with low Toffoli count by decomposing the unitary into a sequence of diagonal phasing operations together with low-cost operations.
We then use that to construct a method for synthesising only a fraction of the columns of the unitary, which we then apply to the method of Ref.~\cite{Sequential} to provide a factor of 7 improvement in Toffoli gates over prior work. For filtering the resulting state, we apply the theory of window functions in order to minimise the probability of estimates with large error. We describe the optimal performance provided by the Slepian prolate spheroidal window \cite{Slepian1965,Imai_2009, patel2024optimal} and compare the costs to the Kaiser window previously reported in Ref.~\cite{BerPRXQ24}. 
In the case of sampling, this suppresses the probability of exceptionally low energy estimates that would make the overall estimate low, overcoming the first problem.
Moreover, we provide a tighter bound on the contribution to the error from excited states, by analysing the interplay between the contribution to errors where the estimated energy is too low versus too high. We find that Kaiser windows can provide substantially improved performance.

We also provide improved scaling with the overlap by using a binary search together with amplitude estimation, similar to Ref.~\cite{LinTong2020near}.
We successively reduce the possible range for the ground state by performing amplitude estimation at each step in order to eliminate a fraction of the range.
In this way we are able to achieve the speedup promised by amplitude amplification without any initial estimate of the ground state energy.
On the other hand, the overhead induced by this procedure means that it is preferable for small overlaps $p\lesssim 0.003$, and for large overlaps the sampling method is preferable.
We apply the optimal windows for phase estimation to the amplitude estimation procedure, and thereby significantly reduce the resources compared to Ref.~\cite{LinTong2020near}.

Armed with a detailed costing of the number of times QPE must be repeated, we estimate the full quantum resources necessary to refine the ground state energy of several Fe-S clusters: [2Fe-2S], [4Fe-4S], and FeMo-cofactor (FeMoco). In order to obtain overlaps of low bond dimension MPS states with the true ground state we introduce an extrapolation protocol that uses two MPS wavefunctions to derive an empirical estimate of the overlap of a fixed bond dimension wavefunction and the infinite (exact) bond dimension MPS.
In FeMoco, at a finite bond dimension we can obtain low-bond dimension MPS that are candidates for different low-energy states of the cluster, although these initial MPS do not give a reliable energy ordering. For each candidate MPS, we can estimate the overlap with the eventual eigenstate, which allows us to cost out energy estimation for FeMoco in the high-confidence regime (95\% and 99\% confidence level) to chemical accuracy.
We also re-analyze the block encoding costs for FeMoco and other Fe-S clusters using symmetry shifting, resulting in a reduction of the LCU 1-norm by a factor of up to 2. We find that very few iterations of QPE are required (2 or 3 iterations) due to the high overlap of low-bond dimension MPS. These additional iterations over single-shot QPE resource estimates provided for FeMoco in Ref.~\cite{PRXQuantum.2.030305} along with symmetry shifting reductions in the LCU 1-norm result in $7.3 \times 10^{10}$ Toffoli gates required for a full resource cost to refine an energy estimate for FeMoco. For a single candidate ground state, this amount of resources is only 2.3 times that in Ref.~\cite{PRXQuantum.2.030305} (which uses the Hamiltonian defined in Ref.~\cite{li2019electronic}) and can likely be reduced further through improved symmetry shifting. Once the accurate energies of different candidates are obtained, they can be ordered  to determine the ground-state energy.
Ultimately, the high extrapolated overlap {achieved in this problem} suggests that
{the combination of generating one (or more) candidate MPS initial states followed by QPE is a practical approach to refine the ground-state energy  in a realistic challenging chemical system.}

In the following we begin in Section \ref{sec:results} by summarising the results that will be presented.
We then give the background for both MPS preparation and phase estimation in Section \ref{sec:background}.
Section \ref{sec:prep} describes our new method for unitary synthesis and preparing MPS states.
Then Section \ref{sec:phase_estimation_with_window_functions} summarises the procedure to perform phase estimation optimally for confidence intervals.
This is a procedure which is used for both approaches to searching for the ground state energy.
In Section \ref{sec:sampling} we describe the sampling approach, and in Section \ref{sec:binary_search} we describe the binary search approach.
These results are used to estimate resource requirements for real systems in Section \ref{sec:resource_estimates}, and we conclude in Section \ref{sec:conc}.

\subsection{Results overview}\label{sec:results}
We provide results for three main areas: preparation of matrix product states, energy estimation with these prepared states, and resource estimates for ground state energy estimates of the FeMoco chemical system.

For MPS preparation, we first derive a new result for the synthesis of general unitary operations with reduced Toffoli count.
This method reduces the synthesis to layers of phase shifts alternating with increment/decrement operations and Hadamard gates.
The layers of phase shifts can be applied with reduced Toffoli count using QROM.
As a result there are at most $\Nunit+1$ layers of QROM for dimension $\Nunit$, and each QROM has complexity $\mathcal{O}(\sqrt{\Nunit})$.

We then use this result to derive a procedure to synthesise columns of a unitary operation.
That is, a unitary operation where the input state is restricted to lie within a subspace.
If only half the columns of the unitary need be synthesised, we reduce the problem to synthesis of an initial unitary of dimension $\Nunit/2$, a controlled qubit rotation, then controlled synthesis of a unitary of dimension $\Nunit/2$.
That controlled unitary has half the complexity of synthesising a unitary operation of dimension $\Nunit$, because only half the layers are required.
Then in order to synthesise $\Nunit/d$ columns of a unitary we can iterate this procedure another $d-2$ times.

The MPS preparation can then be achieved by a sequence of steps where $\Nunit/d$ columns of a unitary need to be synthesised.
Moreover, the initial unitary in the above procedure can be merged with other operations, further reducing the complexity.
As a result, the overall Toffoli complexity is significantly reduced over that obtained for methods in prior work, now making it a small complexity as compared to the complexity of phase estimation.

For energy estimation, there are two approaches we consider: direct sampling, and a binary search with amplitude estimation.
The parameters of the problem are
\begin{itemize}[noitemsep,topsep=0pt]
    \item the initial squared overlap of the prepared state with the ground state, $p$,
    \item the block encoding normalization factor $\lambda$,
    \item the allowable error in the energy estimate $\epsilon$, and
    \item the confidence level $1-q$.
\end{itemize}
For the direct sampling approach, we provide expressions to determine exact costings in terms of special functions.
We then derive asymptotic expressions to provide the expected scaling of the complexity in the above parameters.
The total number of queries to the qubitized walk operator encoding the Hamiltonian is approximately
\begin{equation}\label{eq:eq1_kaiser_total_queries}
    \frac{n\lambda}{2\epsilon}\ln(1/\delta) \approx \frac{\lambda\ln(2/q)}{2p\epsilon}\ln\left[ \frac{\ln(2/q)}{pq}
    \right]  ,
\end{equation}
corresponding to the leading term in Eq.~\eqref{eq:asyn}.
This expression is found by solving
\begin{equation}\label{eq:optimal_delta_wrt_n}
    [1-p(1-\delta/2)]^n + 1-(1-\delta/2)^n =q
\end{equation}
for $\delta$, then minimising with respect to the number of samples $n$.
The optimal choice for $n$ is slightly larger than $(1/p)\ln(1/q)$.

For the binary search approach, under the same conditions, the query complexity is
\begin{equation}
\frac{7.77\lambda}{\sqrt{p}\epsilon}\ln\left(\frac{4}{\sqrt{p}}\right)\ln\left(\frac{\log_{\sqrt{2}}(\lambda/\epsilon)}{q}\right),  
\end{equation}
as in Eq.~\eqref{eq:total_complexity_optimized_binary_search_7.77}.
This has improved scaling as $1/\sqrt{p}$, which is a square root improvement, albeit with a constant factor about 16 times larger.
This constant factor suggests that we would need $p\lesssim 0.003$ for this approach to provide an improvement.

These asymptotic results can be inaccurate for realistic values of parameters, so we provide improved approximations for phase measurements with window functions in Section~\ref{sec:higher}.
We provide series expansions for the error and cost, both for the Kaiser window and for the prolate spheroidal window.
In the process we correct an error in the work of Slepian from 1965 \cite{Slepian1965}, which gave incorrect terms.
We show that the error is asymptotically lower for the prolate spheroidal window, but the cost (the number of oracle calls to achieve a given error) is asymptotically the same for the two windows.

We show how to properly account for the excited states in sampling to obtain the ground state.
These states contribute to two types of error.
\begin{enumerate}[align=left,label=Type \Roman*:,noitemsep,topsep=0pt]
\item The excited states contribute to the probability of a sample with large error, more than $\epsilon$ below the ground state energy.
\item Samples corresponding to excited states contribute to estimates that are too high, because it is an accurate (or high) estimate of that energy, but the excited state energy is higher than the ground state energy.
\end{enumerate}
When excited state energies are close to the ground state, then they increase the probability of a Type I error, but that also reduces the probability of a Type II error.
We provide a careful accounting of the contribution of these two errors to provide a more accurate estimate of the cost of the sampling approach.
Surprisingly, we find that the Kaiser window can provide \emph{better} results than the prolate spheroidal window when accounting for excited states.
Taking all these considerations into account yields results close to those for the approximate asymptotic expression given above.

We provide numerical results for query complexities accounting both for the exact error with window functions and the effect of the excited states.
The results for queries to block encoding of the Hamiltonian are provided in Fig.~\ref{fig:compare_W_queries}, and for the number of calls to the initial state preparation in Fig.~\ref{fig:compare_Uinit_queries}.
This shows that the binary search approach is optimal for small squared overlap $p$, but the sampling approach is optimal for moderate to large values of $p$ (above $p\sim 0.003$ for the case of 95\% confidence intervals).

For FeMoco resource estimates we develop an extrapolation scheme that allows us to estimate the overlap of a fixed bond dimension MPS with an infinite bond dimension MPS--e.g.\ the true eigenstate. The extrapolation protocol is constructed from two empirically observed linear relationships
\begin{align}\label{eq:res_summary_extrapolation}
	\log\left(1 - \left|\langle \Phi(\bondb') | \Phi(\infty) \rangle \right|^2\right) \ & \mathrm{vs.}\ \left(\log(\bondb')\right)^2  \\
	\log\left(\left|\langle \Phi(\bondb') | \Phi(\bondb'') \rangle \right|^2 - \left|\langle \Phi(\bondb') | \Phi(\infty) \rangle \right|^2\right) \ & \mathrm{vs.}\ \left(\log(\bondb'')\right)^2 \ \mathrm{where} \ \bondb' \ll \bondb'' 
\end{align}
and verified on Fe$_{2}$S$_{2}$ and Fe$_{4}$S$_{4}$ systems where accurate estimates of the ground states can be computed, and where the above extrapolation can be verified. We analyze three different MPS wave functions for FeMoco initialized through a procedure similar to Ref.~\cite{li2017spin, li2019electronic} {that is believed to generate candidates for competing low-energy eigenstates (corresponding to different spin couplings) of the $S=3/2$ ground state. The high overlaps estimated with some eigenstate in the low-energy manifold, as produced by our protocol, suggest that the combination of initializing different candidate MPS states, followed by QPE, is a promising computational procedure to map out the lowest eigenenergies and subsequently refine the ground-state energy in this practically challenging chemical simulation problem.}

\section{Background}
\label{sec:background}

\subsection{Matrix product state preparation}
Matrix product states provide a systematically improvable approximation of entangled states thus providing a class of initial states with tunable overlap.
The Toffoli complexity of MPS preparation was previously analysed in Ref.~\cite{fomichev2024initial} using the approach from Ref.~\cite{Sequential} together with the unitary synthesis scheme of Low, Kliuchnikov and Schaeffer (LKS) Ref.~\cite{low2018trading}.
In this work we provide a significantly more efficient unitary synthesis scheme than that of Ref.~\cite{low2018trading}, thereby enabling more efficient MPS preparation.

MPS states for $\subsys$ subsystems of dimension $d$ are of the form
\begin{equation}\label{eq:mps_cartoon}
    \sum_{\{ n \}} {\rm Tr}\left[ A_1^{(n_1)}A_2^{(n_2)}\ldots A_\subsys^{(n_\subsys)} \right] \ket{n_1,n_2,\ldots,n_\subsys} \, .
\end{equation}
The matrices $A_j^{(n_j)}$ are of dimension $\bonda$, called the bond dimension, and the indices $n_j$ range over $d$ values.
The principle of the approach from Ref.~\cite{Sequential} is to use an ancilla of dimension $\bonda$, together with a sequence of unitary operations on this ancilla together with the subsystems.
Using the fact that one is free to represent the MPS of Eq.~\eqref{eq:mps_cartoon} in left canonical form, the matrices of the MPS can be cast as unitaries such that
\begin{equation}
    G[j]_{\alpha_j n_j, \alpha_{j-1} 0} = (A_{j}^{(n_j)})_{\alpha_{j-1},\alpha_j} \, .
\end{equation}
That is, the unitaries $G[j]$ are of dimension $d\bonda$, but only $\bonda$ columns are specified due to the input on the physical leg being zero.
The notation $(A_{j}^{(n_j)})_{\alpha_{j-1},\alpha_j}$ indicates the matrix element $\alpha_{j-1},\alpha_j$ of matrix $A_{j}^{(n_j)}$.
There is a requirement for this technique that the specified columns are orthonormal so they may correspond to columns of a unitary.

\begin{figure}[tbh]
\centering
    \includegraphics[width=8.19cm]{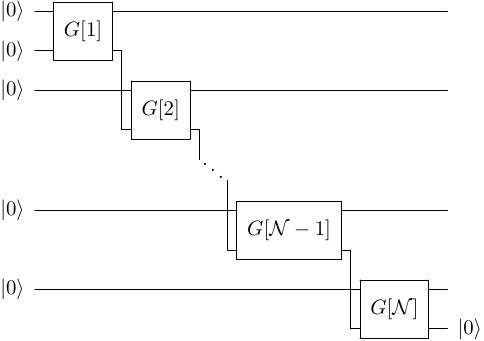}
    \caption{The sequence of operations used to prepare a MPS.
    The input state $\ket{0}$ second from the top is of dimension $\bonda$, as is the final output state at the bottom $\ket{0}$.}
    \label{fig:MPS}
\end{figure}

The sequence of unitary operations used to prepare the MPS is shown in Fig.~\ref{fig:MPS}.
There are three distinct cases where these unitary operations are performed.
\begin{enumerate}
    \item The initial unitary $G[1]$ is on dimension $d\bonda$, but is guaranteed to have input states that are zeroed.
    Therefore it corresponds to simply preparing a state of dimension $d\bonda$, which is simpler than synthesising a general unitary operation on this dimension.
    \item There are $\subsys-2$ unitaries $G[2]$ to $G[\subsys-1]$ on dimension $d\bonda$ where one of the input registers is initialised to $\ket{0}$, but the other is the ancilla which may be in a general state entangled with other registers.
    Therefore, only $\bonda$ columns of the unitary need be synthesised.
    \item The final unitary $G[\subsys]$ is required to reset the ancilla to $\ket{0}$.
\end{enumerate}
For this last operation to be possible, it must be the case that the ancilla prior to the operation has support on dimension $d$.
It must therefore be possible to transform to the required final state (with other qubits being zeroed) via a unitary without introducing an ancilla.
This unitary can therefore be combined with $G[\subsys-1]$ to give a single unitary, and $G[\subsys]$ does not contribute to the cost.

\subsection{Phase estimation}

In phase estimation for ground state energy estimation, a standard approach \cite{BerryNPJ18,Poulin} is to construct a walk operator $W$ from a block encoding of the Hamiltonian, which yields eigenvalues $e^{\pm i\phi_j}$, with
\begin{equation}
    \phi_j = \arccos(\lambda_j/\lambda).
\end{equation}
Here $\lambda$ is the constant in the block encoding of $H$; that is, the block encoding gives $H/\lambda$.
Note that there are two conventions in the definition of $W$ depending on what factor of $i$ is included (corresponding to the two references \cite{BerryNPJ18} and \cite{Poulin}).
The other convention yields eigenvalues $\mp e^{\mp i \arcsin(\lambda_j/\lambda)}$, but use of either convention yields equivalent results.

From measuring $\pm\phi_j$ we can recover $\lambda_j$ via $\lambda_j = \lambda\cos(\phi_j)$.
This method has a number of useful features.
\begin{enumerate}
    \item The cosine function is even so eliminates the $\pm$ sign ambiguity.
    \item The function $\lambda\cos(\phi_j)$ is monotonically decreasing in $\phi_j$, from $\lambda$ for $\phi_j=0$ to $-\lambda$ for $\phi_j=\pi$. (We use the convention that $\arccos$ gives values in the range $[0,\pi]$.) Therefore $\lambda_0$, which is the smallest among all $\lambda_j$, corresponds to the largest among $\phi_j$.
    \item Because $|\cos(x)-\cos(y)|\leq |x-y|$ for all $x,y\in\RR$, if we want to estimate $\lambda_0$ to within additive error $\epsilon$, it suffices to estimate $\phi_0$ to within $\epsilon/\lambda$.
\end{enumerate}

Each time we run the quantum phase estimation circuit, we will get a phase estimate $\hat{\phi}$ that corresponds to $\phi_j$ for some $j$, up to a phase error that we will define later. Each phase estimate then gives us an estimate for the corresponding eigenvalue $\lambda_j$ through
\begin{equation}
    \hat{\lambda} = \lambda\cos(\hat{\phi}).
\end{equation}
One of the problems that we will deal with is that we will not know with certainty which eigenvalue or eigenstate an estimate $\hat{\lambda}$ really corresponds to. For example, an estimate $\hat{\lambda}$ may correspond to an excited state, but the phase error can make it smaller than the ground state energy $\lambda_0$.

\subsubsection{The quantum phase estimation circuit}
\label{sec:the_qpe_circuit}

A main building block of the quantum phase estimation algorithm we are going to use is the controlled walk operator
\begin{equation}
    \ket{0}\bra{0}\otimes W^{\dagger} + \ket{1}\bra{1}\otimes W.
\end{equation}
Compared to the controlled-$W$, this operator needs the same number of gates to implement but doubles the resulting phase difference \cite{babbush2018encoding}.
Using this version of the controlled walk operator, the controlled part in quantum phase estimation becomes
\begin{equation}
    \sum_{k=0}^{\Nphas-1} \ket{k}\bra{k}\otimes W^{2k-N}.
\end{equation}
By using one more controlled $W$ (rather than controlling between $W$ and $W^\dagger$)
and relabelling, the controlled unitary becomes
\begin{equation}
    \sum_{n=0}^{2\Nphas-1} \ket{n}\bra{n}\otimes W^{n-\Nphas}.
\end{equation}

Now suppose that the control register is initialized in the state $\ket{\Gamma}=\sum_{n=0}^{2\Nphas-1}\gamma_n\ket{n}$ and the system register is initialized in a superposition of eigenstates $\ket{\Phi}=\sum_j \Phi_j\ket{\psi_j}$, then the state after applying the controlled unitary is
\begin{equation}
    \sum_{n=0}^{2\Nphas-1} \gamma_n\ket{n}\otimes W^{n-N}\ket{\Phi} = \sum_{j}\Phi_j
    \sum_{n=0}^{2\Nphas-1}\gamma_n e^{i(n-N)\phi_j}\ket{n}\otimes \ket{\psi_j} \, .
\end{equation}
After applying the QFT on the control register, the quantum state becomes
\begin{equation}
\label{eq:final_state_QPE}
    \sum_{j} \Phi_j \sum_{l=0}^{2\Nphas-1} (-1)^l e^{-(\phi_j-{\pi l}/{\Nphas})/2} \, \Gamma\left(\phi_j-\frac{\pi l}{\Nphas}\right) \ket{l}\otimes\ket{\psi_j},
\end{equation}
where $\Gamma(x)$ is a kernel function defined to be
\begin{equation}
\label{eq:define_kernel}
    \Gamma(x) = \frac{1}{\sqrt{2\Nphas}}\sum_{n=0}^{2\Nphas-1}e^{inx}\gamma_{n} \, .
\end{equation}
The phase factor does not affect the probabilities so can be ignored in the following discussion.

\subsubsection{The phase error}
\label{sec:phase_error}

From Eq.~\eqref{eq:final_state_QPE} the probability of getting a phase estimate $\pi l/N$ is
\begin{equation}
    \Pr\left[\hat{\phi}=\frac{\pi l}{\Nphas}\right]=\sum_{j} |\Phi_j|^2 \left|\Gamma\left(\phi_j-\frac{\pi l}{\Nphas}\right)\right|^2.
\end{equation}
Here $\hat{\phi}$ is the random variable corresponding to the phase estimate. Note that the above probability resembles a convolution between two probability measures. From this observation we can write down a decomposition for $\hat{\phi}$ in the following way:
\begin{equation}
\label{eq:phase_estimate_decompose}
    \hat{\phi} = \phi + \Delta \phi,
\end{equation}
where $\phi$ is a random variable satisfying
\begin{equation}
    \Pr[\phi=\phi_j] = |\Phi_j|^2,
\end{equation}
which means that $\phi$ is the output of an exact phase estimation for $W$
and $\Delta \phi$ is described by the following conditional distribution
\begin{equation}
\label{eq:dist_phase_err}
    \Pr\left[\Delta \phi = \frac{\pi l}{\Nphas}-\phi\Big|\phi\right] = \left|\Gamma\left(\phi-\frac{\pi l}{\Nphas}\right)\right|^2.
\end{equation}
We call $\Delta \phi$ the \emph{phase error}.

In QPE, we want the phase error $\Delta \phi$ to be small. One way to characterize this is through the variance of $\Delta \phi$, which is minimised by using amplitudes proportional to a cosine function \cite{cosine}.
The use of this control state for this application was considered in Ref.~\cite{babbush2018encoding}.
But oftentimes we want to make $\Delta \phi$ small with a probability that is arbitrarily close to $1$, thus giving us reliable estimates in many runs. More precisely, we want
\begin{equation}
\label{eq:target_precision_QPE}
    \Pr\big[|\Delta \phi|\geq \epsilon_{\phi}|\phi\big]\leq \delta \, ,
\end{equation}
where $|\Delta \phi|$ is calculated modulo $2\pi$.
The task of optimising the performance is described by the theory of window functions, where the optimal performance is provided by the Slepian prolate spheroidal window \cite{Slepian1965,Imai_2009}.
That is difficult to calculate, but can be approximated by the Kaiser window, which can be calculated by Bessel functions \cite{Kaiser,BerPRXQ24}.
Either of these can be used to obtain scaling $\Or(\epsilon_{\phi}^{-1}\ln(\delta^{-1}))$ with a small constant factor.

\subsubsection{Window functions}
\label{sec:windows}

In the theory of window functions, we would replace $\gamma_n$ with a continuous function $w(z)$, so
$w(n-\Nphas+1/2)=\gamma_n$.
Then the kernel function is approximately
\begin{equation}
    \Gamma(x) = \frac{1}{\sqrt{2\Nphas}}\sum_{n=0}^{2N-1}e^{inx}\gamma_{n} 
    \approx \frac{e^{i(\Nphas-1/2)x}}{\sqrt{2\Nphas}} \int_{-\Nphas}^{\Nphas} e^{izx} w(z) \, dz \, .
\end{equation}
In the phase estimation it is also trivial to adjust the measurement so that a continuous range of outcomes is obtained.
That is achieved by imposing an extra (known) phase shift in addition to $\phi$, and correcting for it in the estimate.
Then we can consider the error probability distribution as a continuous function of $x$ given by $|\Gamma(x)|^2$.
When $\Gamma(x)$ is approximated by the integral rather than the sum, then it is no longer periodic modulo $\pi$.
The integral gives a non-periodic function of $x$ over the whole real line.

The discrete case with $\gamma_n$ corresponds to sampling $w(z)$ at integer spacing at $2\Nphas$ points from $-(\Nphas-1/2)$ to $\Nphas-1/2$.
That corresponds to multiplying $w(z)$ by a comb function, so $\Gamma(x)$ is the periodic function obtained by convolving the Fourier transform of $w(z)$ with a comb function.
As described in Ref.~\cite{BerPRXQ24}, the tail probabilities for this continuous case correspond to the average over the tail probabilities for the discrete case, where the samples are shifted (so starting from $-(\Nphas-\nu)$ for $\nu\in[0,1]$).
This means that some discrete case must give at least the performance (in terms of small tail probabilities) as the continuous case.
A further advantage of using continuous window functions is that they can be scaled to a unit interval and analysed independently of the specific value of $N$.

\begin{figure}[tbh]
\begin{picture}(175mm,55.8mm)
\put(0,0){\includegraphics[width=8.5cm]{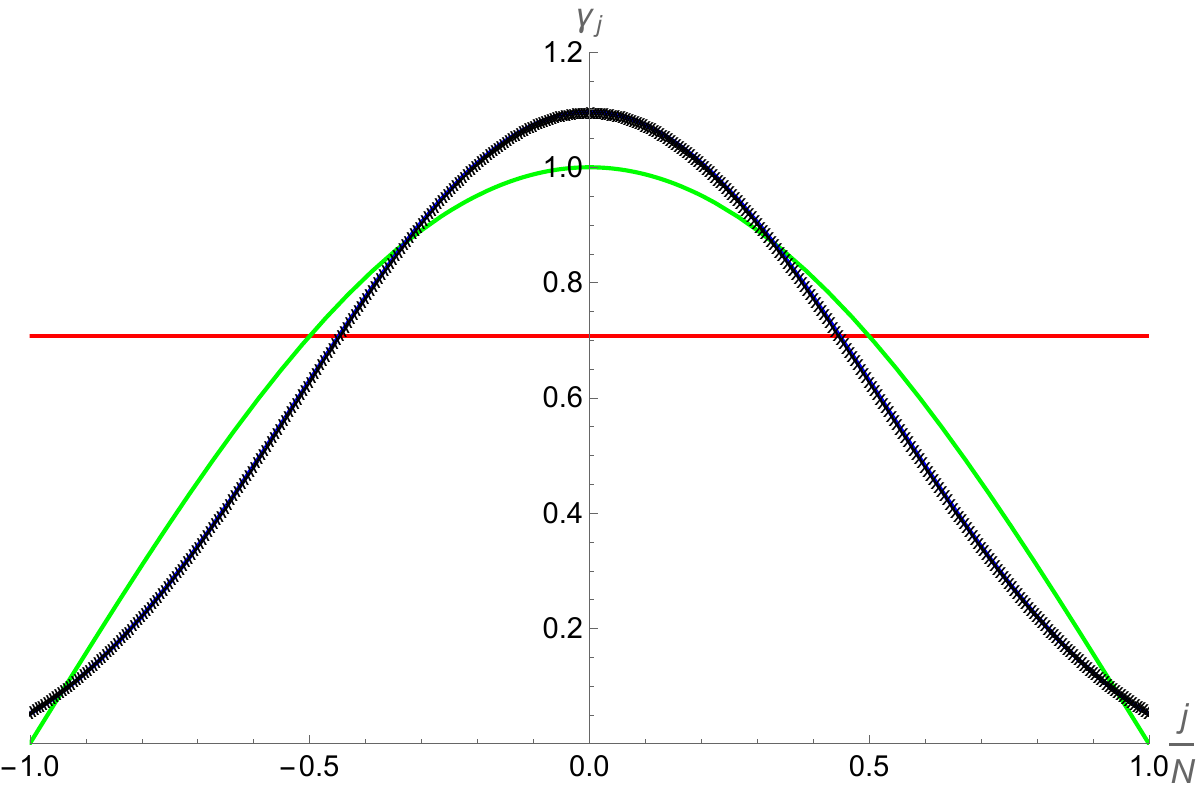}}
\put(90mm,0){\includegraphics[width=8.5cm]{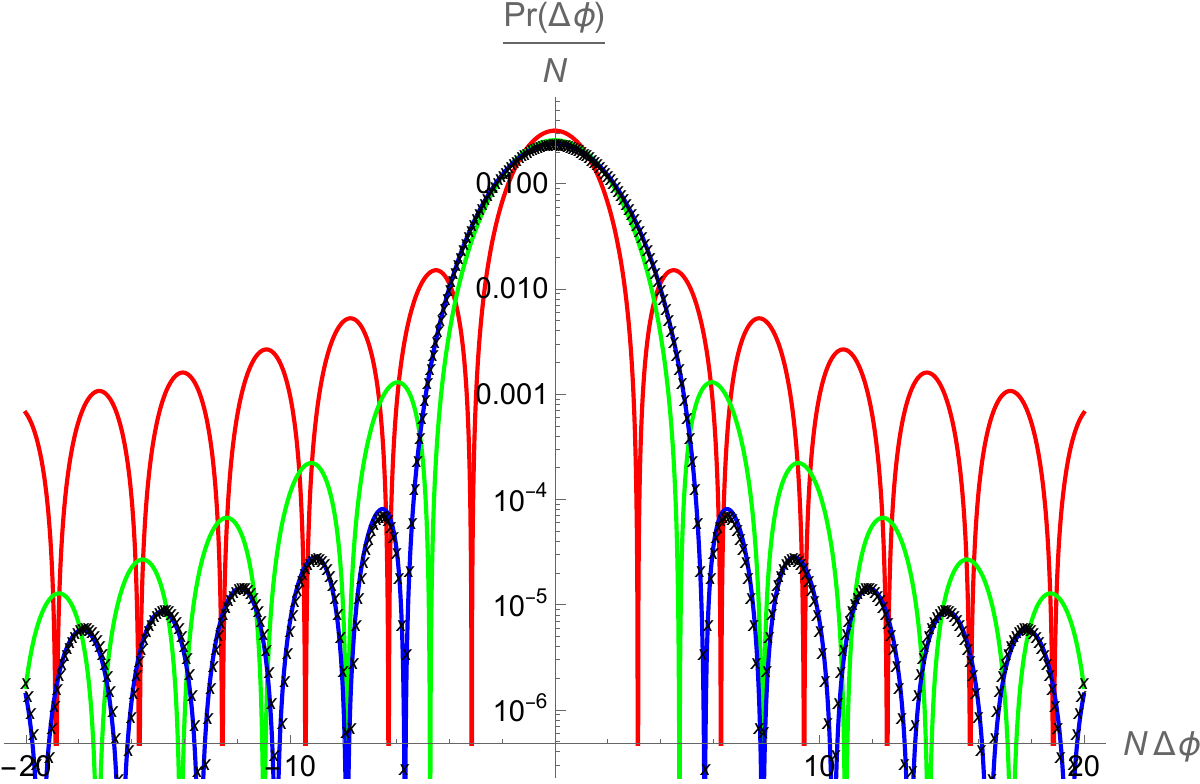}}
\put(0,50mm){(a)}
\put(90mm,50mm){(b)}
\end{picture}
\caption{The windows (a) and error probability distribution (b) for phase measurements.
In red is the flat window, and in green is the cosine window, which provides the minimum phase variance.
In blue is the Kaiser window with $\alpha\approx 1.5$, and the black crosses are the Slepian window with $c=5$.}
\label{fig:phase}
\end{figure}

Examples of the probability distributions for the error for a range of window functions are given in Fig.~\ref{fig:phase}.
The traditional textbook version \cite{nielsen00} of quantum phase estimation uses a flat distribution, which corresponds to a control register of unentangled qubits in $\ket{+}$ states.
That window gives a narrow peak for the error that decays slowly, resulting in both large variance and large tails.
The cosine window gives tails that decay more rapidly to yield excellent performance for the variance.
The Kaiser and Slepian windows give tails that are lower resulting in smaller tail probabilities for confidence intervals.
They do not decay as fast as the sine window, so the variance is larger.
Note that for this example the Kaiser and Slepian windows are almost indistinguishable.

\section{MPS preparation}
\label{sec:prep}

In this section we analyse the Toffoli complexity needed for preparation of matrix product states.
We first introduce a general unitary synthesis method that improves on the approach of LKS, then provide a method to generalise this approach to synthesising columns of a unitary, as is appropriate for MPS preparation.

\subsection{New unitary synthesis method}
Now we provide an alternative method of unitary synthesis that significantly improves on the method of LKS.
We consider the rectangular array of beam splitters for decomposing a multiport interferometer as in Ref.~\cite{Clements}.
This was previously considered for unitary synthesis in the optical context by Ref.~\cite{LopezPastor}, but we provide a significant improvement over that work.

\begin{figure}[tbh]
    (a) \hspace{99mm}  (b)\hspace{51mm}\phantom{.}
    \centering
    \includegraphics[height=4cm]{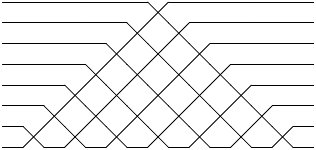}\hspace{2cm} \includegraphics[height=4cm]{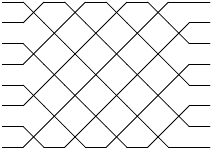}
    \caption{The decomposition of a multiport interferometer for 8 modes, equivalent to a transformation on 3 qubits.
    The triangular Reck-Zeilinger \cite{Reck} decomposition is in (a), and (b) is the rectangular decomposition from Ref.~\cite{Clements}.}
    \label{fig:interferometer}
\end{figure}

The form of the decomposition from Ref.~\cite{Clements} is shown in Fig.~\ref{fig:interferometer}(b) for the example of $\Nunit=8$.
An $M$-port interferometer is equivalent to an $\Nunit\times \Nunit$ unitary, and each layer of beam splitters is equivalent to a block-diagonal matrix with $2\times 2$ blocks.
The first layer of beam splitters corresponds to a block-diagonal matrix where the first block is in rows and columns 1 and 2.
The first layer of beam splitters in Fig.~\ref{fig:interferometer}(b) would correspond to a unitary matrix with nonzero entries shown by the asterisks in
\begin{equation}\label{eq:form1}
    \begin{bmatrix}
        * & * & &&&&& \\
        * & * & &&&&& \\
        && * & * &&&& \\
        && * & * &&&& \\
        &&&& * & * && \\
        &&&& * & * && \\
        &&&&&& * & * \\
        &&&&&& * & * \\
    \end{bmatrix} .
\end{equation}
The next layer has blocks shifted by one, so would correspond to a matrix with nonzero entries
\begin{equation}\label{eq:form2}
    \begin{bmatrix}
        * &&&&&&&   \\
        &* & * & &&&& \\
        &* & * & &&&& \\
        &&& * & * &&& \\
        &&& * & * &&& \\
        &&&&& * & * & \\
        &&&&& * & * & \\
        &&&&&&& * &  \\
    \end{bmatrix} .
\end{equation}
The layers alternate between operators of these two forms.
Because the second form in Eq.~\eqref{eq:form2} is equivalent to that in Eq.~\eqref{eq:form1} except shifted by 1, it can be transformed to the same form by increment and decrement operations.

\begin{figure}[tbh]
\centerline{
\Qcircuit @R=2.3em @C=2.3em {
&\qw\link{1}{1} &  & \qw        \\
&\qw\link{-1}{1}&              &\qw
} \qquad\raisebox{-4.5mm}{=}\qquad
\Qcircuit @R=1em @C=1em {
& \gate{\phi} &\qw\link{1}{2} & & & \gate{\theta} & \qw\link{1}{2} && & \gate{\varphi_0}& \qw       \\
& \qw &\qw\link{-1}{2}& & & \qw & \qw\link{-1}{2} & & & \gate{\varphi_1}&\qw
}}
    \caption{The crossings depicted in Fig.~\ref{fig:interferometer} are general beam splitters with arbitrary phases and reflectivities (left).
    They can be implemented with two 50/50 beam splitters with phase shifts (right).}
    \label{fig:phases}
\end{figure}

In the optical interferometer, each beam splitter may be expressed as two 50/50 beam splitters with a phase shift in between, as in Fig.~\ref{fig:phases}.
The 50/50 beam splitter corresponds to a Hadamard matrix.
That is equivalent to expressing a general qubit unitary as
\begin{equation}
    \begin{bmatrix}
    e^{i\varphi_0} & 0 \\
    0 & e^{i\varphi_1}
    \end{bmatrix}
    \begin{bmatrix}
    1/\sqrt 2 & 1/\sqrt 2 \\
    1/\sqrt 2 & -1/\sqrt 2
    \end{bmatrix}     \begin{bmatrix}
    e^{i\theta} & 0 \\
    0 & 1
    \end{bmatrix}
    \begin{bmatrix}
    1/\sqrt 2 & 1/\sqrt 2 \\
    1/\sqrt 2 & -1/\sqrt 2
    \end{bmatrix}    
    \begin{bmatrix}
    e^{i\phi} & 0 \\
    0 & 1
    \end{bmatrix} .
\end{equation}

Since the layers correspond to $2\times 2$ block-diagonal unitaries, this decomposition may be performed on the blocks individually, so for example Eq.~\eqref{eq:form1} is of the form (with $H$ the Hadamard)
\begin{equation}\label{eq:form3}\scalebox{0.75}{$
    \begin{bmatrix}
        e^{i\varphi_0} &  & &&&&& \\
         & e^{i\varphi_1} & &&&&& \\
        && e^{i\varphi_2} &  &&&& \\
        &&  & e^{i\varphi_3} &&&& \\
        &&&& e^{i\varphi_4} &  && \\
        &&&&  & e^{i\varphi_5} && \\
        &&&&&& e^{i\varphi_6} &   \\
        &&&&&&  & e^{i\varphi_7}  \\
    \end{bmatrix}
    \begin{bmatrix}
        H & & & \\
        & H & & \\
        & & H & \\
        & & & H  \\
    \end{bmatrix}
    \begin{bmatrix}
        e^{i\theta_0} &  & &&&&& \\
         & 1 & &&&&& \\
        && e^{i\theta_1} &  &&&& \\
        &&  & 1 &&&& \\
        &&&& e^{i\theta_2} &  && \\
        &&&&  & 1 && \\
        &&&&&& e^{i\theta_3} &   \\
        &&&&&&  & 1 \\
    \end{bmatrix}
        \begin{bmatrix}
        H & & & \\
        & H & & \\
        & & H & \\
        & & & H  \\
    \end{bmatrix}
    \begin{bmatrix}
        e^{i\phi_0} &  & &&&&& \\
         & 1 & &&&&& \\
        && e^{i\phi_1} &  &&&& \\
        &&  & 1 &&&& \\
        &&&& e^{i\phi_2} &  && \\
        &&&&  & 1 && \\
        &&&&&& e^{i\phi_3} &   \\
        &&&&&&  & 1  \\
    \end{bmatrix}$}.
\end{equation}
This can be further simplified, because the phases $\varphi_j$ can be combined with the next layer.
Then the next layer can be similarly decomposed, and its $\varphi_j$ phases can be combined with the layer after, and so on.
As a result, the phases on \emph{all} basis states are only needed for the very last operation.
As a result we have $2\Nunit$ layers of phase shifts with $\lfloor \Nunit/2\rfloor$ phases each, and one final layer with $\Nunit$ phases.
Then the block diagonal matrices have diagonals with all Hadamards, which can be achieved with just a Hadamard on a single qubit.

Note that the Toffoli cost of QROM is minimised if we are able to output more of the data together.
It is possible to output the data for two layers of phase shifts at once, so there are $\Nunit+1$ uses of QROM with output size $\Nunit$.
Considering the first layer of beam splitters, it is now decomposed into two layers of phase shifts that can be chosen to be only on odd (or even) modes, as well as two layers of 50/50 beam splitters.
The equivalent quantum circuit corresponds to two layers of phase shifts that only depend on the first $n-1$ qubits, and two layers of Hadamards on the last qubit, as shown in Fig.~\ref{fig:phaselayers}.

\begin{figure}[tbh]
\centerline{
\Qcircuit @R=2em @C=1em {
& \qw & \multigate{2}{\rotatebox{-90}{\rm ctrl}} & \qw & \multigate{2}{\rotatebox{-90}{\rm ctrl}} & \qw & \qw \\
& \qw & \ghost{\rotatebox{-90}{\rm ctrl}} & \qw & \ghost{\rotatebox{-90}{\rm ctrl}} &\qw &\qw \\
& \qw & \ghost{\rotatebox{-90}{\rm ctrl}} & \qw & \ghost{\rotatebox{-90}{\rm ctrl}} &\qw  &\qw\\
& \qw & \gate{R(\phi)}\qwx  & \gate{H} & \gate{R(\theta)}\qwx &\gate{H}&\qw
}}
\caption{\label{fig:phaselayers}Two layers of phases controlled by the first $n-1$ qubits and Hadamards on the last qubit.
The boxes labelled $R(\phi)$ and $R(\theta)$ are phase rotations controlled by the first $n-1$ qubits.}
\end{figure}
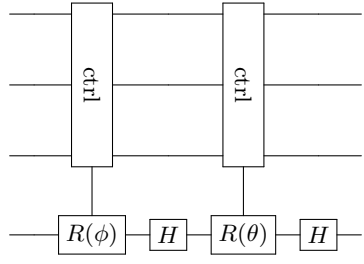

\begin{figure}[tbh]
\centerline{
\Qcircuit @R=2em @C=1em {
& \qw & \multigate{2}{\rotatebox{-90}{\rm QROM$_j$}} & \qw & \qw & \qw & \qw & \multigate{2}{\rotatebox{-90}{\rm QROM$_j^{-1}$}} & \qw \\
& \qw & \ghost{\rotatebox{-90}{\rm QROM$_j$}} & \qw & \qw & \qw & \qw & \ghost{\rotatebox{-90}{\rm QROM$_j^{-1}$}} &\qw \\
& \qw & \ghost{\rotatebox{-90}{\rm QROM$_j$}} & \qw & \qw & \qw & \qw & \ghost{\rotatebox{-90}{\rm QROM$_j^{-1}$}} &\qw \\
& \qw & \qw\qwx & \gate{R(\phi_j)} & \gate{H} & \gate{R(\theta_j)} & \gate{H}&\qw\qwx & \qw  \\
& \qw & \gate{\phi_j}\qwx & \ctrl{-1} & \qw & \qw & \qw & \gate{\phi_j}\qwx & \qw \\
& \qw & \gate{\theta_j}\qwx & \qw & \qw & \ctrl{-2} & \qw & \gate{\theta_j}\qwx & \qw \\
}}
\caption{\label{fig:qromphase}Two layers of phases obtained by using a single QROM.
The bottom two registers are temporary data registers used for the output of the QROM.
The QROM on the left outputs both $\phi_j$ and $\theta_j$, then the QROM on the right is an inverse QROM for erasure.
In the middle the rotations are controlled by the values in the data registers.}
\end{figure}
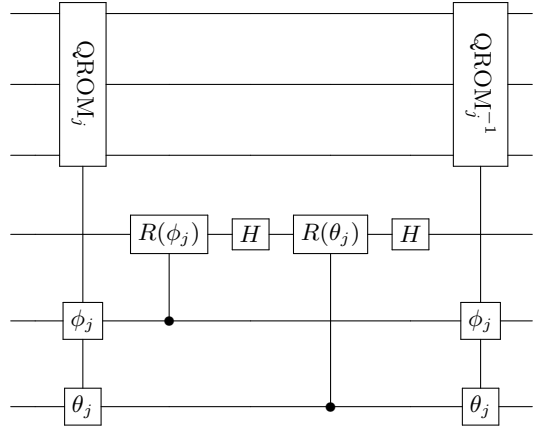

It is therefore possible to perform a QROM on the first $n-1$ qubits, perform the operations on qubit $n$, then erase the QROM because it does not depend on the last qubit.
The quantum circuit is shown in Fig.~\ref{fig:qromphase}.
A further simplification is possible because of the way QROM erasure is performed.
The data qubits are measured in the $X$ basis, and sign corrections need to be performed on the control qubits.
However, these sign corrections are phases that can be combined into $\varphi_j$.
That is, we just need to modify the phases used in the remainder of the circuit to account for the sign fixups needed for the QROM erasure.
The Toffoli complexity is therefore
 \begin{equation}
     \left\lceil \frac{\Nunit}{2\Lambda}\right\rceil + (\Lambda-1)2b -1 \,
 \end{equation}
to output the $\Nunit/2$ items of data of size $2b$ (for two rotations), with $\Lambda$ a power of 2.
In this expression $-1$ accounts for the fact that the cost in the first term is for unary iteration \cite{BerryQUA19}, and the Toffoli cost of unary iteration is 1 less than the number of items \cite{babbush2018encoding}.
This is a cost including a control, and this control wll be needed in the overall scheme so is allowed for here.
The cost for the two rotations is $2(b-2)$, where the $b-2$ cost for a phase rotation is explained in Ref.~\cite{SandersPRQ20}.
In the following we will bundle these two costs together.

A minor issue is the even layers where the blocks are shifted by 1.
This is easily accounted for in the quantum circuit by applying an increment in the computational basis to shift the blocks such that the layer may be implemented in the same way as for the first layer.
Each increment (or decrement) can be performed with $n-2$ Toffolis, and there are $\Nunit-1$ needed, for a total complexity of $(n-2)(\Nunit-1)$.
(Recall that the cost of modular addition is $n-1$ as per Ref.~\cite{Gidney2018halvingcostof}, with a saving of one Toffoli when the number to be added is classical, rather than provided in a quantum register.)
Bringing all these complexities together we have the following.
\begin{itemize}
    \item The complexity for the $2\Nunit$ layers of phase shifts and Hadamards is
    \begin{equation}
        \Nunit \left( \left\lceil \frac{\Nunit}{2\Lambda}\right\rceil + 2\Lambda b -5
      \right) \, .
    \end{equation}
    \item The increments and decrements have complexity $(n-2)(\Nunit-1)$.
    \item The final phase shifts have complexity
    \begin{equation}\label{eq:funalpha}
        \left\lceil \frac{\Nunit}{\Lambda}\right\rceil + \Lambda b
     +\left\lceil \frac{\Nunit}{\Lambda'}\right\rceil + \Lambda'  - 6 \, .
    \end{equation}
\end{itemize}
For Eq.~\eqref{eq:funalpha} there is a term $-6$ which comes from $-1$ for the initial QROM, $-2$ for the addition into the phase gradient, and $-3$ for the sign fixup.
The sign fixup is explained in Fig.~6 of Ref.~\cite{BerryQUA19}, and can be constructed from two unary iterations on subsets of the qubits.
Only one of these need be controlled, for an overall saving of $-3$ Toffolis.
In numerical testing of this approach against the LKS approach, we find that the complexity is reduced by about a factor of 7 over a wide range of parameters.

\subsection{Synthesis of columns of unitary}

We can now use this approach to reduce the cost for the case where it is only necessary to synthesise half the columns of the unitary.
The key to this simplification is that the unitary can be simplified by an initial unitary on dimension $\Nunit/2$ (ignoring the extra qubit), then afterward using the extra qubit to control one of two unitaries on the remaining qubits.
This then diagonalises the two $\Nunit/2\times \Nunit/2$ blocks of the unitary that we need.
That is, we have a decomposition of the unitary as
\begin{equation}
\begin{bmatrix}
    A & ? \\
B & ? 
\end{bmatrix} =
\begin{bmatrix}
    U_1 & 0 \\
    0 & U_2
\end{bmatrix}
\begin{bmatrix}
    D_1 & ? \\
D_2 & ? 
\end{bmatrix}
\begin{bmatrix}
    V & 0 \\
    0 & V
\end{bmatrix} .
\end{equation}
In this expression the question marks indicate blocks that we do not need to specify, so $A$ and $B$ indicate half of the columns of the unitary matrix that need to be correctly produced.
The blocks $D_1,D_2$ are diagonal, and $U_1,U_2,V$ are unitaries.
That is, we require
\begin{equation}
    A = U_1 D_1 V, \qquad B = U_2 D_2 V .
\end{equation}
The matrices $U_1, D_1, V$ are easily determined via a singular value decomposition of $A$.
Then $U_2,D_2$ can be determined from a QR decomposition of $BV^\dagger$.
The QR decomposition guarantees that $D_2$ is upper triangular, and the requirement that $D_1$ and $D_2$ are blocks of a unitary matrix ensures that $D_2$ is diagonal.

So, the procedure to apply the unitary that we need is as follows.
\begin{enumerate}
    \item Perform a unitary of dimension $\Nunit/2\times \Nunit/2$ on the first $n-1$ qubits.
    \item Use $n-1$ qubits to control rotation on the remaining qubit.
    \item Use that qubit to control unitaries on the $n-1$ qubits.
\end{enumerate}
For the costing of this procedure, the cost of the first step is the same as in the formulae for unitary synthesis above, except replacing $\Nunit$ with $\Nunit/2$ and $n$ with $n-1$.

\begin{figure}[tbh]
\centerline{
\Qcircuit @R=2em @C=1em {
& \qw & \gate{V} & \multigate{1}{D_{1,2}} & \qw & \qw \\
& \lstick{\ket{0}} & \qw & \ghost{D_{1,2}} & \gate{U_{1,2}} & \gate{V'} & \multigate{1}{D'_{1,2}} & \qw & \qw  \\
& & & & \lstick{\ket{0}} & \qw & \ghost{D'_{1,2}} & \gate{U'_{1,2}} & \qw  \\
}}
\caption{\label{fig:mpsprep}Two consecutive steps in the MPS state preparation, showing how the unitaries required for consecutive steps can be combined.
The primes show the unitaries needed for the second step.
The input registers with $\ket{0}$ are qubits, and the top two outputs are qubits.
}
\end{figure}
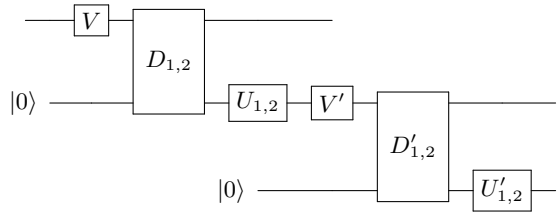

However, in the MPS preparation we have a unitary on these $n-1$ qubits first, and this unitary can be combined with that one so it need not be performed and requires no additional complexity.
This principle is illustrated in Fig.~\ref{fig:mpsprep}, where it can be seen that the dimension $\Nunit/2$ register where $U_1$ and $U_2$ are performed is the same as that where $V'$ is applied for the next step.

Step 2 is just a QROM and rotation, for complexity
\begin{equation}
     \left\lceil \frac{\Nunit}{2\Lambda}\right\rceil + \Lambda b  - 2\, .
\end{equation}
The $-2$ term is because the rotation is implemented by a controlled addition or subtraction on half the phase, which requires one more bit for the addition into the phase gradient register.
The QROM erasure does not introduce a Toffoli cost here, because the sign fixup can be incorporated into the unitaries $U_1$ and $U_2$.
A subtlety in this costing is that the blocks are distinguished by the \emph{least} significant qubit, so we only need consider dimension $\Nunit/2$ for the sub-blocks.
For simplicity we are assuming $\Nunit$ is even for this discussion.

For the controlled unitaries on dimension $\Nunit/2$, in each layer we have the same size of the QROM as for the operations on the full dimension $\Nunit$, with the only difference being that we now only need $\Nunit/2$ layers rather than $\Nunit$.
Also, the increments and decrements are on $n-1$ qubits rather than $n$, slightly decreasing the complexity.
The final phase shifts also have the same cost as before.
The complete cost is therefore
\begin{align}
    & \left\lceil \frac{\Nunit}{2\Lambda}\right\rceil + \Lambda b 
      -2 \qquad &{\rm (controlled~qubit~rotation)}\\
& +\frac \Nunit 2 \left( \left\lceil \frac{\Nunit}{2\Lambda}\right\rceil + 2\Lambda b -5
      \right) \qquad &{\rm (phasing~layers)}\\
     & + (n-3)(\Nunit/2-1)  \qquad &{\rm (increments~and~decrements)} \\
&+\left\lceil \frac{\Nunit}{\Lambda}\right\rceil + \Lambda b
     +\left\lceil \frac{\Nunit}{\Lambda'}\right\rceil + \Lambda' -6 \, . \qquad &{\rm (final~phase~shifts)}
\end{align}
This gives a similar factor of $\sim 7$ improvement over LKS as for synthesis of the complete unitary.

So far we have discussed the case where the dimension on each site is $d=2$ for the MPS.
We can solve the case for $d>2$ by iterating the procedure, and the cost is multiplied by $d-1$.
To explain the principle, we first explain how to obtain the result for $d=3$.
Then we can construct a decomposition of the form
\begin{equation}
\begin{bmatrix}
    A & ? & ? \\
B & ? & ? \\
C & ? & ?
\end{bmatrix} =
\begin{bmatrix}
    I & 0 & 0 \\
    0 & U_{11} & U_{12} \\
    0 & U_{21} & U_{22} \\
\end{bmatrix}
\begin{bmatrix}
    A & ? & ? \\
R_{11} & R_{12} & ? \\
0 & R_{22} & ?
\end{bmatrix} .
\end{equation}
In this expression, $A,B,C$ are the blocks of the unitary we wish to construct, $U_{ij}$ are blocks of a unitary, and $R_{ij}$ are blocks of an upper triangular matrix.
The block $R_{21}$ must be zero because of the upper triangular form, so is just given as zero above.
In this form, $U_{ij}$ and $R_{ij}$ are obtained by a QR decomposition of the 4 blocks of the desired unitary operation including $B,C$ and two unspecified blocks indicated by the question marks.
It doesn't matter that the blocks indicated by question marks are unknown, because they can just be replaced with zero for the purposes of the QR decomposition.
We only need to determine $R_{11}$ and $U_{11},U_{21}$, because the other blocks do not affect $A,B,C$, and those blocks are obtained correctly by the QR decomposition with question marks replaced with zero.

Now it is possible to correctly apply the blocks $U_{11}$ and $U_{21}$ of the first unitary by the above procedure for synthesising half of the columns of a unitary.
Similarly, $A$ and $R_{11}$ have orthonormal columns, so can be regarded as half the columns of some unitary, which can also be synthesised by the above scheme.
Moreover, the construction of the unitary with blocks $U_{11},U_{21}$ requires an initial unitary of dimension $\Nunit/3$.
This unitary can be combined with $R_{11}$ and so need not add to the cost of synthesising $U_{11},U_{21}$.
Similarly, if this unitary synthesis is part of MPS preparation, the initial dimension $\Nunit/3$ unitary for synthesis of $A,R_{11}$ can be combined with other operations and does not add to the cost.

In the general case, we need to construct the correct first $\bonda$ columns of a matrix of size $\Nunit=d\bonda$.
Consider the block consisting of the last $(d-1)\bonda$ rows and first $(d-1)\bonda$ columns (equivalent to $B,C$ and the question marks in the $d=3$ example above).
Apply a QR decomposition, to express it in the form of a unitary operation followed by an upper triangular matrix.
Similar to the example above, the question marks may be replaced with zero in this decomposition, because they only affect blocks that we do not need to specify.

In exactly the same way as in the example above, we need only correctly synthesise the first $\bonda$ columns of the first operation, and the upper-triangular form guarantees that the only non-zero blocks are $A$ and an upper triangular $\bonda\times\bonda$ block (which is $R_{11}$ in the example above).
These blocks may be synthesised by the above procedure for $d=2$, where we are synthesising half the columns of a unitary.

The unitary which we obtained by the QR decomposition is of size $(d-1)\bonda \times (d-1)\bonda$, and we only need to correctly reproduce the first $\bonda$ columns.
In the example above, these were $U_{11},U_{21}$.
Therefore we have reduced the problem to synthesising the first $\bonda$ columns of a $(d-1)\bonda \times (d-1)\bonda$ unitary, which is the same as the initial problem, with $d$ reduced by 1.
We may therefore iterate this procedure to completely reduce the problem to that for $d=2$.
Therefore, we are able to reduce the problem to $d-1$ applications of the scheme for $d=2$.

There is a very small increase in the cost for $d>4$.
Because the schemes for $d=2$ are on a subspace, the QROMs need to be controlled, which is a cost accounted for above.
However, for $d>4$ we also need Toffolis to produce the qubit flagging the control for the QROMs.
That is only performed once for each application of the $d=2$ scheme, so is a negligible contribution to the overall cost.

We can also combine the $d=2$ schemes in a slightly more efficient way.
First, note that at each step there is a controlled unitary performed on two dimension $\bonda$ subspaces, but only one is used in the next $d=2$ scheme.
Instead of performing both, we can instead just perform one unitary of dimension $\bonda\times \bonda$, and then perform controlled unitaries on the $d$ subspaces at the end.
Then the costs of $d-1$ controlled $U_1,U_2$ operations are replaced with $d-2$ unitaries of dimension $\bonda\times \bonda$, followed by selection between $d$ unitaries of this size.

In either case the increase in cost with $d$ is linear in $d$, whereas for the LKS approach the factor is about $\sqrt{d/2}$, so LKS has better scaling with $d$.
For $d=4$ we find that the improvement over LKS is about a factor of $3.5$, and the crossover where LKS is more efficient is for large values of $d$ about 30.

It is also possible to prepare states more efficiently than the procedure of LKS.
The method is to use interspersed layers of Hadamards and diagonal phasing operators, and we find that three phasing layers are sufficient.
That is, prepare an equal superposition state, apply phases in the computational basis, a layer of Hadamards, then more phases.
The initial two layers of phases and Hadamards produce the correct amplitudes.
The phases may be found efficiently by a simple generalisation of the Gerchberg-Saxton algorithm; we have tested this with dimension up to $2^{20}$.
This approach may also be used in the MPS preparation to slightly reduce the cost.
Increasing the number of layers can also be used for unitary synthesis, but solving for the phases becomes computationally intractable for larger dimensions (above about 128).

\section{Phase estimation with window functions}
\label{sec:phase_estimation_with_window_functions}

First we describe how to perform phase estimation that is optimal for confidence intervals.
That is, for a given number of controlled applications of an operator and confidence level, it gives the smallest width of the confidence interval.
This problem is related to that of window functions in classical signal processing theory, and optimal confidence intervals are given by the prolate spheroidal window.
The analysis of the prolate spheroidal window was given in Ref.~\cite{Imai_2009}, but it is difficult to calculate, so Ref.~\cite{BerPRXQ24} gives an analysis of the Kaiser window, which gives near optimal results.
The use of the Kaiser window was also mentioned on page 35 of Ref.~\cite{SandersPRQ20}.
The error for the Kaiser window is in terms of the sinc function, so can be calculated with standard mathematical software.
The prolate spheroidal window can also be calculated using specialist mathematical software.

We will provide both asymptotic results for the cost of phase estimation using these window functions, as well as exact expressions using special functions and numerical results calculated from these functions.
For the Kaiser window, we provide first-order approximations for the error and cost in Eq.~\eqref{eq:erap1} and Eq.~\eqref{eq:kaiser_first_order}, respectively, and higher-order approximations for the error in Eq.~\eqref{eq:errorser}, and cost in Eq.~\eqref{eq:asympNKai}.
For the prolate spheroidal window, we provide a higher-order approximation of the error in Eq.~\eqref{eq:slepnum}, and the cost in Eq.~\eqref{eq:slepian_window_higher_order}.
We show that, although the Slepian window provides asymptotically improved error, the cost is the same up to leading order.

\subsection{The Kaiser window}
\label{sec:Kaiser}

First, we summarise the Kaiser window and its asymptotic scaling.
The standard form of the Kaiser window is proportional to
\begin{equation}
    w(x) = I_0\left( \pi\alpha\sqrt{1-(x/\Nphas)^2}\right) \, ,
\end{equation}
for $|x|\le \Nphas$, and 0 otherwise.
As discussed in Section \ref{sec:windows}, the control state used would correspond to samples of this continuous window at discrete points.
The window function yields a probability distribution for the error $\theta$ proportional to
\begin{equation}
    \frac{\sin^2\left( \sqrt{(\Nphas\theta)^2-(\pi\alpha)^2} \right)}{{(\Nphas\theta)^2-(\pi\alpha)^2}}.
\end{equation}
That is the square of the Fourier transform of $w(x)$.

A simple approximation for the tail probabilities with the Kaiser window is given in Ref.~\cite{BerPRXQ24}.
The method used there is to first approximate the normalisation by approximating the centre of the distribution by a Gaussian.
In that approximation, the integral over $\theta$ then gives
\begin{align}
\int_{-\infty}^\infty \frac{\sinh^2(\pi\alpha)}{\pi^2\alpha^2} e^{-\Nphas^2(\pi\alpha\coth(\pi\alpha)-1)\theta^2/(\pi^2\alpha^2)} d\theta &=
    \frac{\sinh^2(\pi\alpha)}{\Nphas \sqrt\pi\alpha\sqrt{\pi\alpha\coth(\pi\alpha)-1}} \nn 
    &\approx \frac {e^{2\pi\alpha}}{4 \Nphas\sqrt\pi \alpha \sqrt{\pi\alpha-1}}\nn 
    &\approx \frac {e^{2\pi\alpha}}{4 \Nphas\pi \alpha^{3/2} } \, , \label{eq:norm1}
\end{align}
where in the second line we have used $\sinh^2(\pi\alpha)\approx e^{2\pi\alpha}/4$ and $\coth(\pi\alpha)\approx 1$, and in the last line we have made an approximation for large $\alpha$.

Then the tail probabilities can be approximated by replacing $\sin^2$ with $1/2$, and integrating from the first zero at $\theta=(\pi/\Nphas)\sqrt{1+\alpha^2}$ to give
\begin{equation}
    \int_{(\pi/\Nphas)\sqrt{1+\alpha^2}}^\infty \frac{1}{[\Nphas^2\theta^2-(\pi\alpha)^2]} d\theta = \frac{{\rm arcsinh}(\alpha)}{\pi \Nphas \alpha}
    \approx \frac{\ln(2\alpha)}{\pi \Nphas \alpha} \, ,
\end{equation}
using ${\rm arcsinh}(\alpha)\approx \ln(2\alpha)$.
Dividing by the approximation for the normalisation in Eq.~\eqref{eq:norm1} then gives
\begin{equation}\label{eq:erap1}
    \delta \approx 4\ln(2\alpha) \sqrt{\alpha} \, e^{-2\pi\alpha}.
\end{equation}
This is a factor of 2 smaller than the expression in Ref.~\cite{BerPRXQ24}, because we have approximated $\sin^2$ with $1/2$.

Therefore, to obtain confidence level $1-\delta$ (so the probability of error outside the range is $\delta$), we take
\begin{equation}
    \ln (1/\delta) \approx 2\pi\alpha - \ln[4\ln(2\alpha) \sqrt{\alpha}].
\end{equation}
Solving for $\alpha$ then gives the approximation
\begin{equation}\label{eq:alaprx1}
    \alpha \approx \frac{1}{2\pi}\ln (1/\delta) +  \frac{1}{4\pi}\ln(8\ln (4/\delta)/\pi) + \frac 1{2\pi} \ln\left[ \ln(\ln(4/\delta)/\pi) \right]\, .
\end{equation}
The size of the confidence interval is $(\pi/N)\sqrt{1+\alpha^2}$, so if that needs to be $\epsilon$, we should take
\begin{equation}\label{eq:kaiser_first_order}
    \Nphas = \frac{\pi}{\epsilon}\sqrt{1+\alpha^2} = \frac{1}{2\epsilon} \ln(1/\delta) + \mathcal{O}(\epsilon^{-1} \ln \ln (1/\delta)).
\end{equation}
The higher-order $\ln \ln$ term for $\alpha$ is larger than the correction term for approximating $\sqrt{1+\alpha^2}$ by $\alpha$.

\subsection{Higher-order terms}
\label{sec:higher}

It is possible to obtain more accurate approximations by taking advantage of special functions.
First the integral can be given as, using the Plancherel theorem,
\begin{equation}
    \int \frac{\sin^2\left( \sqrt{\Nphas^2\theta^2-(\pi\alpha)^2} \right)}{\Nphas^2\theta^2-(\pi\alpha)^2} d\theta
    = \frac {\pi}{2\Nphas} \int_{-1}^{1} I_0^2 \left( \pi \alpha \sqrt{1-x^2} \right) dx \, .
\end{equation}
Ignoring the factor of $1/N$ for simplicity, this integral can be evaluated as
\begin{align}
    \frac {\pi}{2}\int_{-1}^{1} I_0^2 \left( \pi \alpha \sqrt{1-x^2} \right) dx = 
    \pi\int_{0}^{1} I_0^2 \left( \pi \alpha y \right) \frac {y \, dy}{\sqrt{1-y^2}} = 
     \frac {\pi}{2} [ I_0(2\pi\alpha) (2+\pi L_1(2\pi\alpha)) - \pi I_1(2\pi\alpha)L_0(2\pi\alpha)] ,
\end{align}
where $L_0,L_1$ are modified Struve functions.
Expanding about $\alpha=\infty$ then gives
\begin{equation}\label{eq:normser}
\frac{e^{2\pi\alpha}}{4\pi \alpha^{3/2}}
    \left(
    1 
     + \frac{5}{2^4 \pi \alpha}
 + \frac{129}{2^9 \pi^2 \alpha^2}
 + \frac{2655}{2^{13} \pi^3 \alpha^3}
 + \frac{301035}{2^{19} \pi^4 \alpha^4} 
    + \frac{10896795}{2^{23} \pi^5 \alpha^5}
    + \frac{961319205}{2^{29} \pi^6 \alpha^6} + \mathcal{O}(\alpha^{-7})
    \right) \, .
\end{equation}

For the integral over the tails, we can use
\begin{align}
    2\int_{(\pi/\Nphas)\sqrt{1+\alpha^2}}^\infty \frac{\sin^2\sqrt{\Nphas^2\theta^2-\pi^2\alpha^2}}{\Nphas^2\theta^2-\pi^2\alpha^2} d\theta
    &= \frac{2}{\Nphas}\int_{\pi\sqrt{1+\alpha^2}}^\infty \frac{\sin^2\sqrt{\theta^2-\pi^2\alpha^2}}{\theta^2-\pi^2\alpha^2} d\theta \nn
    &= \frac{2}{\Nphas}\int_{\pi}^\infty \frac{\sin^2 x}{x\sqrt{x^2+\pi^2\alpha^2}} dx \nn
    &= \frac{1}{\Nphas}\int_{\pi}^\infty \frac{1-\cos 2x}{x\sqrt{x^2+\pi^2\alpha^2}} dx \nn
    &= \frac{\,{\rm arcsinh}(\alpha)}{\pi \Nphas \alpha} - \frac{1}{\Nphas}\int_{\pi}^\infty \frac{\cos 2x}{x\sqrt{x^2+\pi^2\alpha^2}} dx \, .
\end{align}
For the remaining integral, we can use integration by parts to give, where ${\rm Ci}$ is the cosine integral,
\begin{align}\label{eq:intbypart1}
    -\int_{\pi}^\infty \frac{\cos 2x}{x\sqrt{x^2+\pi^2\alpha^2}} dx 
    &= \frac{{\rm Ci}(2\pi)}{\pi \sqrt{1+\alpha^2}} - \int_\pi^\infty \frac{x\,{\rm Ci}(2x)}{(x^2+\pi^2\alpha^2)^{3/2}} dx \nn
    &= \frac{{\rm Ci}(2\pi)}{\pi \sqrt{1+\alpha^2}} + \frac{4\pi^2\,{\rm Ci}(2\pi)-1}{8\pi^3(1+\alpha^2)^{3/2}} + \int_\pi^\infty \frac{3x[\cos(2x)+2x\sin(2x)-4x^2\,{\rm Ci}(2x)]}{8(x^2+\pi^2\alpha^2)^{5/2}} dx \, .
\end{align}
Note that the final integral here is of order $\alpha^{-5}$, and repeating the integration by parts makes the remaining integral higher and higher order in $\alpha$.
We give further details of the integration by parts in Appendix \ref{app:higher}.
Dividing by the asymptotic expression above for the normalisation factor in Eq.~\eqref{eq:normser} gives, with $C_\alpha=\ln(2\alpha)+{\rm Ci}(2\pi)$,
\begin{align}\label{eq:errorser}
   \delta =  4C_\alpha \sqrt\alpha e^{-2\pi\alpha} \left[ 1 - \frac{5 }{16 \pi \alpha}
  + \mathcal{O}(\alpha^{-2})
    \right] .
\end{align}
The leading-order term here only differs from what we obtained in Eq.~\eqref{eq:erap1} in that it has $C_\alpha$ with ${\rm Ci}(2\pi)$ in addition to $\ln(2\alpha)$.

Given error $\delta$, expanding in a series solution for $\alpha$ yields
\begin{equation}
    \alpha = \frac{\ln(1/\delta)}{2\pi} + \frac{\ln(8\ln(4/\delta)/\pi)}{4\pi} + \frac 1{2\pi} \ln\left[ {\rm Ci}(2 \pi) + \ln(\ln(4/\delta)/\pi) \right] + \mathcal{O} \left( \frac{\ln\ln(1/\delta)}{\ln(1/\delta)} \right) .
\end{equation}
The leading-order terms here only differ from Eq.~\eqref{eq:alaprx1} in that there is an extra ${\rm Ci}(2 \pi)$ in the third term, so the previous approximation given can be expected to be accurate.
The expression for $\Nphas$ is then
\begin{equation}\label{eq:asympNKai}
    \Nphas = \frac{\pi}{\epsilon}\sqrt{1+\alpha^2} = \frac{\ln(1/\delta)}{2\epsilon} + \frac{\ln(8\ln(4/\delta)/\pi)}{4\epsilon} + \frac 1{2\epsilon} \ln\left[ {\rm Ci}(2 \pi) + \ln(\ln(4/\delta)/\pi) \right] + \mathcal{O} \left( \frac{\ln\ln(1/\delta)}{\epsilon\ln(1/\delta)} \right) .
\end{equation}
To this order of approximation $\sqrt{1+\alpha^2}\approx\alpha$, and the correction is in the order term above.

The performance of the Kaiser window can be improved by adjusting the width from $\pi\sqrt{1+\alpha^2}$ to $\pi\sqrt{\Delta^2+\alpha^2}$ for some general $\Delta\ne 1$.
Then the integral over the tails can be adjusted to
\begin{align}
    2\int_{(\pi/\Nphas)\sqrt{\Delta^2+\alpha^2}}^\infty \frac{\sin^2\sqrt{\Nphas^2\phi^2-\pi^2\alpha^2}}{(\Nphas^2\phi^2-\pi^2\alpha^2)} d\phi
    &= \frac{\,{\rm arcsinh}(\alpha/\Delta)}{\pi \Nphas \alpha} - \frac{1}{\Nphas}\int_{\pi\Delta}^\infty \frac{\cos 2x}{x\sqrt{x^2+\pi^2\alpha^2}} dx \, .
\end{align}
For the remaining integral,
\begin{align}
    &-\int_{\pi\Delta}^\infty \frac{\cos 2x}{x\sqrt{x^2+\pi^2\alpha^2}} dx \nn
    &= \frac{{\rm Ci}(2\pi\Delta)}{\pi \sqrt{\Delta^2+\alpha^2}} - \int_{\pi\Delta}^\infty \frac{x\,{\rm Ci}(2x)}{(x^2+\pi^2\alpha^2)^{3/2}} dx \nn
    &= \frac{{\rm Ci}(2\pi\Delta)}{\pi \sqrt{\Delta^2+\alpha^2}} + \frac{4\pi^2\Delta^2\,{\rm Ci}(2\pi)-\cos(2\pi\Delta)-2\pi\Delta\sin(2\pi\Delta)}{8\pi^3(\Delta^2+\alpha^2)^{3/2}} + \int_{\pi\Delta}^\infty \frac{3x[\cos(2x)+2x\sin(2x)-4x^2\,{\rm Ci}(2x)]}{8(x^2+\pi^2\alpha^2)^{5/2}} dx \, ,
\end{align}
with further details given in Appendix \ref{app:higher}.
Dividing by the normalisation factor in Eq.~\eqref{eq:normser} and using $C_{\alpha,\Delta}=\ln(2\alpha/\Delta)+{\rm Ci}(2\pi\Delta)$, we obtain
\begin{align}\label{eq:errorser2}
   \delta =  4C_{\alpha,\Delta} \sqrt\alpha e^{-2\pi\alpha} \left[ 1 - \frac{5 }{16 \pi \alpha}
  + \mathcal{O}(\alpha^{-2})
    \right] ,
\end{align}
which is the same as in Eq.~\eqref{eq:errorser} except for the more general expression for $C_{\alpha,\Delta}$ and the higher-order terms not shown here but given in Appendix \ref{app:higher}.

In practice we would consider a confidence interval of fixed width $c=\pi\sqrt{\Delta^2+\alpha^2}$ and aim to minimise the cost.
If we expand in a series in $c$, then we obtain
\begin{align}
    \delta = 4 C_{c,\Delta} \sqrt{c/\pi} e^{-2c} \left[ 
    1 - \frac{(5-16\pi^2\Delta^2)}{16 c}+ \mathcal{O}(c^{-2}) 
    \right] \, ,
\end{align}
where $C_{c,\Delta}=\ln(2c/\pi\Delta)+{\rm Ci}(2\pi\Delta)$.
Now $C_{c,\Delta}$ takes its minimum value at $\Delta=1$, but the second term in square brackets above increases with $\Delta$.
This implies that for $c$ not too large, the optimal value of $\Delta$ will be less than $1$, but in the limit of large $c$ the optimal value of $\Delta$ approaches 1.
That is indeed what is found numerically.
This also implies that the first terms for $N$ in terms of $\epsilon$ in Eq.~\eqref{eq:asympNKai} are appropriate even when optimising $\Delta$.
In fact, optimising $\Delta$ would make the third term \emph{larger}, with the improvement only in the higher-order terms.
This is because the third term is equivalent to $\ln(C_{c,\Delta})/2\epsilon$.
It is more accurate to continue using the third term as given, since the true value of $N$ is reduced when optimising $\Delta$, and using a larger value of $C_{c,\Delta}$ in this term would make the estimate of $N$ larger.

\subsection{The prolate spheroidal window}
\label{sec:prolate_spheroidal_window}

For comparison with the optimal window,
the error is given in Eq.~(13) of \cite{Imai_2009} as
\begin{equation}\label{eq:slepy}
    \delta = 4\sqrt{\pi c}\, e^{-2c} \left[ 1-\frac 3{32c} +\mathcal{O}(c^{-2})\right] \, .
\end{equation}
That result is from Eq. (1.36) of Ref.~\cite{Slepian1965}, with $n=0$.
In fact, Eq.~(4.4) of that work gives
\begin{equation}\label{eq:slepy2}
    \delta = 4\sqrt{\pi c}\, e^{-2c} \left[ 1-\frac 3{32c} -
    \frac{389}{2^{11}c^2}+
    \mathcal{O}(c^{-3})\right] \, ,
\end{equation}
where we have used $n=0$ and simplified $2334/3!=389$ from the expression in that work.
It turns out that this expression is not correct.
By repeating the derivation as given on page 138 of Ref.~\cite{Slepian1965}, we obtain
\begin{equation}\label{eq:slepnum}
    \delta = 4\sqrt{\pi c} \, e^{-2c} \left[ 1-\frac 7{16c} -
    \frac{91}{2^{9}c^2}+
    \mathcal{O}(c^{-3})\right] \, .
\end{equation}
See Appendix \ref{app:slepian} for the details of this derivation.

In this expression the factor of $\ln c$ that was obtained for the Kaiser window error has been eliminated, so this error has asymptotically better scaling.
According to the above analysis for the asymptotic expansion for $\Nphas$, we would get
\begin{equation}
\label{eq:slepian_window_higher_order}
    \Nphas = \frac{c}{\epsilon} = \frac{\ln(1/\delta)}{2\epsilon} + \frac{\ln(8\pi\ln(4\sqrt\pi/\delta))}{4\epsilon}+  \mathcal{O} \left( \frac{\ln(\ln(1/\delta))}{\epsilon\ln(1/\delta)} \right) .
\end{equation}
That is, despite the optimal window giving asymptotically smaller error, the cost $\Nphas$ is reduced by only removing the third term in Eq.~\eqref{eq:asympNKai}.
That term is triple-logarithmic in $1/\delta$, meaning that the performance is only marginally improved.

The optimal window is the
angular spheroidal function of the first kind
$\text{PS}_{0,0}(c ,z)$ for $z\in [-1,1]$.
Integrating then gives (see Appendix \ref{app:slepian} for explanation of why this integral is used)
\begin{align}\label{eq:intps}
    (1-\delta)\text{PS}_{0,0}(c ,0) &=
    \frac c{\pi} \int_{-1}^1 \text{sinc}(c z) \, \text{PS}_{0,0}(c ,z) \, dz \nn
    &= \frac {2c}{\pi}  \, 
    \text{PS}_{0,0}(c ,0) \,
    [S_{0,0}^1(c , 1)]^2 \, ,
\end{align}
so
\begin{equation}\label{eq:exactslep}
    \delta = 1-\frac {2c}{\pi} \, [S_{0,0}^1(c , 1)]^2 \, ,
\end{equation}
where the function $S_{0,0}^1(c ,1)$
is the radial spheroidal function of the first kind.
Thus it is possible to compute the error in terms of special functions.
Various methods are discussed in Appendix \ref{app:calculate}.

In comparison, the value of $\Nphas$ for root-mean-square (RMS) error $\epsilon$ is \cite{babbush2018encoding}
\begin{equation}
     \Nphas \approx \frac{\pi}{2\epsilon} \, .
\end{equation}
That is, to leading order the expression for $\Nphas$ for the confidence interval replaces $\pi$ with $\ln(1/\delta)$.
For a 95\% confidence interval, for example, $\ln(1/\delta)$ is less than $\pi$, but calculation using the exact expression in Eq.~\eqref{eq:exactslep} shows that the complexity is about 63\% larger than for achieving RMS error $\epsilon$.
For a 90\% confidence interval the complexity is only about 35\% larger than for achieving RMS error $\epsilon$.

\begin{figure}[tbh]
    \centering
    \includegraphics[width=0.5\textwidth]{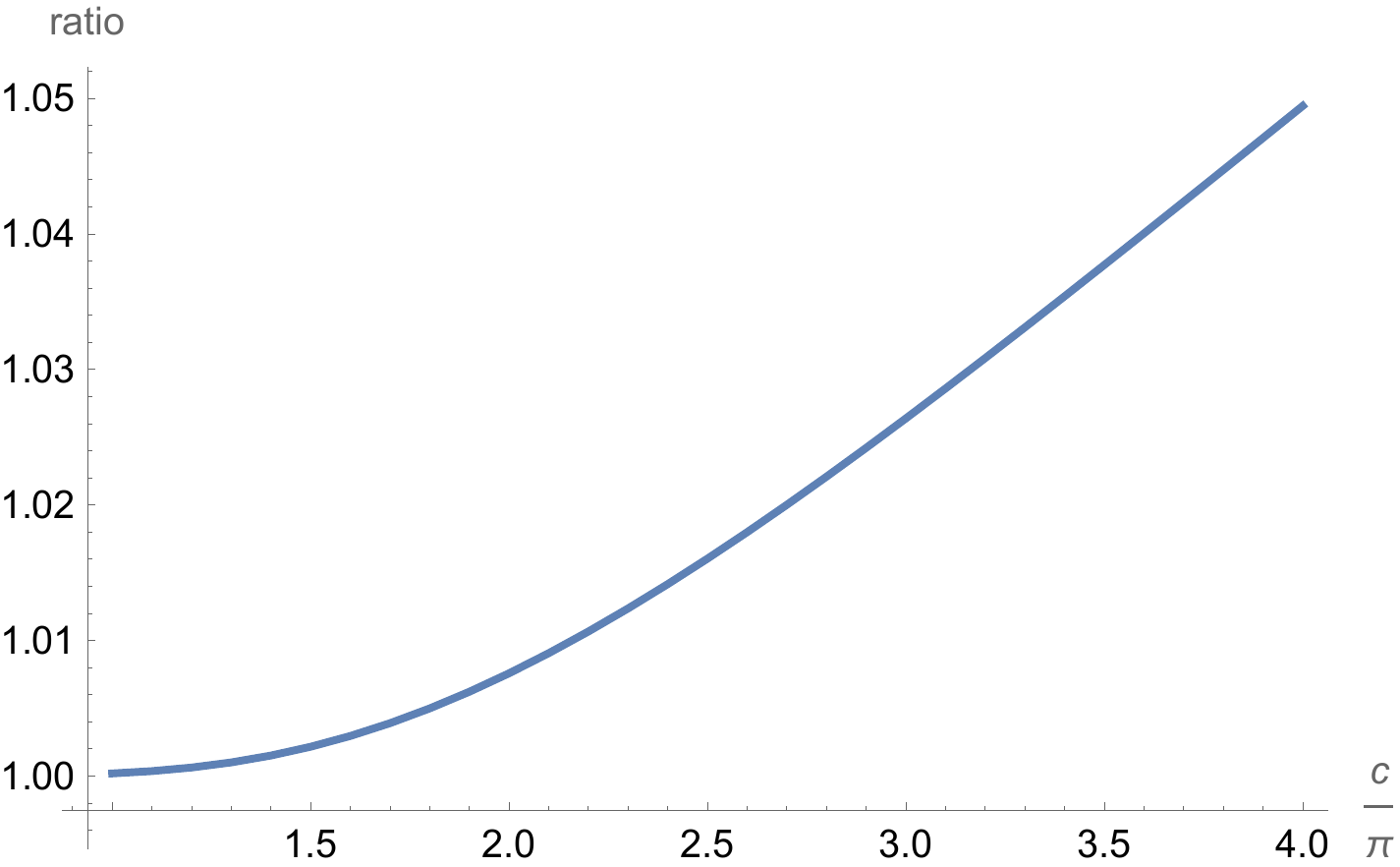}
    \caption{The ratio of the error for the Kaiser window to that for the optimal Slepian window as a function of $c$.
    For the Kaiser window the value of $\Delta$ is optimised to minimise the error.}
    \label{fig:Kratio}
\end{figure}

Next we numerically compare the error outside the confidence interval for the Kaiser window versus that for the optimal window.
According to the above asymptotic analysis for the error, the ratio of the errors should increase with $c$, and we find that occurs even optimising for $\Delta$.
In Fig.~\ref{fig:Kratio} we show the ratio, and it is very close to 1 for $c=\pi$, and increases to be about 5\% larger for $c=4\pi$.

\begin{figure}[tbh]
    \centering
    \includegraphics[width=0.5\textwidth]{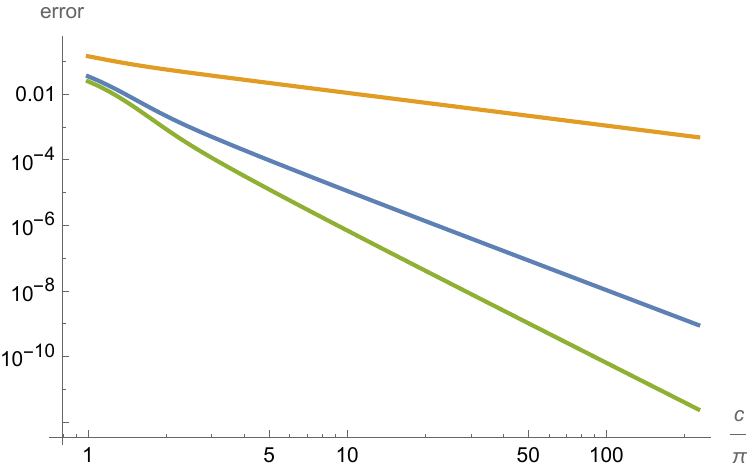}
    \caption{The differences between the various asymptotic series and exact error, divided by $4\sqrt{\pi c} \, e^{-2c}$.
    Results for the series of Slepian [see Eq.~\eqref{eq:slepy2}] are shown in orange, our series in Eq.~\eqref{eq:slepnum} is given in blue, and our numerically estimated series in Eq.~\eqref{eq:moreacc} is shown in green.}
    \label{fig:slepy}
\end{figure}

Now we test the series of Slepian compared to the one we have given in Eq.~\eqref{eq:slepnum}.
In Fig.~\ref{fig:slepy} we take the difference between the exact error and asymptotic approximation divided by $4\sqrt{\pi c} \, e^{-2c}$.
It can be seen that our expression is far more accurate, verifying that our expression is correct.
Numerically we have estimated the next higher order term as
\begin{equation}\label{eq:moreacc}
    \delta\approx 4\sqrt{\pi c} \, e^{-2c} \left( 1-\frac 7{16c} -
    \frac{91}{2^{9}c^2}-\frac{2657}{2^{13}c^3}\right) \, .
\end{equation}
The error for this expression is also shown in Fig.~\ref{fig:slepy}, and is even smaller.

\section{The sampling method}
\label{sec:sampling}

\subsection{Asymptotic approximations}

When we are performing phase estimation on a block encoded Hamiltonian, the Hamiltonian is encoded as $H/\lambda$, and the eigenvalues of the corresponding qubitised operator are $\pm e^{\pm i \arcsin(E_k/\lambda)}$ for eigenvalues $E_k$ of $H$.
This means that restricting the error to $\le\epsilon/\lambda$ for the phase estimation means that the error in $E_k$ is no more than $\epsilon$.
In the following we denote the ground state energy by $E_0$, so we aim to have an estimate in the region $[E_0-\epsilon,E_0+\epsilon]$.
Then the number of queries for phase estimation with either the Kaiser or Slepian window becomes, to leading order
\begin{equation}
\label{eq:Kaiser_window_single_sample_complexity}
    \frac{\lambda}{2\epsilon} \ln(1/\delta) + \mathcal{O}((\lambda/\epsilon) \ln \ln (1/\delta)) \, .
\end{equation}
If you the take the amplitude for the ground state to be $\gamma$ and $p=\gamma^2$, then the probability of failing to have the ground state once in sampling the energies $n$ times is $(1-p)^n \approx e^{-pn}$.
If we want the probability of failing to be less than $q$, we would need to take 
\begin{equation}
    \label{eq:num_Uinit_queries_QPE}
    n \approx (1/p)\ln(1/q)  \, .
\end{equation}

If we are taking the minimum result for the eigenvalue, then the probability of the error in that measurement result being outside the $\epsilon$ interval is now bounded by $n\delta$.
This suggests that we should divide $\delta$ by $n$ in order to obtain an appropriately bounded error probability.
However, the importance of the errors is asymmetric.
If the measurement corresponds to an excited eigenstate, but the measurement error yields an estimate of the energy that is exceptionally low, then that estimate could be taken to be the smallest out of all samples, yielding an inaccurate result.
On the other hand, if the measurement error gives an estimate of the energy that is exceptionally high, then it is far less likely to be taken as the smallest estimate out of all samples.
This means that the measurement errors in each direction need to be quantified differently.

To do this, let the confidence level for each individual estimate of the phase be $1-\delta$, so the probability of error on \emph{one side} is $\delta/2$.
That is because we are choosing a measurement technique where the error distribution is symmetric.
Given squared overlap with the ground state $p$, then for each estimate there is probability $1-p$ of it corresponding to the wrong eigenstate, and $p\delta/2$ of it corresponding to the correct eigenstate but the true eigenvalue being below the confidence interval on the lower side (so the estimate is too large).
We will group these possibilities together as a `high' error, which has total probability $1-p(1-\delta/2)$.
The probability of there being high errors on all $n$ samples is then $[1-p(1-\delta/2)]^n$.

The probability of a `low' error outside the confidence interval is $\delta/2$ for any individual measurement, so the probability of any low error occurring in the $n$ samples is $1-(1-\delta/2)^n$.
Note that it is possible for measurements corresponding to excited states to give a low error that is still not below the desired confidence interval for the measurement of the ground state.
The estimate above is not taking that into account, so is a fairly loose upper bound on the error.
Our total upper bound on the error is then
\begin{equation}\label{eq:toterr}
    [1-p(1-\delta/2)]^n + 1-(1-\delta/2)^n \, .
\end{equation}
Given an allowable error $q$ (confidence level $1-q$ in the final estimate), we can then solve for $\delta$.
This approach results in the value of $\delta$ being about twice what it would be if we did not take account of the asymmetry.

\begin{figure}[tbh]
    \centering
    \includegraphics[width=0.5\textwidth]{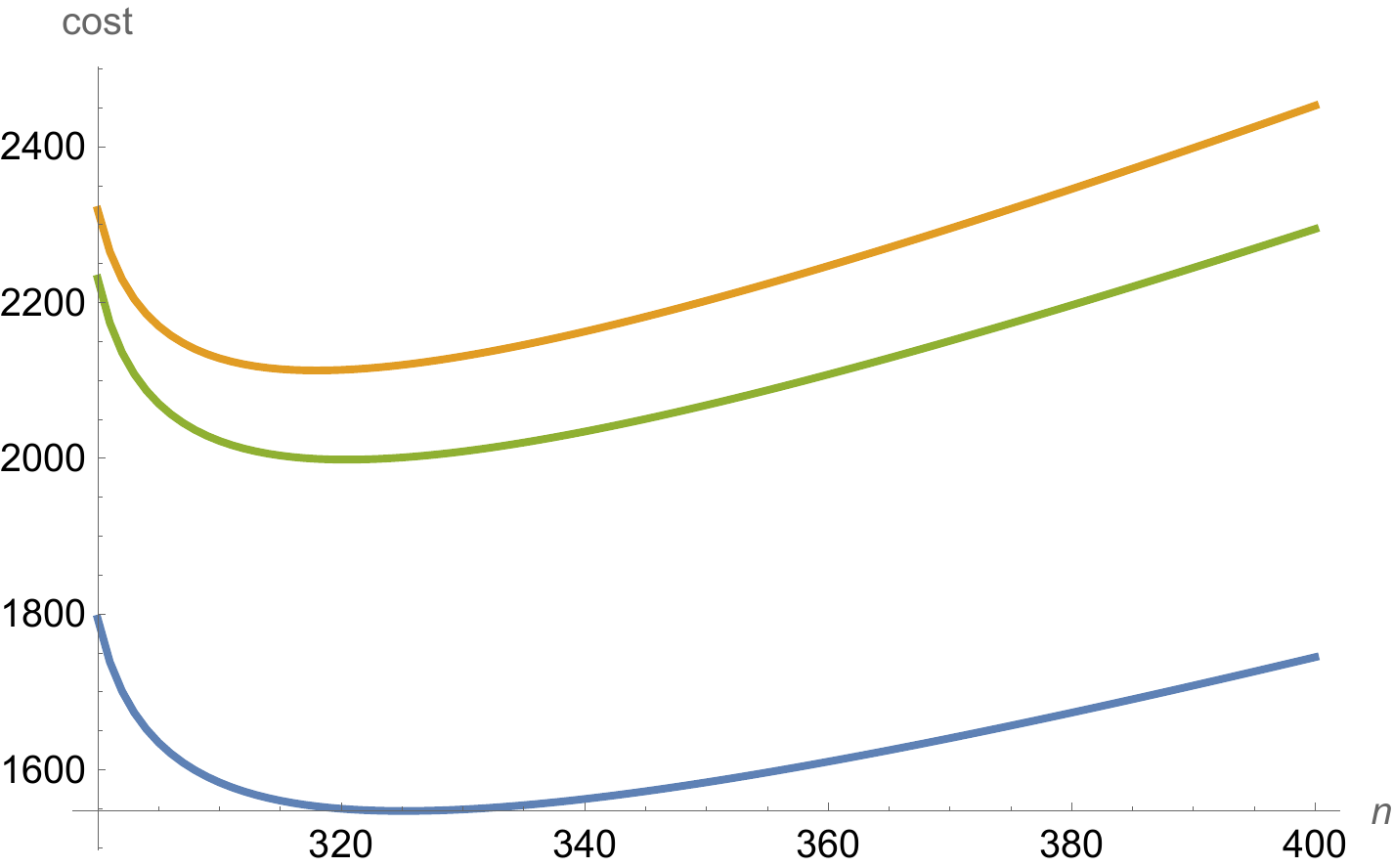}
    \caption{The factor in the complexity as a function of $n$ for $p=0.01$ and $q=0.05$, with $\delta$ obtained by solving for $q$ equal to Eq.~\eqref{eq:toterr}.
    The blue curve is the approximation $n\ln(1/\delta)/2$, the orange curve is the factor $n\pi\sqrt{1+\alpha^2}$ for the Kaiser window, and the green curve is $n\pi\sqrt{\Delta^2+\alpha^2}$ with $\Delta=0.3239$.}
    \label{fig:costfactor}
\end{figure}

The overall complexity will then be approximately
\begin{equation}
\label{eq:overall_complexity_direct_sampling_n_free}
    \frac{n\lambda}{2\epsilon} \ln(1/\delta) \, ,
\end{equation}
with $\delta$ as chosen by solving for $q$ equal to Eq.~\eqref{eq:toterr}, and $n$ chosen as at least $(1/p)\ln(1/q)$, resulting in a total complexity
\begin{equation}
    \label{eq:total_complexity_direct_sampling}
    \frac{\lambda}{2p\epsilon}\ln(1/q) \ln(1/\delta).
\end{equation}
For any specific example we can tweak the value of $n$ in order to minimise the overall complexity.
As an example, let us consider $\gamma=0.1$ so $p=0.01$, and require a 95\% confidence interval so that $q=0.05$.
In that case we have the factor $n\ln(1/\delta)/2$ as a function of $n$ shown in Fig.~\ref{fig:costfactor}.
That is, this is the factor in the complexity that is multiplied by $\lambda/\epsilon$ to give the overall complexity.
Here it can be seen that the optimal value of $n$ is 325, which is moderately above $(1/p)\ln(1/q)\approx 300$.
Note that this choice of the optimal value of $n$ is independent of $\lambda$ and $\epsilon$.

We can further develop asymptotic approximations for the solutions in the case of small $p,q$ to estimate the optimal values of $n$ and $\delta$.
First, we linearise Eq.~\eqref{eq:toterr} and set it equal to $q$ to give
\begin{equation}
(1-p)^n + \frac{\delta n}2 [1+(1-p)^{n-1}p] \approx q \, .
\end{equation}
Solving for $\delta$ then gives
\begin{equation}
    \delta \approx \frac 2n \frac{q-(1-p)^n}{1+(1-p)^{n-1} p}  \, .
\end{equation}
If we expand $\delta$ to second order, then we obtain the next term in the expansion
\begin{equation}
    \delta \approx \frac 2n \frac{q-(1-p)^n}{1+(1-p)^{n-1} p}
    + \frac{(n-1)[1-(1-p)^{n-2}]}{n^2[1+(1-p)^{n-1}p]^3}(q-(1-p)^n)^2
    \, .
\end{equation}
Up to terms that are smaller by factors of $q$ or $p$, we can give
\begin{equation}
    \frac 1{\delta} \approx  \frac {n}{2[q-(1-p)^n]} = \frac {n}{2(q-e^{-n\rho})} \, ,
\end{equation}
where $\rho=-\ln(1-p)\approx p$.

Next, given the solution for $\delta$ the task is to choose $n$ to minimise $n\ln(1/\delta)$.
Let us first consider this expression with only the leading order in the solution for $\delta$, which is (ignoring the factor of 2 for the moment)
\begin{equation}\label{eq:costdelsol}
    n \ln \left( \frac {n}{2(q-e^{-n\rho})} \right)  .
\end{equation}
The higher-order terms in the solution for $\delta$ will result in corrections that are at least a factor of $q$ smaller, so can be safely ignored in the following analysis where the terms in the expansions are significantly larger.
Taking the derivative with respect to $n$ gives
\begin{equation}
    1 - \frac{e^{-n\rho} n\rho}{q- e^{-n\rho}} + \ln \left( \frac {n}{2(q-e^{-n\rho})} \right)  .
\end{equation}
Setting this to zero and rearranging gives
\begin{equation}
    h = g - \ln g \, ,
\end{equation}
where
\begin{align}
    g &= \frac{e^{-n\rho}n\rho}{q-e^{-n\rho}} , \\
    h &= 1+n\rho-\ln(2\rho) .
\end{align}
We can then solve for $g$ as a series in $h$ as
\begin{equation}
    g \approx h + \ln h + \frac{\log h} h + \frac{2\ln h-\ln^2 h}{2h^2} + \frac{6 \ln h - 9 \ln^2 h + 2 \ln^3 h}{6h^3} + \mathcal{O}\left( \frac{\ln^4 h}{h^4}\right) \, .
\end{equation}
Now $q$ can be given in terms of $g$ as
\begin{align}
    q &= e^{-n\rho} + \frac{e^{-np}n\rho}{g}
    \, .
\end{align}
We can then rewrite this as
\begin{align}
    \ln(2/q) = n\rho - \ln\left( \frac 12 + \frac{n\rho}{2g}\right) \, .
\end{align}
By substituting a series for $g$ in terms of $h$, and inverting to obtain a series for $n\rho$, we obtain
\begin{align}
    n\rho &= \ln(2/q) - \frac{Q+1}{2\ln(2/q)}\left(1 - \frac{3Q-1}{4\ln(2/q)}  
    - \frac{5 + 4 Q - 7 Q^2}{12\ln^2(2/q)}\right) +
    \mathcal{O}\left( \frac{Q^4}{\ln^4(2/q)} \right)\, , \\
    Q &:= \ln(\ln(2/q)/2\rho) \, .
\end{align}
The difficulty with using this expression is that the successive terms are not smaller if $Q \sim \ln(2/q)$.
That will be the case if $q>p$.
This expression does give accurate results if $q$ is small compared to $p$.

A more accurate approximation may be given by
\begin{align}
    n\rho &= \ln(2/q) - \frac{Q+1}{2\ln(2/q)} \left( 1+\frac{3Q-1}{4\ln(2/q)}+\frac{23-2Q-Q^2)}{48\ln^2(2/q)}\right)^{-1}    
    + \mathcal{O}\left( \frac{Q^4}{\ln^4(2/q)} \right)\, .
\end{align}
Using this expression with $p=0.01$ and $p=0.05$ gives $n\approx 325$, close to the true value.
This series then gives the leading-order terms for $n\ln(1/\delta)/2$ as
\begin{equation}\label{eq:asyn}
    \frac n2\ln(1/\delta) = \frac 1{2\rho}\ln(2/q)\ln\left[ \frac{\ln(2/q)}{\rho q}\right]
    -\frac{(1 + Q)^2}{8\rho \ln(2/q)} + \frac{(Q-1) (1 + Q)^2}{
 16 \rho \ln^2(2/q)} + \frac{(1 + Q)^2 (5 + 10 Q - 7 Q^2)}{192 \rho \ln^3(2/q)}
 + \mathcal{O}\left( \frac{Q^5}{\rho\ln^4(2/q)} \right)
    \, .
\end{equation}
For this example the estimated value using the terms shown is 1525, within 1.5\% of the true value of 1547.
Using only the leading term gives 1634, still within 6\% of the correct value.
Note also that using only the leading order term for $\delta$ in estimating the cost affects the result by very little, less than 0.06\% for this example.
This justifies omitting higher-order terms for $\delta$ in the above analysis.
In contrast, using the approximation $\rho \approx p$ affects the result by about 0.6\%, so it is useful to make that correction.

\subsection{Real cost for window functions}

Next we consider the actual cost for the Kaiser and prolate spheroidal window functions, rather than just the asymptotic approximation.
We can explicitly integrate the probability distribution for the error in the Kaiser window.
That gives a factor of $n\pi\sqrt{1+\alpha^2}$ rather than $n\ln(1/\delta)/2$, with $\alpha$ solved to give error $\delta$.
The resulting factor is also shown in Fig.~\ref{fig:costfactor}.
The factor has been increased by about 37\% above that for the asymptotic approximation given above, from about 1547 to 2113.
It is also possible to adjust the cutoff used in the Kaiser window to $\sqrt{\Delta^2+\alpha^2}$ for $\Delta\ne 1$, which can give improved results.
The result for $\Delta=0.3239$ gives approximately the optimal result, and is shown in Fig.~\ref{fig:costfactor}.
Now the constant factor is only increased by about 29\% to 1998.
The accuracy of the asymptotic approximation is better for smaller $q$, so $q=0.01$ results in the cost being about 25\% above the asymptotic value, but it would take very small $q$ for the approximation to be accurate.

The prolate spheroidal window further reduces the cost, but only by a very small amount.
The curve is indistinguishable from that for optimised $\Delta$ in Fig.~\ref{fig:costfactor}, so is not shown separately.
In this case we find the constant factor is reduced very slightly to 1997 with $n=320$, a reduction of only $0.06\%$.
This shows that it is possible to accurately approximate results for the optimal window using the Kaiser window and adjusting $\Delta$.

\subsection{Contribution to cost from excited states}
\label{sec:exciting}

We can give a tighter bound on the cost by more accurately accounting for the contribution to the error from excited states.
In practice, there will be a small contribution to the probability of low estimates of the ground state energy $E_0$ from excited states.
When they are distant from $E_0$, there is very little probability of them yielding an estimate of the eigenvalue below the desired confidence interval, and if they are close to $E_0$ they will also increase the probability of having a result within the desired confidence interval.

To gauge the effect, let us consider just a single excited state with energy $\beta\epsilon$ above the ground state.
That is, when $\beta<1$ the energy is actually within the desired confidence interval for estimating the ground state.
It can be shown that the error is maximised for a single excited state, so our analysis for a single excited state is sufficient to bound the worst case for any spectrum of excited states; see Appendix \ref{app:excited}.
We denote by $\delta_1$ the probability for the estimate above the confidence interval for the excited state, and $\delta_2$ for the estimate below the desired confidence interval.
It will be expected that $\delta<\delta_1+\delta_2$, and in typical cases where the eigenvalue is \emph{much} higher than the ground value, there will be $\delta_1\sim 1$ and $\delta_2\sim 0$.

We can then replace $1-p$ for the probability of error due to the incorrect eigenstate being obtained with $(1-p)\delta_1$; that is, the incorrect eigenstate is obtained \emph{and} the estimate is too high.
Then the probability of all $n$ samples being too high is
\begin{equation}
    [(1-p)\delta_1+p\delta/2)]^n .
\end{equation}
Then the probability of a single sample being too low is $p\delta/2 + (1-p)\delta_2$, so the probability of \emph{any} of the samples being too low is $1-\{1-[p\delta/2 + (1-p)\delta_2]\}^n$.
Adding together these two probabilities of error then gives
\begin{align}\label{eq:excitederr}
     P_{\rm err} = [p\delta/2+(1-p)\delta_1]^n + 1-\{1-[p\delta/2 + (1-p)\delta_2]\}^n \, .
\end{align}
This expression can be expected to be smaller than that given before, because $p\delta/2+(1-p)\delta_1$ will be smaller than $p\delta/2+(1-p)$.

\begin{figure}[tbh]
    \centering
    \includegraphics[width=0.5\textwidth]{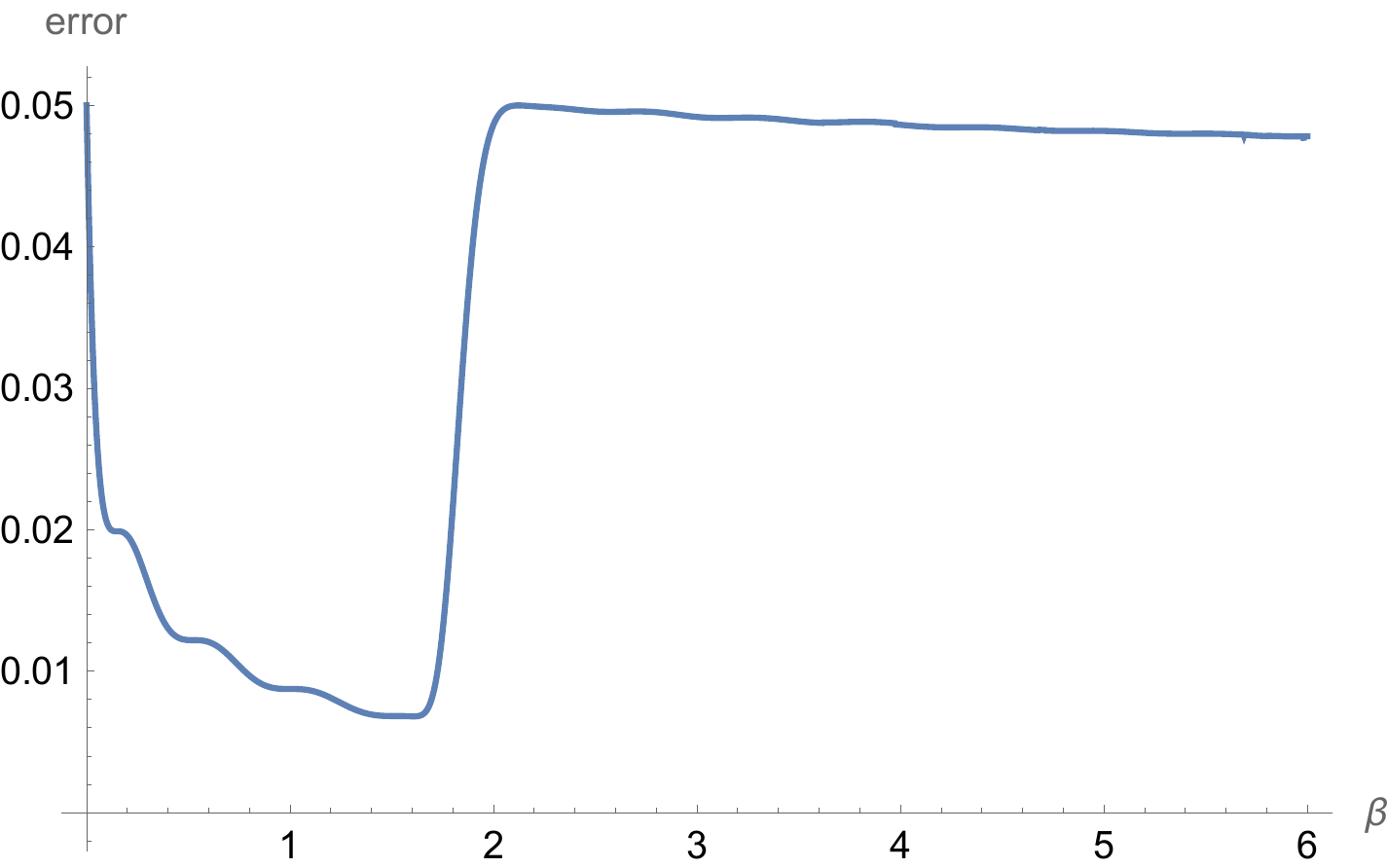}
    \caption{The error according to Eq.~\eqref{eq:excitederr} as a function of $\beta$ with $p=0.01$, $\alpha=1.70116$, $\Delta=0.074476$, and $n=309$.
    The error does not exceed $q=0.05$.}
    \label{fig:exerr}
\end{figure}

In numerical testing, we choose values of $\alpha,\Delta,n$ \emph{without} knowing $\beta$ (the excited state energy), so the above error needs to be no greater than $q$ for all choices of $\beta$.
We find that for $p=0.01$, $q=0.05$, we can choose $\alpha=1.70116$, $\Delta=0.074476$, and $n=309$.
These values minimise the constant factor while keeping the error below $q$; see Fig.~\ref{fig:exerr}.
The constant factor is approximately 1673, so it is significantly smaller than the results not taking into account this factor, and similar to the result with the very simple asymptotic approximation we gave first.
In testing with multiple excited state eigenvalues, the results are no worse than with just one excited state eigenvalue, as predicted.

An interesting feature of the results is that the error is larger for larger $\beta$ (around 2 and above), then is smaller for small $\beta$, but with a spike near $\beta=0$.
The reason for this is that, as $\beta$ is reduced below 2, there is greater overlap between the distribution of measurement results obtained for the excited state and the desired interval $[E_0-\epsilon,E_0+\epsilon]$ for estimates of $E_0$.
That is, both of these are of half-width $\epsilon$, so reducing the gap below $\beta\epsilon$ means the two regions overlap.
This means that measurements of energy for the excited state being within the desired region is reducing the overall error probability.
It reduces quite quickly, because we are taking the minimum of the measurement results, and with many samples there is a high probability of a measurement result on the excited state being around $\epsilon$ below its true energy.

\begin{figure}[tbh]
    \centering
    \includegraphics[width=0.5\textwidth]{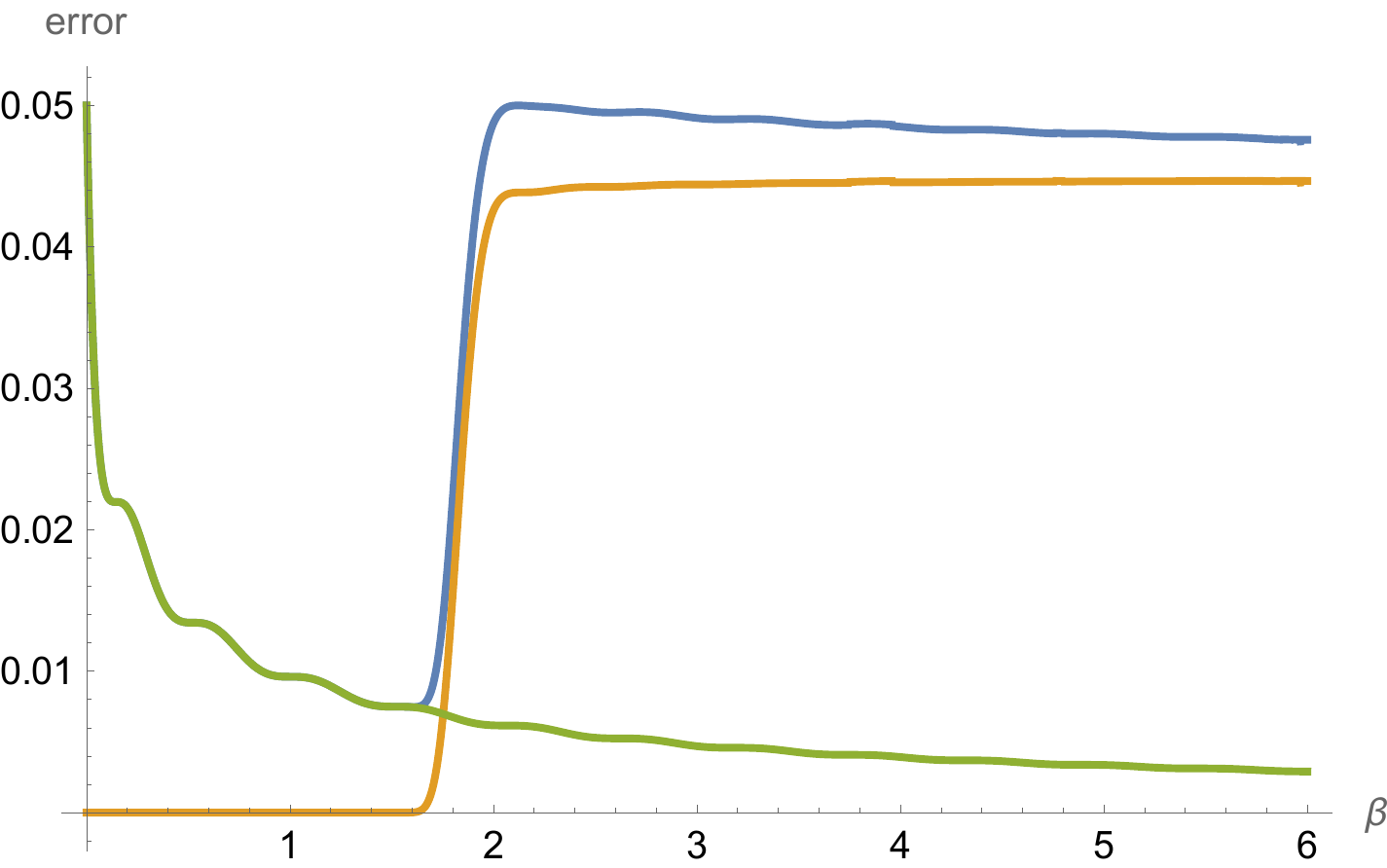}
    \caption{The error according to Eq.~\eqref{eq:excitederr} as a function of $\beta$ with $p=0.01$, $\alpha=1.70116$, $\Delta=0.074476$, and $n=309$.
    The green line is the probability of an error below the desired interval, the orange line is for errors above the interval, and blue is the sum of the two.}
    \label{fig:exer2}
\end{figure}

But then there is a separate spike for the error probability as the gap closes to zero.
This is because the measurements of energy on the excited state now have a significant probability of being \emph{below} the desired confidence interval.
To bound the error by $q$ we need to ensure the error for both $\beta=0$ and $\beta$ around 2 is bound by $q$.
We find that the best results (in terms of the lowest constant factor) tends to be those where the error reaches $q$ for both.
This is illustrated in Fig.~\ref{fig:exer2}.
It shows that the probability of errors more than $\epsilon$ above the true ground state rapidly approaches zero as $\beta$ is reduced below 2, and the probability of errors that are too low gradually increases.
This behaviour is for small $p$, whereas for $p$ close to 1 the error is less than $q$ for $\beta=0$.

It turns out that using the prolate spheroidal wavefunction now gives \emph{worse} results.
For this example we find that the performance is optimised for $n=318$, $\alpha=1.71229$, where the constant factor is about 1711, or 2\% \emph{worse} than for the Kaiser window.
To see why, consider the same parameters as for the Kaiser window, with $\beta=2.12103$.
We then find that most of the probabilities are the same, but $\delta_2$ is significantly lower for the Kaiser window.
It is about $0.0000184942$, versus $\delta_2\approx 0.0000336569$ for the prolate spheroidal window, which is about 82\% larger.

This is very significant in this case with small $p$, because it contributes to the chance of the excited state yielding an estimate of the energy that is too \emph{low}.
In this case, with the prolate spheroidal wavefunction, the probability of an estimate that is too low with the excited state is only about $1/5$ that for the ground state.
Since the excited state is about 100 times more common in this example, that is a significant contributor to the error.
In contrast, the Kaiser window more strongly suppresses the tails, so estimates that are too low coming from the excited state are less of a problem.
Again the behaviour is different for $p$ close to 1, where the prolate spheroidal window provides better performance.
That is because there are few repetitions, so the effect of the contribution to the error from $\delta_2$ being amplified by repetitions is less.

\section{The binary search approach}
\label{sec:binary_search}

In the previous sections we have focused on estimating the ground state energy by directly reading off the energy samples from the quantum phase estimation algorithm and taking the minimum among these samples. As shown in Eq.~\eqref{eq:num_Uinit_queries_QPE}, the number of samples needed scales as $\wt{\Or}(1/p)$. This is however not the optimal scaling. As shown in \cite[Theorem 8]{LinTong2020near}, one can improve the dependence on $p$ to $\wt{\Or}(1/\sqrt{p})$. In this section we will propose a method based on \cite{LinTong2020near} and incorporating the window functions to reduce the resources needed in practical implementation. We also show numerically that this method is beneficial when $p<10^{-3}$. We will refer to this method as \textit{the binary search approach} henceforth.

\subsection{From the fuzzy bisection problem to amplitude estimation}

From the previous sections, we can see that in a single run of the QPE algorithm, in order to ensure that the phase error is below $\eta$ with probability at least $1-\delta$, the query complexity is given by a function $Q(\eta,\delta)$, which in the case of the Kaiser or Slepian window, scales as
\begin{equation}
\label{eq:kaiser_window_asymp_expression}
    Q(\eta,\delta) \approx \frac{1}{2\eta}\ln(1/\delta)
\end{equation}
for small $\eta$ and $\delta$, according to Eq.~\eqref{eq:asympNKai}.
In this notation we use $\eta$ for the phase estimation error to distinguish it from the error $\epsilon$ for estimation of eigenvalues.
We will use this expression in our computation of the asymptotic complexity of the binary search approach, but in numerics we will numerically compute the function $Q(\eta,\delta)$ in a more accurate manner.

In the binary search process in Ref.~\cite{LinTong2020near}, we gradually shrink an interval $[\lambda_L,\lambda_R]$ in which the ground state energy is located.
In particular, in the last search step, in order to estimate $\lambda_0$ to $\epsilon$ precision, we have $\lambda_L,\lambda_R$ such that $\lambda_L\leq \lambda_0\leq \lambda_R$, with $\lambda_R-\lambda_L= 3\epsilon$. 
We want to distinguish between two cases:
    \begin{equation}
\lambda_0>\frac{1}{3}\lambda_L+\frac{2}{3}\lambda_R,\quad \text{or}\quad \lambda_0<\frac{2}{3}\lambda_L+\frac{1}{3}\lambda_R.
    \end{equation}
If $\frac{2}{3}\lambda_L+\frac{1}{3}\lambda_R\leq \lambda_0\leq \frac{1}{3}\lambda_L+\frac{2}{3}\lambda_R$ then we can output anything.
Solving the problem of distinguishing the cases will give us an interval $[\lambda_L^{\mathrm{new}},\lambda_R^{\mathrm{new}}]\ni \lambda_0$ of size $2\epsilon$. 
That interval corresponds to an estimate of $\lambda_0$ up to $\epsilon$ error. We call this problem the \textit{fuzzy bisection problem}.

Because $\arccos$ is monotonically decreasing in $[-1,1]$, we only need to distinguish between
    \begin{equation}
\arccos\left(\frac{\lambda_0}{\lambda}\right)<\arccos\left(\frac{\lambda_L+2\lambda_R}{3\lambda}\right),\ \text{or}\ \arccos\left(\frac{\lambda_0}{\lambda}\right)>\arccos\left(\frac{2\lambda_L+\lambda_R}{3\lambda}\right).
    \end{equation}
We then perform QPE with a Kaiser or Slepian window on $W$. We denote the phase output by $\hat{\phi}\in[-\pi,\pi]$. We want the phase error to be at most 
    \begin{equation}
\frac{\epsilon}{2\lambda}=\frac{\lambda_R-\lambda_L}{6\lambda}\leq \frac{1}{2}\left(\arccos\left(\frac{2\lambda_L+\lambda_R}{3\lambda}\right)-\arccos\left(\frac{\lambda_L+2\lambda_R}{3\lambda}\right)\right)
    \end{equation}
with probability at least $1-\delta_1$.

If $\arccos\left(\frac{\lambda_0}{\lambda}\right)<\arccos\left(\frac{\lambda_L+2\lambda_R}{3\lambda}\right)$, then all eigenvalues $e^{\pm i\arccos(\lambda_k/\lambda)}$ of $W$ satisfy  $\arccos\left(\frac{\lambda_k}{\lambda}\right)<\arccos\left(\frac{\lambda_L+2\lambda_R}{3\lambda}\right)$. 
Therefore if 
    \begin{equation}
|\hat{\phi}|> \bar{\phi} := \frac{1}{2}\left(\arccos\left(\frac{2\lambda_L+\lambda_R}{3\lambda}\right)+\arccos\left(\frac{\lambda_L+2\lambda_R}{3\lambda}\right)\right),
    \end{equation}
then a phase error larger than 
    \begin{equation}
\bar{\phi} - \arccos\left(\frac{\lambda_L+2\lambda_R}{3\lambda}\right) \geq \frac{\epsilon}{2\lambda}
    \end{equation}
must have occurred. This event has probability at most $\delta_1$. Consequently
\begin{equation}
\label{eq:prob_case_small}
    \mathrm{Pr}[|\hat{\phi}|> \bar{\phi}] \leq \delta_1.
\end{equation}

If $\arccos\left(\frac{\lambda_0}{\lambda}\right)>\arccos\left(\frac{2\lambda_L+\lambda_R}{3\lambda}\right)$, suppose $\hat{\phi}$ comes from eigenvalues $e^{\pm i\arccos(\lambda_0/\lambda)}$, then it will satisfy $|\hat{\phi}|> \bar{\phi}$ with probability at least $1-\epsilon_1$. This is because
    \begin{equation}
\arccos\left(\frac{2\lambda_L+\lambda_R}{3\lambda}\right)-\bar{\phi}\geq \frac{\epsilon}{2\lambda}.
    \end{equation}
The phase $\hat{\phi}$ comes from eigenvalues $e^{\pm i\arccos(\lambda_0/\lambda)}$ with probability at least $p$. Therefore
\begin{equation}
\label{eq:prob_case_large}
    \mathrm{Pr}[|\hat{\phi}|> \bar{\phi}] \geq p(1-\delta_1).
\end{equation}

We therefore only need to distinguish between two cases in Eq.~\eqref{eq:prob_case_small} and Eq.~\eqref{eq:prob_case_large}.
This is an amplitude estimation problem. We define
    \begin{equation}
\gamma_1= \sqrt{\delta_1},\quad \gamma_2 = \sqrt{p(1-\delta_1)},
    \end{equation}
and aim to distinguish the cases where the amplitude corresponding to $|\hat{\phi}|> \bar{\phi}$ is at most $\gamma_1$ or at least $\gamma_2$. We choose the parameters so that $\gamma_1<\gamma_2$.
To generate a single sample of $\hat{\phi}$ we need to run a coherent QPE circuit that involves 
\begin{equation}
    d_1 = Q(\eta,\delta_1) 
\end{equation}
queries to $W$, where $\eta=\epsilon/(2\lambda)$ is the allowed phase error.

\subsection{Amplitude estimation with the Kaiser window}
\label{sec:amplitude_estimation_kaiser_window}

From the previous section we can see that to solve the fuzzy bisection problem, it suffices to estimate 
\begin{equation}
    A = \sqrt{\mathrm{Pr}[|\hat{\phi}|> \bar{\phi}]}.
\end{equation}
We will do so using amplitude amplification. 
If this amplitude can be estimated with error at most $(\gamma_2-\gamma_1)/2$, then we will be able to distinguish between $A>\gamma_2$ and $A<\gamma_1$.
In amplitude estimation, we construct a walk operator $\mathcal{W}$ using the QPE circuit (two applications of it), such that 
\begin{equation}
    \mathcal{W}\ket{\Phi^{\pm}} = e^{\pm i2\arcsin(A)}\ket{\Phi^{\pm}},
\end{equation}
and $\ket{\Phi}=\frac{1}{2}(\ket{\Phi^{+}}+\ket{\Phi^{-}})$ can be prepared using the QPE circuit.

With this walk operator $\mathcal{W}$, we can then run QPE with the Kaiser or Slepian window to estimate $\arcsin(A)$. We only need to estimate $A$ to precision $(\gamma_2-\gamma_1)/2$, which means it suffices to estimate the phase $2\arcsin(A)$ to precision $\gamma_2-\gamma_1$.
To do this with probability at least $1-\delta_2$ requires running $\mathcal{W}$ $d_2$ times, where 
\begin{equation}
    d_2 = Q(\gamma_2-\gamma_1,\delta_2).
\end{equation}
For the last search step, we need to use $W$ for a total of
\begin{equation}
    d_1(2d_2+1) = Q\left(\frac{\epsilon}{2\lambda},\delta_1\right)\left(2Q(\gamma_2-\gamma_1,\delta_2)+1\right)
\end{equation}
times, which in the context of the Kaiser window and using Eq.~\eqref{eq:Kaiser_window_single_sample_complexity} is
\begin{equation}
    d_1(2d_2+1)\approx 2d_1d_2 \approx \frac{\lambda}{(\gamma_2-\gamma_1)\epsilon}\ln(1/\delta_1)\ln(1/\delta_2)
\end{equation}
times, up to the leading order. Note that in $2d_2+1$ the $+1$ comes from preparing the initial state $\mathcal{W}\ket{0}$ for amplitude estimation (see \cite[Figure~1]{BrassardHoyerMoscaTapp2002quantum}).
The number of times we need to use $U_{\mathrm{init}}$ is $2d_2+1$, which in the context of the Kaiser window is
\begin{equation}
    2d_2+1 \approx 2d_2 = \frac{1}{(\gamma_2-\gamma_1)}\ln(1/\delta_2).
\end{equation}

The value of $\delta_1$ can be chosen to minimise the cost as follows.
Using the expressions $\gamma_1= \sqrt{\delta_1}$, $\gamma_2 = \sqrt{p(1-\delta_1)}$ we have a factor in the complexity
\begin{equation}
    \frac{\ln(1/\delta_1)}{\sqrt{p(1-\delta_1)}-\sqrt{\delta_1}} .
\end{equation}
Approximating $\sqrt{1-\delta_1}\approx 1$ and using $x=\sqrt{\delta_1}$ this factor is approximately proportional to
\begin{equation}
    \frac{\ln(1/x)}{\sqrt{p}-x} \, .
\end{equation}
Taking the derivative with respect to $x$ then yields
\begin{equation}
    \frac{-1/x}{\sqrt{p}-x}+ \frac{\ln(1/x)}{(\sqrt{p}-x)^2} \, .
\end{equation}
For this to be zero, we should have
\begin{equation}
    x = \frac{\sqrt{p}-x}{\ln(1/x)} \, .
\end{equation}
Starting with $x=\sqrt{p}/\ln(1/\gamma)$ then iterating $x\mapsto (\sqrt{p}-x)/\ln(1/x)$ quickly gives the solution.

\subsection{Query complexity of all search steps}
\label{sec:all_search_steps}

Previously we focused on the cost of the last search step.
Here we account for the costs in all search steps:
the binary search terminates in $L=\lceil\log_{3/2}(\lambda/\epsilon)\rceil$ steps. If we want a final success probability of $q$, we need $\delta_2=q/L$.
For the number of queries to $W$, we observe that each search step requires $2/3$ of the resolution of the next step, and therefore the query complexity is also $2/3$ of that of the last step.
Therefore the total number of queries to $W$ is
\begin{align}
\label{eq:num_queries_to_W_prelim_general}
& Q\left(\frac{\epsilon}{2\lambda},\delta_1\right)\left(2Q(\gamma_2-\gamma_1,\delta_2)+1\right)\left(1+\frac{2}{3}+\left(\frac{2}{3}\right)^2+\cdots\right) \nn
&\leq 3Q\left(\frac{\epsilon}{2\lambda},\delta_1\right)\left(2Q(\gamma_2-\gamma_1,\delta_2)+1\right).
\end{align}
For the Kaiser window the total number of queries needed is then
\begin{align}
\label{eq:num_queries_to_W_prelim_Kaiser}
\frac{3\lambda}{(\gamma_2-\gamma_1)\epsilon}\ln(1/\delta_1)\ln(1/\delta_2).
\end{align}
The total number of queries to $U_{\mathrm{init}}$ is the same for each search step. Therefore it is
\begin{equation}
    \label{eq:num_queries_to_Uinit_prelim}
    \frac{L}{\gamma_2-\gamma_1}\ln(1/\delta_2).
\end{equation}

We recall that
\begin{equation}
    L\approx \log_{3/2}(\lambda/\epsilon),\quad \delta_2=q/L.
\end{equation}
At the end of the previous section we have discussed how to choose $\delta_1$ by solving an optimization problem, but here to get a concise expression we will choose a sub-optimal $\delta_1$, which does not have much effect on the final cost.
Here $\delta_1$ is chosen to be
\begin{equation}
    \delta_1 = p/16,
\end{equation}
and then
\begin{equation}
\label{eq:choice_gamma1_gamma2}
    \gamma_2 = \sqrt{p(1-\delta_1)}= \sqrt{p(1-p/16)},\quad \gamma_1=\sqrt{\delta_1}=\sqrt{p}/4.
\end{equation}
Consequently
\begin{equation}
\gamma_2-\gamma_1\approx (3/4)\sqrt{p}.
\end{equation}
Substituting these values into Eq.~\eqref{eq:num_queries_to_W_prelim_Kaiser} and Eq.~\eqref{eq:num_queries_to_Uinit_prelim},
we can see that in order to estimate the ground state energy to precision $\epsilon$ with probability at least $1-q$, with an initial squared overlap of $p$,
we need to use $W$
\begin{equation}
\label{eq:num_W_queries_binary_search}
\frac{8\lambda}{\sqrt{p}\epsilon}\ln\left(\frac{4}{\sqrt{p}}\right)\ln\left(\frac{\log_{3/2}(\lambda/\epsilon)}{q}\right)
\end{equation}
times, and
we need to use $U_{\mathrm{init}}$
\begin{equation}
    \label{eq:num_Uinit_queries_binary_search}
    \frac{4\log_{3/2}(\lambda/\epsilon)}{3\sqrt{p}}\ln\left(\frac{\log_{3/2}(\lambda/\epsilon)}{q}\right)
\end{equation}
times.

\subsection{Optimizing the shrinking factor}

In the binary search algorithm we discussed above, we shrink the interval in which $\lambda_0$ is located by $2/3$ in each iteration. This is however not the optimal shrinking factor.
If we shrink by a factor of $\omega$ instead of $2/3$ at each search step, then using the leading order in the asymptotic expression for the Kaiser and Slepian windows in Eq.~\eqref{eq:asympNKai},
we need to use $W$
\begin{align}
\label{eq:total_cost_optimize_chi}
    &\underbrace{\frac{1}{1-\omega}}_{\text{search steps}}\times \underbrace{\frac{\omega}{2(2\omega-1)}\frac{\lambda}{\epsilon}\ln\left(\frac{16}{p}\right)}_{\text{QPE}}\times\underbrace{\frac{4}{3\sqrt{p}}\ln\left(\frac{\log_{1/\omega}(\lambda/\epsilon)}{q}\right)}_{\text{amplitude estimation}} \nn
&=\frac{4\omega}{3(2\omega-1)(1-\omega)}\frac{\lambda}{\sqrt{p}\epsilon}\ln\left(\frac{4}{\sqrt{p}}\right)\ln\left(\frac{\log_{1/\omega}(\lambda/\epsilon)}{q}\right)
\end{align}
times.

We will explain how we arrived at the above expressions. For each binary search step, the two cases we wish to distinguish are now
    \begin{equation}
\lambda_0>(1-\omega)\lambda_L+\omega\lambda_R,\quad \text{or}\quad \lambda_0<\omega\lambda_L+(1-\omega)\lambda_R.
    \end{equation}
If we have eliminated the first case, then the value of $\lambda_R$ is mapped as
\begin{equation}
    \lambda_R \mapsto (1-\omega)\lambda_L+\omega\lambda_R \, ,
\end{equation}
with $\lambda_L$ unchanged, so the range of values is mapped as
\begin{equation}
    \lambda_R - \lambda_L \mapsto (1-\omega)\lambda_L+\omega\lambda_R - \lambda_L = \omega (\lambda_R - \lambda_L) \, .
\end{equation}
The reduction in the range is equivalent if we eliminate $\lambda_0<\omega\lambda_L+(1-\omega)\lambda_R$.

We now need phase error at most
    \begin{equation}
\frac{(2\omega-1)(\lambda_R-\lambda_L)}{2\lambda}\leq \frac{1}{2}\left[\arccos\left(\frac{\omega\lambda_L+(1-\omega)\lambda_R}{\lambda}\right)-\arccos\left(\frac{(1-\omega)\lambda_L+\omega\lambda_R}{\lambda}\right)\right].
    \end{equation}
At each step the range is shrunk by $\omega$, and we start with $\lambda_R-\lambda_L=2\lambda$, so at step $j$ the error is
\begin{equation}
    (2\omega-1) \omega^{j-1} \, .
\end{equation}
The number of queries for the QPE circuit is then approximately, using Eq.~\eqref{eq:Kaiser_window_single_sample_complexity},
\begin{equation}
    \frac{1}{2(2\omega-1)} (1/\omega)^{j-1} \ln(1/\delta_1)  \, .
\end{equation}
Now the number of steps $L$ is calculated so that
\begin{equation}
    (1/\omega)^L \lambda = \epsilon \, .
\end{equation}
If we sum the cost above over $j=1$ to $L$, then we obtain
\begin{equation}
\label{eq:cost_qpe_optimize_chi}
    \frac{1}{2(2\omega-1)} \times \frac{(1/\omega)^L - 1}{1/\omega-1} \ln(1/\delta_1) \approx \frac{\omega}{2(2\omega-1)(1-\omega)}\frac{\lambda}{\epsilon} \ln(1/\delta_1) \, .
\end{equation}
To minimise the cost we therefore aim to maximise $(2\omega-1)(1/\omega-1)$.
That has a derivative of $1/\omega^2-2$, so the turning point is $\omega=1/\sqrt 2$.

Note that an extra factor needs to be taken into account for amplitude estimation. This is independent of the choice of $\omega$ and therefore identical to the case we studied before for $\omega=2/3$, which is given in Eq.~\eqref{eq:num_queries_to_W_prelim_Kaiser}. With the parameters $\gamma_1$ and $\gamma_2$ given in Eq.~\eqref{eq:choice_gamma1_gamma2}, we thus arrive at the expressions in Eq.~\eqref{eq:total_cost_optimize_chi} by multiplying this extra factor by Eq.~\eqref{eq:cost_qpe_optimize_chi}.

Optimizing $\omega$ for \eqref{eq:total_cost_optimize_chi} outside the $\ln\ln$ we have $\omega=1/\sqrt{2}$, and the number of queries are roughly
\begin{equation}
\label{eq:total_complexity_optimized_binary_search_7.77}
\frac{7.77\lambda}{\sqrt{p}\epsilon}\ln\left(\frac{4}{\sqrt{p}}\right)\ln\left(\frac{\log_{\sqrt{2}}(\lambda/\epsilon)}{q}\right)  .
\end{equation}
We need to use $U_{\mathrm{init}}$
\begin{align}
&\underbrace{\log_{1/\omega}(\lambda/\epsilon)}_{\text{search steps}}\times\underbrace{\frac{4}{3\sqrt{p}}\ln\left(\frac{\log_{1/\omega}(\lambda/\epsilon)}{q}\right)}_{\text{amplitude estimation}} \nn
&=\frac{4\log_{1/\omega}(\lambda/\epsilon)}{3\sqrt{p}}\ln\left(\frac{\log_{1/\omega}(\lambda/\epsilon)}{q}\right)
\end{align}
times, which is monotonically increasing with respect to $\omega$.

\begin{figure}
    \centering
    \includegraphics[width=0.5\textwidth]{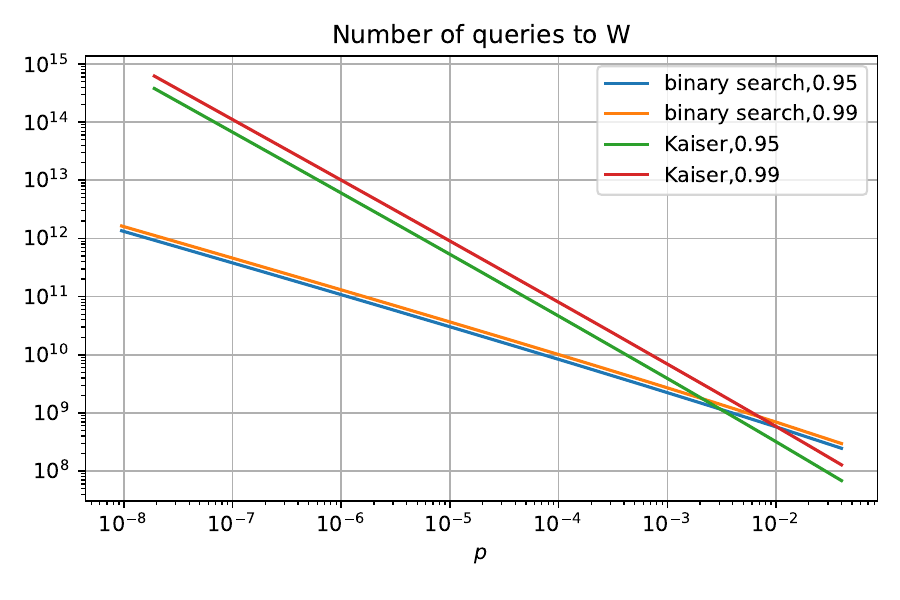}
    \caption{Number of queries to $W$ as a function of the squared overlap $p$ for FeMoco with $\lambda=306$ and $\epsilon=0.0016$, using both the binary search method discussed in this section and pure QPE+Kaiser window. At $p=0.01$, the number of queries using the binary search method are 571 million and 691 million for $95\%$ confidence and $99\%$ confidence respectively, compared to 320 million and 587 million using pure QPE+Kaiser window.
    The binary sampling cost is affected very little by whether the Kaiser or optimal window is used for phase estimation.
    For the $95\%$ and $99\%$ confidence intervals the cost is only reduced by $0.04$\% and $0.06$\%, respectively. The value of $\omega$ is chosen to be $1/\sqrt{2}$.}
    \label{fig:compare_W_queries}
\end{figure}

\begin{figure}
    \centering
    \includegraphics[width=0.5\textwidth]{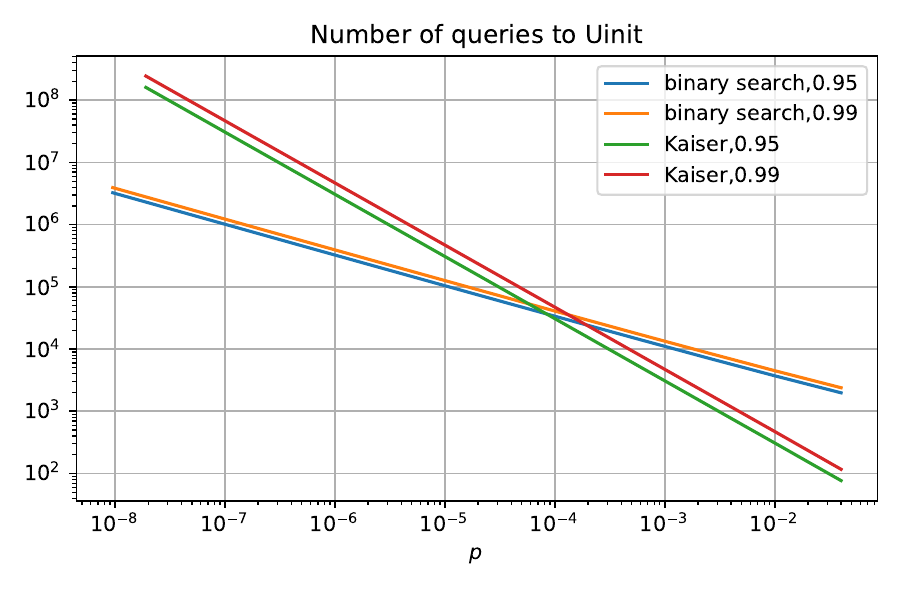}
    \caption{Number of queries to $U_{\mathrm{init}}$ as a function of the squared overlap $p$ for FeMoco with $\lambda=306$ and $\epsilon=0.0016$, using both the binary search method discussed in this section and pure QPE+Kaiser window. At $p=0.01$, the number of queries using the binary search method are 3683 and 4457 for $95\%$ confidence and $99\%$ confidence respectively, compared to 309 and 472 using pure QPE+Kaiser window. The value of $\omega$ is chosen to be $1/\sqrt{2}$.}
    \label{fig:compare_Uinit_queries}
\end{figure}

In Figs~\ref{fig:compare_W_queries} and \ref{fig:compare_Uinit_queries} we numerically compute the query complexities of the binary search method and the direct sampling approach, both based on the Kaiser window. The results suggest that a crossover of the query complexity takes place between $p=10^{-3}$ and $p=10^{-2}$ for $95\%$ and $99\%$ confidence levels.
This agrees well with the estimated crossover of around $p\sim 0.003$ based on the constant factors in the scaling.

\section{Total resources for ground state energy estimation in chemical systems}
\label{sec:resource_estimates}
As previously discussed, a key quantity in the cost of applying QPE to ground state energy estimation is the overlap of the initial input state with the ground state wavefunction. Prior work investigated overlaps for simple small molecules~\cite{tubman2018postponing} and found high overlaps with product state wavefunctions substantially above the low $p$ settings discussed in this work.  More recently, it was shown that in realistic examples of more complicated molecules, product states can have a small overlap, while adiabatic state preparation from the lowest energy mean-field state can be more costly than the phase estimation itself, motivating the development of more sophisticated state preparation protocols \cite{Lee2023}. 
However, determining an overlap, or a faithful estimate of it, in problems where classical algorithms have difficulty computing an accurate ground state (and thus which are of most interest for quantum applications) is by definition challenging. 

In this section we describe an extrapolation protocol that allows us to estimate the overlap with the ground state by extrapolating with respect to the bond dimensions of two MPS wavefunctions. Although this extrapolation does not provide any rigorous guarantees, we provide benchmark data to support the procedure; see Appendix~\ref{app:valov}. 
In addition, in the most challenging systems such as FeMoco, it is not possible to classically determine MPS wavefunctions with sufficiently large bond dimension to distinguish between the ground state and nearby excited states. In this case, under the assumption that the MPS we are preparing has good overlap with some low-energy eigenstate, we provide an estimation of the overlap with this eigenstate, and QPE can then be used to resolve the energies of the different eigenstates associated with the different MPS initial states. Leveraging these estimates, we account for the full resources needed for resolving the energy landscape of competing spin structures in FeMoco.

\subsection{Model generation}
In this study, we employed active space models for Fe-S simulations suggested in previous studies for the Fe$_{2}$S$_{2}$, Fe$_{4}$S$_{4}$, and FeMoco iron-sulfur systems~\cite{li2017spin, li2019electronicFemco}.
The active spaces for the Fe$_{2}$S$_{2}$ and Fe$_{4}$S$_{4}$ models consist of Fe-3d, S-3p, and $\sigma$-bonding orbitals between the Fe and thiolate ligands, defined as complete active spaces CAS(30e,20o) for 2Fe(III), CAS(54e,36o) for 2Fe(III)2Fe(II), and CAS(52e,36o) for 4Fe(III). 
In this notation CAS($n$e,$m$o) indicates the complete active space with $n$ being the number of electrons and $m$ the spatial orbitals in the active space (with e and o labelling electrons and spatial orbitals).
The active space model of FeMoco also includes additional Mo-4d and central C-2s and C-2p orbitals, resulting in a CAS(113e,76o) for 4Fe(III)3Fe(II)Mo(III).

The energy landscape of ansatz approximations to FeMoco ground-state wavefunctions with total spin $S=3/2$ is characterized by numerous local electronic minima.
Therefore, obtaining the correct ground state requires starting from a good initial guess.
To achieve this, we used a density matrix renormalization group (DMRG) initialization procedure as described in earlier studies~\cite{li2017spin, li2019electronic} within the implementation in \textsc{Block2}~\cite{zhai2023block2,zhai2021low}.
For the initial guess in the spin-adapted DMRG calculations, we first performed a spin-projected MPS calculation, initiated by a spin-projected broken-symmetry determinant.
We explored 35 different spin-projected determinants corresponding to the broken-symmetry configuration of \{2Fe(III)$\uparrow$,2Fe(III)$\downarrow$,2Fe(II)$\uparrow$,Fe(II)$\downarrow$,Mo(III)$\downarrow$\}, which were previously studied using broken-symmetry density functional theory in Ref.~\cite{cao2018influence}.
The resulting spin-projected MPS was then optimized up to a bond dimension of 50.

\begin{figure}[tbh]
    \centering
    \includegraphics[width=0.85\linewidth]{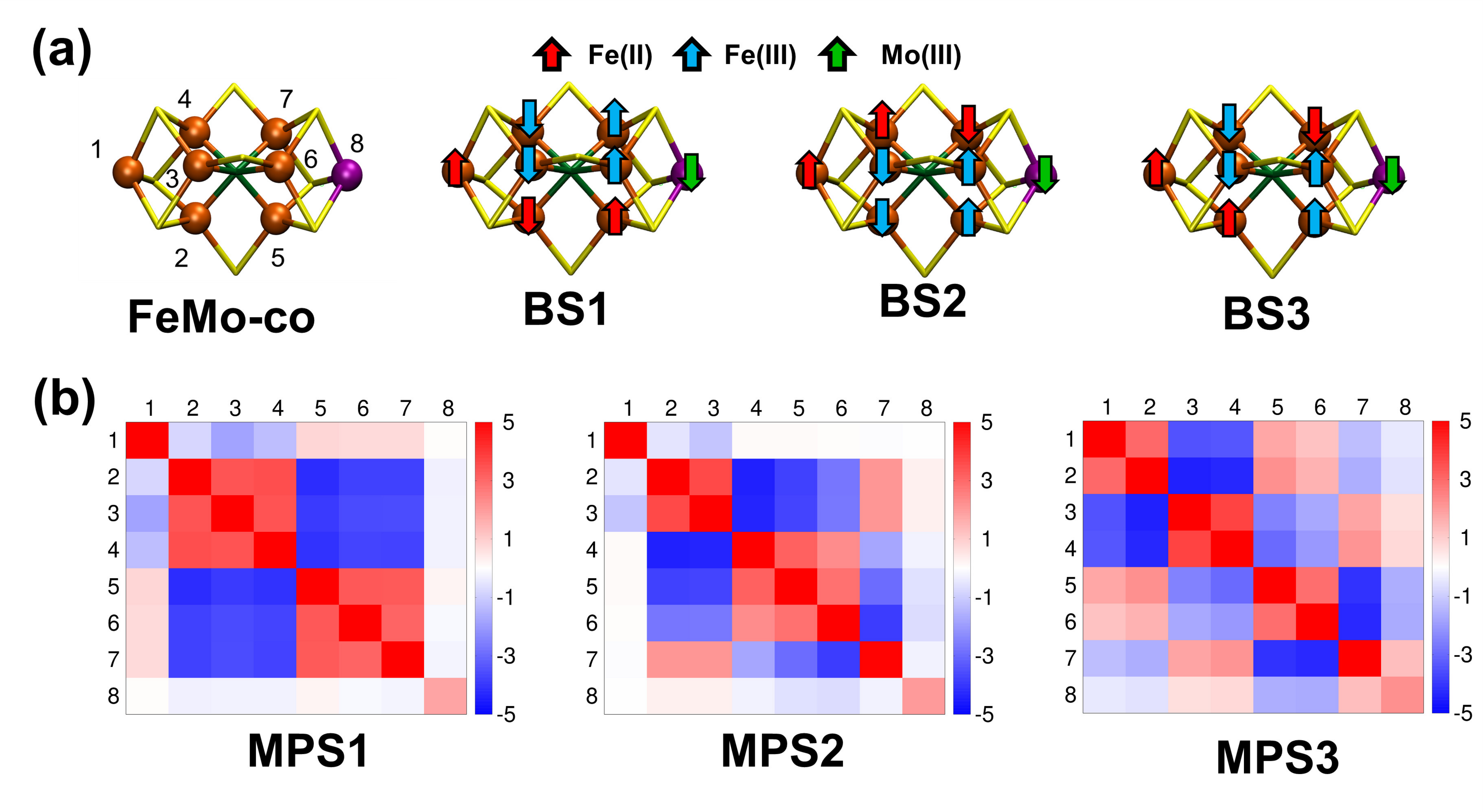}
    \caption{
    (a) A schematic representation of the metal site numbering in FeMoco, along with the spin-projected broken-symmetry determinants used in the spin-projected MPS calculation. Out of 35 broken-symmetry determinants, the three with the lowest DMRG energies at a bond dimension of 2000 are represented.
    (b) Spin correlation matrices for the three chosen MPSs. MPS1, MPS2 and MPS3 start from the BS1, BS2, and BS3 guesses, respectively, and converged to different states.}
    \label{fig:femocoDMRG}
\end{figure}

These initial MPSs were subsequently optimized using spin-adapted DMRG calculations, with the bond dimension increased to 2000.
Out of these 35 MPSs, we selected the three lowest-energy states at this bond dimension as example initial states, denoted MPS1, MPS2, and MPS3.
The character of the spin-couplings of these states, as represented by the spin-projected determinants used to initialize them, is shown in Fig.~\ref{fig:femocoDMRG}(a).
The three MPSs were further optimized, increasing the bond dimension to 7000 for MPS1 and 4000 for MPS2 and MPS3.
Figure~\ref{fig:femocoDMRG}(b) shows the spin correlation matrix $\langle S_{A}\cdot S_{B}\rangle$ between metal centers $\{A,B\}$ of these further optimized MPSs, which is defined by 
\begin{align}
S_{A}\cdot S_{B} &= \sum_{\mu\in \{x,y,z\}}S_{A}^{\mu}S_{B}^{\mu} \quad \mathrm{with} \\
S_{A}^{\mu} &= \sum_{p\in A}s^{\mu}_{p}\, , \\
s_{p}^{x} &= \frac{1}{2}\left(a_{p\uparrow}^{\dagger}a_{p\downarrow} + a_{p\downarrow}^{\dagger}a_{p\uparrow}\right),\\
s_{p}^{y} &= \frac{1}{2i}\left(a_{p\uparrow}^{\dagger}a_{p\downarrow} - a_{p\downarrow}^{\dagger}a_{p\uparrow}\right),\\
s_{p}^{z} &= \frac{1}{2}\left(a_{p\uparrow}^{\dagger}a_{p\uparrow} - a_{p\downarrow}^{\dagger}a_{p\downarrow}\right),
\end{align}
such that $p$ indexes the orbitals local to metal center $A$.

\begin{figure}[tbh]
    \centering
    \includegraphics[width=0.9\linewidth]{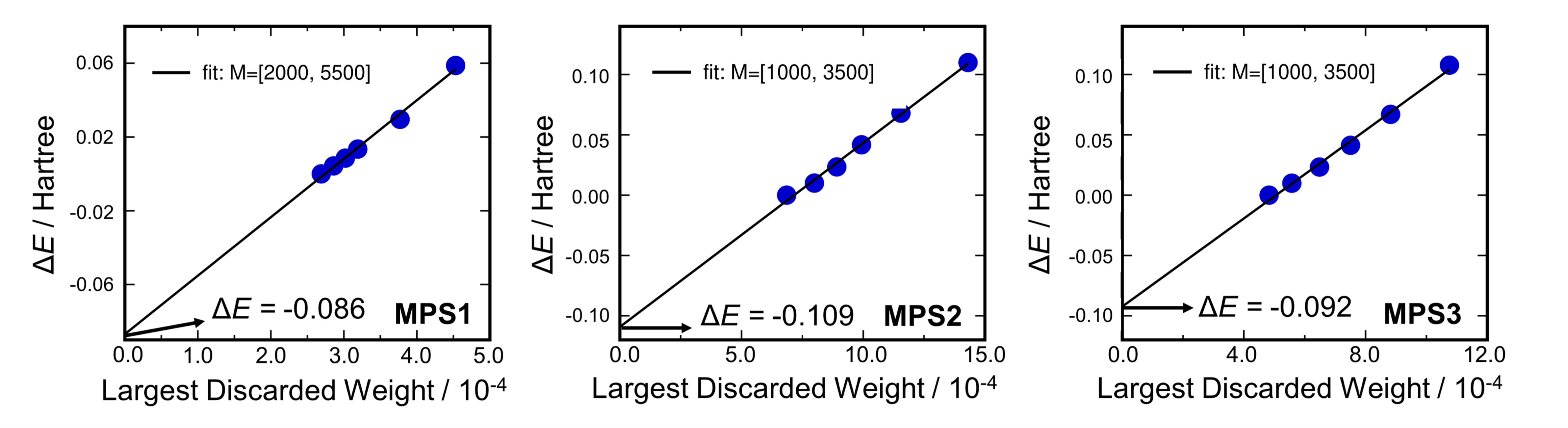}
    \caption{Extrapolated DMRG energy for FeMoco with MPS1, MPS2, and MPS3 with respect to discarded weight, assuming a linear relationship between discarded weight and energy.
    The energy differences ($\Delta E$) in Hartree are represented relative to the DMRG energy at the highest bond dimension used for each extrapolation, which are $-22140.270$, $-22140.249$, and $-22140.223$ for MPS1 ($\bondb=5500$), MPS2 ($\bondb=3500$), and MPS3 ($\bondb=3500$), respectively.
    }
    \label{fig:femoco_linear_discarded_weight}
\end{figure}

We used a well-known energy extrapolation scheme to estimate the energy errors of these MPSs \cite{legeza1996accuracy,olivares2015ab,li2019electronic,xiang2024distributed}.
Figure~\ref{fig:femoco_linear_discarded_weight} shows the extrapolations for the three energies using data obtained with the reverse-schedule DMRG.
In the zero discarded weight limit, which represents the exact MPS, the energies are expected to be 86, 109, and 92 milliHartree lower than the DMRG energies of MPS1 ($\bondb=5500$), MPS2 ($\bondb=3500$), and MPS3 ($\bondb=3500$), respectively.
Because these three states contain qualitatively different correlations, we consider these estimates to correspond to different low-energy eigenstates in the system.
In this section we use $\bondb$ for the bond dimension for consistency with the literature on DMRG, in contrast to the notation $\bonda$ used for MPS preparation.

\subsection{Overlap extrapolation protocol}\label{sec:overlap_mps_extrapolation}
\begin{figure}[tbh]
    \centering
    \includegraphics[width=0.95\linewidth]{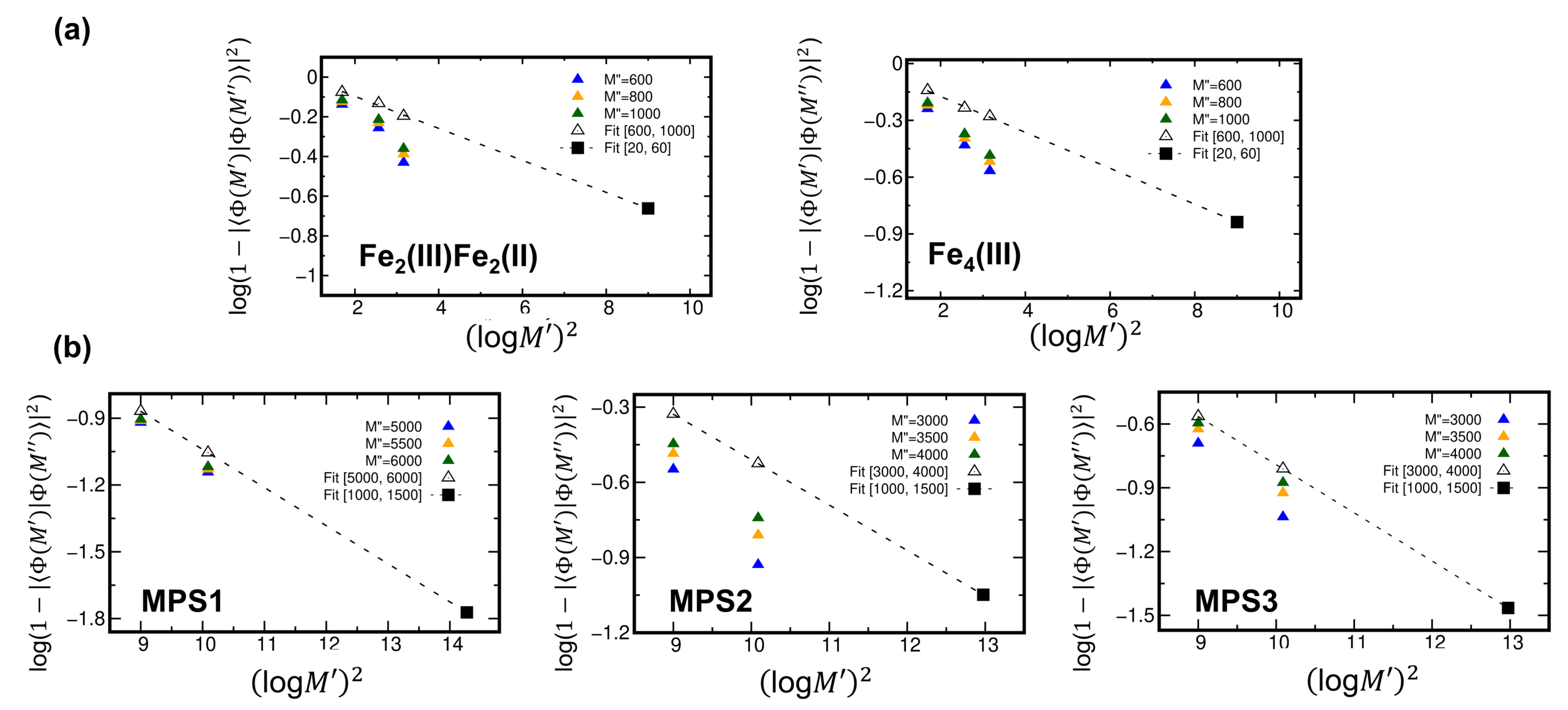}
    \caption{Extrapolated overlap for (a) 2Fe(II)2Fe(III), 4Fe(III), and (b) FeMoco.  }
    \label{fig:cubane}
\end{figure}

Here, we present an extrapolation scheme to predict the squared overlap between a spin-adapted DMRG state with a given bond dimension $\bondb$ and the exact wave function, $ |\langle \Phi(\bondb) | \Phi(\infty) \rangle|^2$. For the extrapolation, we utilized the following two empirical linear relations:
\begin{align}
	\log\left(1 - \left|\langle \Phi(\bondb') | \Phi(\infty) \rangle \right|^2\right) \ & \mathrm{vs.}\ \left(\log(\bondb')\right)^2 \label{eq:lin1} \\
	\log\left(\left|\langle \Phi(\bondb') | \Phi(\bondb'') \rangle \right|^2 - \left|\langle \Phi(\bondb') | \Phi(\infty) \rangle \right|^2\right) \ & \mathrm{vs.}\ \left(\log(\bondb'')\right)^2 \ \mathrm{where} \ \bondb' \ll \bondb'' \label{eq:lin2}
\end{align}
In Appendix~\ref{app:valov}, we show that these linear relations are satisfied in the \ce{Fe_2S_2} system where the exact wave function for the complete active space model of CAS(30e,20o) is accessible.
Building on these empirical linear relations, here we demonstrate the extrapolation scheme in detail.
Specifically, we show an example of estimating $|\langle \Phi(\bondb=1000)|\Phi(\infty) \rangle|^2$ for the 2Fe(II)2Fe(III) system. 
The main objective of this scheme is to accurately determine the squared overlap using data from MPSs with bond dimensions less than $\bondb=1000$.
To achieve this, we first generated MPSs with bond dimensions of 800, 600, 60, 40, and 20 using a reverse sweep  DMRG calculation. 
We then estimated $|\langle \Phi(\bondb'=20) | \Phi(\infty) \rangle|^2$ using the values of \mbox{$|\langle \Phi(\bondb'=20) | \Phi(\bondb'') \rangle|^2$} for $\bondb''=600,800,1000$ based on the second empirical linear relation of Eq.~\eqref{eq:lin2}.
We perform a linear fit of 
\begin{equation}
  \log\left(\left|\langle \Phi(20) | \Phi(\bondb'') \rangle \right|^2 - \left|\langle \Phi(20) | \Phi(\infty) \rangle \right|^2\right) \ \mathrm{vs.}\ \left(\log(\bondb'')\right)^2  
\end{equation}
to determine the value of $|\langle \Phi(20) | \Phi(\infty) \rangle|^2$.
In a similar fashion, we can estimate $|\langle \Phi(\bondb') | \Phi(\infty) \rangle|^2$ for the other bond dimensions of $\bondb'=40$ and $60$.
The empty black triangles in Fig.~\ref{fig:cubane}(a) represent the infidelities for these estimated values using $\bondb'=20, 40, 60$.
Each empty black triangle was estimated from the blue, yellow, and green triangles directly below it.
In Appendix~\ref{app:valov}, we discuss the validity of this extrapolation in detail.

Finally, these estimated values were used to obtain a linear fit represented by the dotted line in Fig.~\ref{fig:cubane}(a), based on the first empirical linear relationship of Eq.~\eqref{eq:lin1}.
Using this line, we estimate the value of $|\langle \Phi(1000) | \Phi(\infty) \rangle|^2$ at the black square.

We used the same extrapolation scheme to predict the overlap for the 4Fe(III) and FeMoco systems.
For 4Fe(III), we utilized MPSs with $\bondb' = 20, 40, 60$ and $\bondb'' = 600, 800, 1000$. For FeMoco, we used MPSs with $\bondb' = 1000, 1500$ and $\bondb'' = 5000, 5500, 6000$ for MPS1 and $\bondb'' = 3000, 3500, 4000$ for MPS2 and MPS3. 
It is noted that we used MPS1 with $\bondb=6000$ for the overlap estimate although MPS1 with $\bondb=7000$ was obtained from the DMRG calculation. This was because the site of the renormalized wavefunction with $\bondb=7000$ differed from the MPSs obtained through the reversed sweep, making it difficult to compute the overlap.
The predicted overlap values are summarized in Table~\ref{tab:ov_est_cost} along with the associated state preparation costs using the unitary synthesis techniques described earlier.
\begin{table}[H]
    \centering
    \begin{tabular}{|c|c|c|c|c|c|}
    \hline
    \hline
   System  &  Estimated $|\langle \mathrm{MPS}|\psi_{0}\rangle|$ & Bond Dimension & Spatial Orbitals & MPS Toffolis  & qubits \\
   \hline
   Fe$_{2}$(III)Fe$_{2}$(II) &  0.88 & 1000 & 36 & \num{42204873} & 359 \\
   Fe$_{4}$(III) &  0.92 & 1000 & 36 & \num{42204873} & 359 \\
   FeMoco~\cite{li2019electronicFemco} MPS1 &  0.99  &  6000 & 76 & \num{1360397118} & 833 \\
   FeMoco~\cite{li2019electronicFemco} MPS2 &  0.95  &  4000 & 76 & \num{732946035} & 682 \\
   FeMoco~\cite{li2019electronicFemco} MPS3 &  0.98  &  4000 & 76 & \num{732946035} & 682 \\
    \hline
    \hline
\end{tabular}
\caption{Extrapolated MPS overlaps as the output of the protocol described in Section~\ref{sec:prep}. The physical dimension of each subsystem for the MPS is taken to be $d=4$ corresponding to $\{|\emptyset\rangle, |{\uparrow}\rangle, |{\downarrow}\rangle, |{\uparrow\downarrow}\rangle\}$. In the case of FeMoco, the overlap is with respect to a low-energy eigenstate, which may not be the ground state. Toffoli and qubit counts are calculated using Qualtran~\cite{harrigan2024expressinganalyzingquantumalgorithms}.} \label{tab:ov_est_cost}
\end{table}

\subsection{MPS preparation costs versus phase estimation costs}
With the costs of performing MPS preparation, extrapolated overlap values, and number of samples using phase estimation, we are in a position to perform resource estimation for the full task of ground state energy estimation.  The cost of performing block encodings for two [4Fe-4S] systems and FeMoco are determined using the tensor hypercontraction factorization of the two-electron integral tensor~\cite{PRXQuantum.2.030305} and double factorization~\cite{PhysRevResearch.3.033055} using number operator symmetry shifting described in Ref.~\cite{loaiza2023reducing} on the one-body and two-body components of the Hamiltonian. For THC resource estimates we apply symmetry shifting only to the one-body component. We provide symmetry shifted and non-symmetry shifted block encoding costs along with high-spin coupled cluster correlation energies (obtained using PySCF~\cite{sun2020recent,sun2018pyscf}) for a variety of cutoff parameters in Appendix~\ref{app:threshold_DF_THC}.  For resource estimates we use the 1 milliHartree threshold in correlation energy differences (with respect to the non-truncated integrals) previously used in Ref.~\cite{PRXQuantum.2.030305}.  We find that for all iron-sulfur clusters studied the 1-norm value is approximately halved by symmetry shifting.

For the cost of sampling to obtain the confidence intervals, one could use Eq.~\eqref{eq:eq1_kaiser_total_queries}, but in this high-overlap regime that expression tends to overestimate the cost.
For a more accurate estimate one can perform the following procedure.
\begin{enumerate}
    \item For a given number of samples $n$, solve Eq.~\eqref{eq:optimal_delta_wrt_n} for $\delta$.
    \item Given that $\delta$, solve Eq.~\eqref{eq:exactslep} for $c$.
    \item Take the cost as $nc\lambda/\epsilon$.
\end{enumerate}
This procedure is used for a number of values of $n$ to find the one that gives the minimum cost.
Here we are using the prolate spheroidal window, which we find to give the best performance in the high-overlap regime.
In the results below we also use the method in Section \ref{sec:exciting} to obtain a more accurate estimate accounting for the excited states.
In this high overlap regime it only gives a small correction.

Given these sampling costs, we then use this overhead multiplied by the block encoding cost.
We then add $n$ times the MPS state preparation cost to obtain the total complexity.
For consistency with previous studies we select $\epsilon=1.0\times 10^{-3}$ Hartree, though here we are taking this to be the confidence interval half-width, rather than the root-mean square (RMS) error as in previous studies.
As a result, the accuracy requirement is somewhat more demanding, as confidence intervals are significantly wider than the RMS error.

In our estimates, we reduced the value of $\lambda_{DF}$ with a symmetry shift of $(\alpha_2/ 2)\hat N\hat N + (\alpha_1 - \alpha_2 / 2)\hat N$ where $\hat N$ is the number operator and $\alpha_1$ and $\alpha_2$ are optimized to minimize $\lambda_{DF}$. With this shift, the 1-norm of the effective electron repulsion integral (ERI) tensor depends only on $\alpha_2$, which is optimized first subject to the constraint that the ERI tensor remains positive semi-definite. After $\alpha_2$ has been determined, $\alpha_1$ is chosen to minimize the 1-norm of the effective 1-electron part of the Hamiltonian which depends on both $\alpha_1$ and $\alpha_2$.
For THC, the LCU 1-norm $\lambda$ is computed with a number operator symmetry shift computed as the median of $\{f_{i}\}$ where $f_{i}$ are eigenvalues of the one-body operator being block encoded.

Using these results for $\lambda$-values, we analyze the cost of Fe-S cluster ground state energy estimation with 95\% and 99\% confidence levels.  The total Toffoli and qubit requirements are listed in Table~\ref{tab:totalFeMoco_Costs}.  Due to the overlaps being so high, few QPE samples are needed, only 2 in the example of FeMoco.
We should note that this corresponds to only one candidate MPS initial guess corresponding to a specific structure of spin-coupling. Including all the BS-DFT derived MPS initial guesses considered in this work would multiply this cost by 35. This number may be reduced by pre-filtering some of the MPS initial guesses by the DMRG extrapolated energies.

\begin{table}[H]
    \centering
    \footnotesize
    \begin{tabular}{|c|c|c|c|c|c|c|}
    \hline
    \hline
   System  & Method & $\lambda$ & BE Toffolis & Qubits & QPE Cost 95\% & QPE Cost 99\%  \\
   \hline
   Fe$_{2}$(III)Fe$_{2}$(II) & THC & 168.7143 & 9120 & 1149 & \num{1.33263}$\times 10^{10}$ & \num{2.44708}$\times 10^{10}$ \\
   & DF
   & 154.7362& 15545 & 3111 &  \num{2.07614}$\times 10^{10}$ & \num{3.81595}$\times 10^{10}$ \\ \hline
   Fe$_{4}$(III) & THC & 164.1287 & 8573 & 1081 & \num{8.36543}$\times 10^{9}$ & \num{1.66679}$\times 10^{10}$ \\
   & DF
   &  150.2923 & 15602 & 3113 & \num{1.38845}$\times 10^{10}$ &  \num{2.76923}$\times 10^{10}$ \\ \hline
   FeMoco~\cite{li2019electronicFemco} & THC & 781.8172 & 16923 & 2194 & \num{7.2731}$\times 10^{10}$ & \num{1.37641}$\times 10^{11}$ \\
   & DF &  582.4211 & 35006 & 6402 & \num{1.11284}$\times 10^{11}$ & \num{2.1131}$\times 10^{11}$\\
    \hline
    \hline
\end{tabular}
\caption{The combined QPE and MPS preparation costs for confidence intervals of half-width $\epsilon=1.0\times 10^{-3}$ Hartree at 95\% and 99\% confidence levels. The QPE costs for FeMoco are calculated using only the smallest 0.95 overlap for MPS2.\label{tab:totalFeMoco_Costs}}
\end{table}

For comparison to the costings given in Ref.~\cite{PRXQuantum.2.030305}, first note that we are using the Li Hamiltonian as from Ref.~\cite{li2019electronicFemco}, whereas Ref.~\cite{PRXQuantum.2.030305} also considered the Reiher Hamiltonian from Ref.~\cite{Reiher}.
The Li Hamiltonian has higher cost to simulate, and it is those costs in Ref.~\cite{PRXQuantum.2.030305} we should compare to.
The relevant costs to compare to are therefore the right column in Table III of Ref.~\cite{PRXQuantum.2.030305}.
For THC, the Toffoli cost is $3.2\times 10^{10}$, so the estimated cost here for a 95\% interval is about $2.3$ times higher.
This factor comes from three main considerations.
\begin{enumerate}
    \item Here we have used symmetry shifting to reduce $\lambda$ to $781.8$, which \emph{reduces} the complexity by a factor of about $1.54$ as compared to the $\lambda$ in Ref.~\cite{PRXQuantum.2.030305} of $1201.5$.
    For the value of $\lambda$, see the line for $M=450$ in Table V of Ref.~\cite{PRXQuantum.2.030305}.
    \item Requiring two samples doubles the complexity.
    \item Requiring the measurements to provide confidence intervals of half-width $\epsilon$ rather than RMS error of $\epsilon$ is more demanding, and also increases the cost.
\end{enumerate}

For DF, the estimated cost from Ref.~\cite{PRXQuantum.2.030305} is $6.4\times 10^{10}$, so the estimated cost for 95\% confidence intervals here is about $1.7$ times larger.
The reduction in $\lambda$ due to symmetry shifting for DF is more significant than for THC.
The value of $\lambda$ is given in the line for $L=394$ in Table XIV of Ref.~\cite{PRXQuantum.2.030305} as 1171.2, so the improvement in $\lambda$ here is more than a factor of 2.
The other considerations for the sampling cost for DF are the same as for THC.

\section{Conclusions}
\label{sec:conc}

In this work we develop improved methods for ground state energy estimation both via improved initial state preparation, and improved filtering of the initial state.
For initial state preparation we developed an improved method for preparing matrix product states.
This preparation is based on our new technique for synthesising general unitary operations, which improves the Toffoli count by about a factor of 7 over prior work.
We then use that to construct an iterative procedure to prepare matrix product states with a substantially reduced complexity. The method for synthesising general unitary operations is of independent interest, as this is a very common task in quantum computing.
Moreover, we have found that it may be possible to further improve the complexity by a procedure interspersing phasing with Hadamard gates.
The drawback to that approach is that we do not have an efficient procedure to determine the phases required, so the unitaries are restricted to lower dimension (no more than about 8 qubits).
In future work it may be possible to develop more efficient procedures to solve for the phases, making that a more viable approach.

For improved filtering, we have proposed two approaches for ground state energy estimation, both using quantum phase estimation whose efficiency is boosted through window functions.
These window functions are chosen to minimize the error in a confidence interval, as opposed to RMS error which is more commonly considered.
This choice for the phase measurements is useful as we need to perform multiple phase measurements, all of which need to avoid large error.
The two methods are direct sampling and a binary search using amplitude estimation.
For both we provided both asymptotic expressions and numerical estimates of the complexity.
The advantage of the binary search with amplitude estimation is that it provides a square-root speedup in the overlap, though it has a larger constant factor than the direct sampling approach.
This means that the direct sampling approach is preferable for the case where the initial guess has large overlap with the exact ground state, whereas the binary search approach is more advantageous in the small overlap situation.
The asymptotic expressions suggest that the crossover is at $p\sim 0.003$, and that prediction is borne out by the numerics. 

Building on the efficient MPS state preparation results along with the optimal window function analysis, we analyzed the cost of refining energy estimates using QPE initialized with an MPS wavefunction. In order to determine total costs for energy refinement in the high confidence regime we determined the overlap through an extrapolation. The extrapolation protocol uses two MPS wavefunctions to determine the infinite bond dimension overlap of a finite bond dimension wavefunction. The extrapolation is empirical but is supported in this set of systems by verifying against true overlaps computed in the smaller FeS cluster where exact ground states can be found through large bond-dimension MPS calculations. In the case of FeMoco where the ground state energy manifold has many competing spin configurations we estimate the overlap for different MPS intial state wavefunctions that are candidates for different low-energy eigenstates. The role of QPE in this setting is then to refine the energy ordering of the states, enabling the determination of the ground-state energy.
Ultimately, due to the high extrapolated overlap, achieving 95\% or 99\% confidence intervals only requires two samples from QPE.
Improvements to block encoding LCU 1-norm through symmetry shifting results in total complexities that are only 2.3 times those of naive QPE assuming perfect overlap (and with the less demanding requirement of RMS error $\epsilon$)~\cite{PRXQuantum.2.030305}. 

The extrapolations and overlaps estimated here provide a concrete numerical example of a complex chemical problem where classical precomputation can be used to prepare initial states of high overlap, enabling efficient QPE. As discussed previously~\cite{Lee2023,chan2024quantum}, the degree of quantum advantage can then be evaluated from the relative cost of classical and quantum refinement from such an initial state. While the MPS wavefunctions considered here are attractive candidates for strongly correlated molecules up to a given finite size, other types of ansatz and techniques may also be used, particularly in the study of even larger strongly correlated molecular problems.
In conclusion, our work finds that algorithms based on classical state preparation and QPE provide a practical approach in real-world examples of challenging molecular chemistry.

\section*{Acknowledgements}

DWB worked on this project under a sponsored research agreement with Google Quantum AI. DWB is also supported by Australian Research Council Discovery Projects DP210101367 and DP220101602. Y.T., L.L., and G.K.C. were supported by the U.S. Department of Energy, Office of Science, National Quantum Information Science Research Centers, Quantum Systems Accelerator.

\bibliographystyle{apsrev4-1}

\appendix

\section{Further details of higher-order approximations}
\label{app:higher}

Here we give further details of the asymptotic expansions used to give approximations for the Kaiser window from Subsection \ref{sec:higher}.
First, continuing the integration by parts from Eq.~\eqref{eq:intbypart1} gives
\begin{align}
    &-\int_{\pi}^\infty \frac{\cos 2x}{x\sqrt{x^2+\pi^2\alpha^2}} dx \nn
    &= -\left[ \frac{{\rm Ci}(2x)}{\sqrt{x^2+\pi^2\alpha^2}}\right]_\pi^\infty - \int_\pi^\infty \frac{x\,{\rm Ci}(2x)}{(x^2+\pi^2\alpha^2)^{3/2}} dx \nn
    &= \frac{{\rm Ci}(2\pi)}{\pi \sqrt{1+\alpha^2}} - \int_\pi^\infty \frac{x\,{\rm Ci}(2x)}{(x^2+\pi^2\alpha^2)^{3/2}} dx \nn
    &= \frac{{\rm Ci}(2\pi)}{\pi \sqrt{1+\alpha^2}} + \left[ \frac{\cos(2x)+2x\sin(2x)-4x^2\,{\rm Ci}(2x)}{8(x^2+\pi^2\alpha^2)^{3/2}} \right]_\pi^\infty + \int_\pi^\infty \frac{3x[\cos(2x)+2x\sin(2x)-4x^2\,{\rm Ci}(2x)]}{8(x^2+\pi^2\alpha^2)^{5/2}} dx \nn
    &= \frac{{\rm Ci}(2\pi)}{\pi \sqrt{1+\alpha^2}} + \frac{4\pi^2\,{\rm Ci}(2\pi)-1}{8\pi^3(1+\alpha^2)^{3/2}} + \int_\pi^\infty \frac{3x[\cos(2x)+2x\sin(2x)-4x^2\,{\rm Ci}(2x)]}{8(x^2+\pi^2\alpha^2)^{5/2}} dx \nn
    &= \frac{{\rm Ci}(2\pi)}{\pi \sqrt{1+\alpha^2}} + \frac{4\pi^2\,{\rm Ci}(2\pi)-1}{8\pi^3(1+\alpha^2)^{3/2}} - \left[ \frac{3[
    (2 x^2-3) \cos(2 x)-  2 x (2 x^2+3) \sin(2 x) + 8 x^4 {\rm Ci}(2 x)    ]}{64(x^2+\pi^2\alpha^2)^{5/2}}\right]_\pi^\infty \nn
    &\quad - \int_\pi^\infty
    \frac{15x[
    (2 x^2-3) \cos(2 x)-  2 x (2 x^2+3) \sin(2 x) + 8 x^4 {\rm Ci}(2 x)    ]}{64(x^2+\pi^2\alpha^2)^{9/2}} dx \nn
    &= \frac{{\rm Ci}(2\pi)}{\pi \sqrt{1+\alpha^2}} + \frac{4\pi^2\,{\rm Ci}(2\pi)-1}{8\pi^3(1+\alpha^2)^{3/2}} +  \frac{3[
    (2 \pi^2-3) + 8 \pi^4 {\rm Ci}(2 \pi)    ]}{64\pi^5(1+\alpha^2)^{5/2}} \nn
    &\quad - \int_\pi^\infty
    \frac{15x[
    (2 x^2-3) \cos(2 x)-  2 x (2 x^2+3) \sin(2 x) + 8 x^4 {\rm Ci}(2 x)    ]}{64(x^2+\pi^2\alpha^2)^{7/2}} dx \, .
\end{align}
Then dividing by the normalisation and expanding in a series gives
\begin{align}\label{eq:errorserall}
   & 4\sqrt\alpha e^{-2\pi\alpha} \left[ C_\alpha - \frac{5 C_\alpha}{16 \pi \alpha}
 - \frac{(64 + 79 C_\alpha - 128 \pi^2)}{ 2^9 \pi^2 \alpha^2}       
  + \frac{5 (64 - 323 C_\alpha - 128 \pi^2)}{ 2^{13} \pi^3 \alpha^3}\right. \nn
  &\left.
  - \frac{ 195253 C_\alpha + 128 (497 - 994 \pi^2 + 384 \pi^4)}{2^{19} \pi^4 \alpha^4}
 - \frac{7579195 C_\alpha - 640 (899 - 1798 \pi^2 + 384 \pi^4)}{2^{23} \pi^5 \alpha^5}\right. \nn
  &\left.
-  \frac{2131315809 C_\alpha +   64 (6514017 - 13028034 \pi^2 + 4733184 \pi^4  - 655360 \pi^6)}{ 2^{28} 3 \pi^6 \alpha^6}    + \mathcal{O}(\alpha^{-7})
    \right] ,
\end{align}
where we are giving further terms beyond that in Eq.~\eqref{eq:errorser}.

Integrating over the tails for $\Delta\ne 1$ gives
\begin{align}
    2\int_{(\pi/\Nphas)\sqrt{\Delta^2+\alpha^2}}^\infty \frac{\sin^2\sqrt{\Nphas^2\phi^2-\pi^2\alpha^2}}{(\Nphas^2\phi^2-\pi^2\alpha^2)} d\phi
    &= \frac{2}{\Nphas}\int_{\pi\sqrt{\Delta^2+\alpha^2}}^\infty \frac{\sin^2\sqrt{\phi^2-\pi^2\alpha^2}}{(\phi^2-\pi^2\alpha^2)} d\phi \nn
    &= \frac{2}{\Nphas}\int_{\pi\Delta}^\infty \frac{\sin^2 x}{x\sqrt{x^2+\pi^2\alpha^2}} dx \nn
    &= \frac{1}{\Nphas}\int_{\pi\Delta}^\infty \frac{1-\cos 2x}{x\sqrt{x^2+\pi^2\alpha^2}} dx \nn
    &= \frac{\,{\rm arcsinh}(\alpha/\Delta)}{\pi \Nphas \alpha} - \frac{1}{\Nphas}\int_{\pi\Delta}^\infty \frac{\cos 2x}{x\sqrt{x^2+\pi^2\alpha^2}} dx \, .
\end{align}
Integration by parts gives
\begin{align}
    &-\int_{\pi\Delta}^\infty \frac{\cos 2x}{x\sqrt{x^2+\pi^2\alpha^2}} dx \nn
    &= -\left[ \frac{{\rm Ci}(2x)}{\sqrt{x^2+\pi^2\alpha^2}}\right]_{\pi\Delta}^\infty - \int_{\pi\Delta}^\infty \frac{x\,{\rm Ci}(2x)}{(x^2+\pi^2\alpha^2)^{3/2}} dx \nn
    &= \frac{{\rm Ci}(2\pi\Delta)}{\pi \sqrt{\Delta^2+\alpha^2}} - \int_{\pi\Delta}^\infty \frac{x\,{\rm Ci}(2x)}{(x^2+\pi^2\alpha^2)^{3/2}} dx \nn
    &= \frac{{\rm Ci}(2\pi\sqrt\Delta)}{\pi \sqrt{\Delta^2+\alpha^2}} + \left[ \frac{\cos(2x)+2x\sin(2x)-4x^2\,{\rm Ci}(2x)}{8(x^2+\pi^2\alpha^2)^{3/2}} \right]_{\pi\Delta}^\infty + \int_{\pi\Delta}^\infty \frac{3x[\cos(2x)+2x\sin(2x)-4x^2\,{\rm Ci}(2x)]}{8(x^2+\pi^2\alpha^2)^{5/2}} dx \nn
    &= \frac{{\rm Ci}(2\pi\Delta)}{\pi \sqrt{\Delta^2+\alpha^2}} + \frac{4\pi^2\Delta^2\,{\rm Ci}(2\pi)-\cos(2\pi\Delta)-2\pi\Delta\sin(2\pi\Delta)}{8\pi^3(\Delta^2+\alpha^2)^{3/2}} + \int_{\pi\Delta}^\infty \frac{3x[\cos(2x)+2x\sin(2x)-4x^2\,{\rm Ci}(2x)]}{8(x^2+\pi^2\alpha^2)^{5/2}} dx \nn
    &= \frac{{\rm Ci}(2\pi\Delta)}{\pi \sqrt{\Delta^2+\alpha^2}} + \frac{4\pi^2\Delta^2\,{\rm Ci}(2\pi)-\cos(2\pi\Delta)-2\pi\Delta\sin(2\pi\Delta)}{8\pi^3(\Delta^2+\alpha^2)^{3/2}} \nn & \quad - \left[ \frac{3[
    (2 x^2-3) \cos(2 x)-  2 x (2 x^2+3) \sin(2 x) + 8 x^4 {\rm Ci}(2 x)    ]}{64(x^2+\pi^2\alpha^2)^{5/2}}\right]_{\pi\Delta}^\infty \nn
    &\quad - \int_{\pi\Delta}^\infty
    \frac{15x[
    (2 x^2-3) \cos(2 x)-  2 x (2 x^2+3) \sin(2 x) + 8 x^4 {\rm Ci}(2 x)    ]}{64(x^2+\pi^2\alpha^2)^{9/2}} dx \nn
    &= \frac{{\rm Ci}(2\pi\Delta)}{\pi \sqrt{\Delta^2+\alpha^2}} + \frac{4\pi^2\Delta^2\,{\rm Ci}(2\pi)-\cos(2\pi\Delta)-2\pi\Delta\sin(2\pi\Delta)}{8\pi^3(\Delta^2+\alpha^2)^{3/2}} \nn
    & \quad + 
    \frac{3[
    (2 \pi^2\Delta^2-3) \cos(2 \pi\Delta)-  2 \pi\Delta (2 \pi^2\Delta^2+3) \sin(2 \pi\Delta) + 8 \pi^4\Delta^4 {\rm Ci}(2 \pi\Delta)    ]}{64\pi^5 (\Delta^2+\pi^2\alpha^2)^{5/2}} \nn
    &\quad - \int_{\pi\Delta}^\infty
    \frac{15x[
    (2 x^2-3) \cos(2 x)-  2 x (2 x^2+3) \sin(2 x) + 8 x^4 {\rm Ci}(2 x)    ]}{64(x^2+\pi^2\alpha^2)^{7/2}} dx \, .
\end{align}
Dividing by the normalisation gives the error as
\begin{align}
   & 4\sqrt\alpha e^{-2\pi\alpha} \left[ C_{\alpha,\Delta} - \frac{5 C_{\alpha,\Delta}}{16 \pi \alpha}
 - \frac{(64\cos(2\pi\Delta)+128\pi\Delta\sin(2\pi\Delta) + 79 C_{\alpha,\Delta} - 128 \pi^2\Delta^2)}{ 2^9 \pi^2 \alpha^2} \right.     \nn &  \left.
  + \frac{5 (64\cos(2\pi\Delta)+128\pi\Delta\sin(2\pi\Delta) - 323 C_{\alpha,\Delta} - 128 \pi^2\Delta^2)}{ 2^{13} \pi^3 \alpha^3}\right. \nn
  &\left.
  - \frac{ 195253 C_{\alpha,\Delta}+ 128 [
  2 \pi^2 \Delta^2 (79 + 192 \pi^2 \Delta^2) + (497 - 
    1152 \pi^2 \Delta^2) \cos(
   2 \pi {\Delta}) + 
 2 \pi {\Delta} (497 - 
    384 \pi^2 \Delta^2) \sin(2 \pi {\Delta})  
  ]}{2^{19} \pi^4 \alpha^4}\right. \nn
  &\left.
 - \frac{7579195 C_{\alpha,\Delta} + 640 [
 2 \pi^2 \Delta^2 (323 - 
    192 \pi^2 \Delta^2) - (899 - 
    1152 \pi^2 \Delta^2) \cos(
   2 \pi \Delta) - 
 2 \pi \Delta (899 - 
    384 \pi^2 \Delta^2) \sin(2 \pi \Delta)]
 }{2^{23} \pi^5 \alpha^5}\right. \nn
  &\left.
  + \mathcal{O}(\alpha^{-6})
    \right] ,
\end{align}
where $C_{\alpha,\Delta}=\ln(2\alpha/\Delta)+{\rm Ci}(2\pi\Delta)$.
If we take $c=\pi\sqrt{\Delta^2+\alpha^2}$ and expand in a series in $c$, then we obtain
\begin{align}
    &4\sqrt{c/\pi} e^{-2c} \left[ 
    C_{c,\Delta} - \frac{(5-16\pi^2\Delta^2)C_{c,\Delta}}{16 c} \right. \\
    & \left. - \frac{[64\cos(2\pi\Delta)+128\pi\Delta\sin(2\pi\Delta) + (79+288  \pi^2  \Delta^2 - 256  \pi^4  \Delta^4) C_{c,\Delta} - 128 \pi^2\Delta^2]}{ 2^9 c^2}
      \right.     \nn &  \left.
  + \frac{64(5 - 16 \pi^2 \Delta^2)[\cos(2\pi\Delta)+2\pi\Delta\sin(2\pi\Delta)] - (
  1615 + 1904 \pi^2 \Delta^2 + 
 1280 \pi^4 \Delta^4 - 2^{12} \pi^6 \Delta^6/3
  ) C_{c,\Delta}}{ 2^{13} c^3}\right.  \nn
  &\left.
  + \frac{640 \pi^2\Delta^2-2^{11}\pi^4\Delta^4}{ 2^{13} c^3} + \mathcal{O}(c^{-4})
    \right] ,
\end{align}
where $C_{c,\Delta}=\ln(2c/\pi\Delta)+{\rm Ci}(2\pi\Delta)$.

\section{Slepian expansion}
\label{app:slepian}
Here we explain the details of how to derive the higher-order terms to correct Eq.~(4.4) of Slepian \cite{Slepian1965}.
We start from Eq.~(4.3) of Slepian, which is
\begin{equation}
    \frac 1{\lambda_n} \frac{\partial \lambda_n}{\partial c} = \frac 2c \left[ \psi_{0,n}(1) \right]^2 \, .
\end{equation}
Now the error for the prolate spheroidal window is given by $1-\lambda_0$, so we need $n=0$.
According to the expression below Eq.~(4.3) of \cite{Slepian1965}, $\left[ \psi_{0,0}(1) \right]^2= N^2_{0,0}k_3^2$.
Now $k_3$ is given in Eq.~(1.12) of \cite{Slepian1965} as
\begin{equation}
    k_3 = e^{-c} c^{(l+1)/2} 2^{(3l+2)/2} \sqrt{\pi} P(c) Q(c).
\end{equation}
Now $l=n-m$, as per the expression below Eq.~(1.10) of \cite{Slepian1965}, and $n\ge m\ge 0$, so with $n=0$ we have $m=l=0$.
Therefore the expression for $k_3$ simplifies to
\begin{equation}
    k_3 = 2 e^{-c} \sqrt{\pi c} 2^{(3l+2)/2} P(c) Q(c).
\end{equation}
For $m=0$ we have $Q(c)=1$, according to the explanation below Eq.~(1.15) of \cite{Slepian1965}.
The function $P(c)$ is given in terms of $g$ coefficients in Eq.~(1.14) of \cite{Slepian1965}, which is
\begin{equation}
    P(c) = \frac{1+g_1^1/c + g_2^1/c^2+\cdots}{1+g_1^2/c + g_2^2/c^2+\cdots}.
\end{equation}
The values of $g$ are given as per Table IV on page 103 of that work, which have the simplified form for $l=m=0$
\begin{align}
    g_1^1 &= -\frac{24}{2^8} \nn
g_2^1 &= -\frac{51840}{3!\, 2^{17}} \nn
g_3^1 &= -\frac{857226240}{3!\, 6!\,  2^{21}} \nn
g_1^2 &= \frac{64}{2^8} \nn
g_2^2 &= \frac{165888}{3!\times 2^{17}} \nn
g_3^2 &= \frac{902430720}{6!\, 2^{22}}\, .
\end{align}

Next, $N^2_{0,0}$ is given by Eq.~(1.16) of \cite{Slepian1965} as
\begin{equation}
    \frac 1 {N^2_{0,0}} = \sqrt{\pi/c} \left( 1 + \frac{3}{2^7 c^2} + \cdots \right) \, .
\end{equation}
This expression is missing the third-order term, meaning we cannot obtain that term in the final expression.
If we assumed that third-order term was zero, we would obtain
\begin{equation}
    \frac 2c \left[ \psi_{0,n}(1) \right]^2 = 2\sqrt{\pi c}\, e^{-2c} \left( 1- \frac{11}{16 c}  - \frac{147}{512 c^2} - \frac{3269}{8192 c^3} - \cdots \right) \, .
\end{equation}
Integrating (from $c$ to infinity) yields
\begin{equation}
    4\sqrt{\pi c}\,  e^{-2c} \left( 1 + \frac{1505}{1536 c} - \frac{3269}{6144 c^2} \right) - \frac{2177 \pi \, {\rm erfc}(\sqrt{2c})}{192 \sqrt 2} \, .
\end{equation}
Expanding an asymptotic series for the erfc function then gives
\begin{equation}
    4\sqrt{\pi c} \, e^{-2c} \left( 1-\frac 7{16c} -
    \frac{91}{2^{9}c^2}+
    \frac{2177}{2^{13} c^3} \right) \, .
\end{equation}
The third-order term here is different when a nonzero third-order term for $N^2_{0,0}$ is used.
Since the term found numerically is quite different, this indicates that the third-order term for $N^2_{0,0}$ is nonzero.

Next we describe how to derive Eq.~\eqref{eq:intps}.
In the discrete case the probability distribution for the phase error is
\begin{equation}
    \frac 1{2\pi} \sum_{n,m=-N}^N f(n)f(m) e^{i(n-m)\theta} \, .
\end{equation}
In the continuous limit, we replace $x=n/(N+1/2)$ and $z=m/(N+1/2)$, and use $\vartheta=(N+1/2)\theta$, to give
the probability distribution for the error
\begin{equation}
    \frac 1{2\pi} \int_{-1}^1 dx \int_{-1}^1 dz \,  f(x)f(z) \, e^{i(x-z)\vartheta} \, .
\end{equation}
with the convention that the continuous function $f(x)$ is normalised over the interval $[-1,1]$.
Then the integral for the confidence level is
\begin{equation}
    \frac 1{2\pi} \int_{-c}^c d\vartheta \int_{-1}^1 dx \int_{-1}^1 dz \,  f(x)f(z) \, e^{i(x-z)\vartheta}
    = \frac c{\pi}\int_{-1}^1 dx \int_{-1}^1 dz \,  f(x)f(z) \, {\rm sinc}(c(x-z)) \, .
\end{equation}
The solution for $f(x)$ with maximum confidence level then corresponds to an eigenfunction of maximum eigenvalue, so that
\begin{equation}
   (1-\delta) \,  f(x) = \frac c{\pi}\int_{-1}^1 dz \, f(z) \, {\rm sinc}(c(x-z)) \, .
\end{equation}
In particular, we have
\begin{equation}
   (1-\delta) \,  f(0) = \frac c{\pi}\int_{-1}^1 dz \, f(z) \, {\rm sinc}(cz) \, .
\end{equation}
We can use this expression for the function $f(z)=\text{PS}_{0,0}(c ,z)$ to give Eq.~\eqref{eq:intps}.
The function $\text{PS}_{0,0}(c ,z)$ is not normalised to 1, but that is unimportant because it appears on both sides of the equation so the normalisation cancels.

\section{Methods of calculating prolate spheroidal functions}
\label{app:calculate}

The prolate spheroidal functions are given in various mathematical software as follows.
\begin{enumerate}
    \item In Mathematica $\text{PS}_{0,0}(c ,z)$ is given as $\text{SpheroidalPS}[0, 0, c, z]$, and $S_{0,0}^1(c ,1)$ is given as $\text{SpheroidalS1}[0, 0, c, 1]$.
    These are normalised according to the Meixner-Schäfke scheme.
    \item Matlab gives the discrete prolate spheroidal sequences (i.e.\ for finite $N$) via the function dpss.
    \item In Python, scipy.signal.windows.dpss has similar functionality as dpss in Matlab.
    Python also provides scipy.special with pro\_ang1 and pro\_rad1 for the angular and radial prolate spheroidal functions, respectively.
    Unfortunately, pro\_rad1 only outputs nan in our testing, making it unusable.
\end{enumerate}

Very similar results to those obtained in Mathematica are obtained using dpss in Matlab, which is a useful independent verification.
To be more specific, consider the control state for phase estimation of the form
\begin{equation}
    \sum_{n=-\Nphas}^{\Nphas} f(n) \ket{n} \, .
\end{equation}
Applying a phase shift and taking an inner product with the phase state gives
\begin{equation}
    \frac 1{\sqrt{2N+1}} \sum_{n=-\Nphas}^{\Nphas} f(n) e^{in(\phi-\hat\phi)} \, .
\end{equation}
The inner product squared is
\begin{equation}
    \frac 1{2N+1} \sum_{n,m=-\Nphas}^{\Nphas} f(n)f(m) e^{i(n-m)\theta} \, ,
\end{equation}
where we have replaced $\phi-\hat\phi$ with $\theta$.
The usual convention for phase measurements is to consider a slightly different normalisation convention with a continuous range of $\theta$ values, so the probability distribution for the error is
\begin{equation}
    \frac 1{2\pi} \sum_{n,m=-\Nphas}^{\Nphas} f(n)f(m) e^{i(n-m)\theta} \, .
\end{equation}
Integrating $\theta$ over $[-\pi,\pi]$ then gives 1.

\begin{figure}[tbh]
    \centering
    \includegraphics[width=0.5\textwidth]{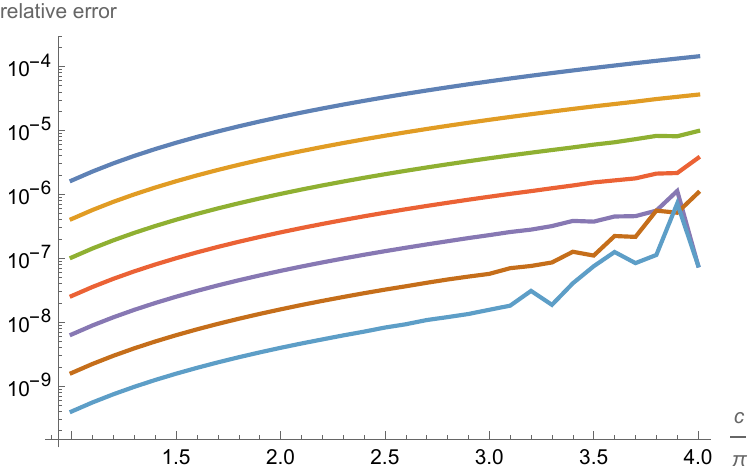}
    \caption{The relative error for the values of $\delta$ estimated with various values of $N$ using dpss in Matlab.
    The lines show the results using $\Nphas=2^{10}$ to $\Nphas=2^{16}$ from top to bottom.}
    \label{fig:dpss}
\end{figure}

To obtain the probability in the confidence interval $[-c/(N+1/2),c/(N+1/2)]$, the integral is
\begin{equation}\label{eq:conf}
    \frac 1{2\pi} \int_{-c/(\Nphas+1/2)}^{c/(\Nphas+1/2)} d\theta \sum_{n,m=-\Nphas}^{\Nphas} f(n)f(m) e^{i(n-m)\theta} = \frac{c}{\pi(\Nphas+1/2)} \sum_{n,m=-\Nphas}^{\Nphas} f(n)f(m)\, {\rm sinc}\left(\frac{c(n-m)}{\Nphas+1/2}\right) \, .
\end{equation}
Given $f(n)$ from dpss, this expression can be used to determine the value of $\delta$ for given $\Nphas$.
The function dpss also gives $1-\delta$ as an output, which can be used instead of performing the explicit sum.
The relative error for the value of $\delta$ estimated with various values of $\Nphas$ is shown in Fig.~\ref{fig:dpss}.
It can be seen that the results using dpss are very accurate, even for moderate values of $\Nphas$.
The results for $c>4\pi$ are obtained less accurately, because dpss gives $1-\delta$ to double precision accuracy, so cannot give accurate values for $\delta$ below about $10^{-14}$.
Nearly identical results are obtained using scipy.signal.windows.dpss in Python, though the function breaks down for $\Nphas>2^{15}$.

Note that dpss is giving the error for a specific value of $\Nphas$, so the relative errors seen in Fig.~\ref{fig:dpss} correspond to the difference between using the continuous window and the sampling of the continuous window for finite $\Nphas$.
For phase estimates of relevance to quantum computing, $\Nphas$ would be well above $2^{10}$, and even for that value the relative error is less than about $0.01\%$.
This difference is less than the significant figures usually reported for complexities, so is reasonable to ignore in the analysis.

These results can be further checked by determining the maximum eigenvalue of a matrix.
The maximisation of the confidence level in Eq.~\eqref{eq:conf} corresponds to finding the maximum eigenvalue of a matrix with entries
\begin{equation}
    A_{n,m} = \frac{c}{\pi(\Nphas+1/2)} {\rm sinc}\left(\frac{c(n-m)}{\Nphas+1/2}\right) \, ,
\end{equation}
for $n,m$ in the range $-\Nphas,\ldots,\Nphas$.
The maximum eigenvalue of this matrix can be determined from the spectral norm.
Numerical testing with a range of values of $c$ and $\Nphas$ yields results equal to those from dpss within numerical precision.

\section{Combinations of excited states}
\label{app:excited}

To show that the probability of error should not be increased for combinations of excited states with different $\beta$, let us first consider the case that we just have a linear combination of the excited state at $\beta$ around 2 and $\beta=0$.
That is equivalent to just considering the case where $p$ is increased.
As we are limiting the probability of error to $q$ for \emph{all} $\beta$, we have the probability of error $\le q$ for $\beta=0$, which corresponds to $p=1$.
We therefore consider Eq.~\eqref{eq:excitederr} for the error with $p\in[p_0,1]$ for $p_0$ the minimum squared overlap.
If there were a larger probability of error for some intermediate value of $p$, then we would need the first derivative of Eq.~\eqref{eq:excitederr} to pass through zero (so there is a turning point) with the second derivative being negative (so there is a maximum).

The first derivative gives
\begin{equation}
    \frac{d P_{\rm err}}{dp} = n(\delta/2-\delta_1)[p\delta/2+(1-p)\delta_1]^{n-1} + n(\delta/2-\delta_2)\{1-[p\delta/2 + (1-p)\delta_2]\}^{n-1} ,
\end{equation}
so the first derivative being zero implies
\begin{equation}
    \frac{(\delta_1-\delta/2)}{(\delta/2-\delta_2)} = \frac{\{1-[p\delta/2 + (1-p)\delta_2]\}^{n-1}}{[p\delta/2+(1-p)\delta_1]^{n-1}}.
\end{equation}
The second derivative gives
\begin{equation}
    \frac{d^2 P_{\rm err}}{dp^2} = n(n-1)(\delta/2-\delta_1)^2[p\delta/2+(1-p)\delta_1]^{n-2} - n(n-1)(\delta/2-\delta_2)^2\{1-[p\delta/2 + (1-p)\delta_2]\}^{n-2} ,
\end{equation}
so for the second derivative to be negative we would need
\begin{equation}
    \frac{(\delta_1-\delta/2)^2}{(\delta/2-\delta_2)^2} < \frac{\{1-[p\delta/2 + (1-p)\delta_2]\}^{n-2}}{[p\delta/2+(1-p)\delta_1]^{n-2}}.
\end{equation}
But, combining that with the first equality gives
\begin{equation}\label{eq:ineqp}
    {\delta_1-\delta/2}<{\delta/2-\delta_2} \, .
\end{equation}
The definitions trivially give that $\delta_1>\delta/2$ and $\delta/2<\delta_2$, because $\delta_1$ is the probability of the estimate being more than $E_0+\epsilon$ for measurements of the excited state.
Moreover, for $\beta>2$ we have $\delta_1\sim 1$ and both $\delta$ and $\delta_2$ very small.
This means that the LHS of Eq.~\eqref{eq:ineqp} is about 1, whereas the RHS is small, so this inequality cannot be satisfied.
As this inequality was needed to obtain a larger probability of error for intermediate values of $p$, that is not possible.

We can perform similar reasoning for the case where there are two excited states both with $\beta>0$.
In that case, let us take the value of $\delta_1$ to be a weighted linear combination of the values for the two excited states as $\delta_1 = s \delta_1^a + (1-s) \delta_1^b$, and similarly $\delta_2 = s \delta_2^a + (1-s) \delta_2^b$.
Now we can perform similar reasoning for the weighting factor $s$ as we used above for $p$.
To have a maximum of the probability of error for $s\in(0,1)$ we must have the first derivative pass through zero when the second derivative is negative.

The first derivative gives
\begin{equation}
    \frac{d P_{\rm err}}{ds} = n(1-p)(\delta_1^a-\delta_1^b)[p\delta/2+(1-p)\delta_1]^{n-1} + n(1-p)(\delta_2^a-\delta_2^b)\{1-[p\delta/2 + (1-p)\delta_2]\}^{n-1} ,
\end{equation}
so the first derivative being zero implies
\begin{equation}\label{eq:lt1}
    \frac{\delta_1^b-\delta_1^a}{\delta_2^a-\delta_2^b} = \frac{\{1-[p\delta/2 + (1-p)\delta_2]\}^{n-1}}{[p\delta/2+(1-p)\delta_1]^{n-1}}.
\end{equation}
The second derivative gives
\begin{equation}
    \frac{d^2 P_{\rm err}}{ds^2} = n(n-1)(p-1)^2(\delta_1^a-\delta_1^b)^2[p\delta/2+(1-p)\delta_1]^{n-2} - n(n-1)(p-1)^2(\delta_2^a-\delta_2^b)^2\{1-[p\delta/2 + (1-p)\delta_2]\}^{n-2} ,
\end{equation}
so for the second derivative to be negative we would need
\begin{equation}
    \left(\frac{\delta_1^b-\delta_1^a}{\delta_2^a-\delta_2^b}\right)^2 < \frac{\{1-[p\delta/2 + (1-p)\delta_2]\}^{n-2}}{[p\delta/2+(1-p)\delta_1]^{n-2}}.
\end{equation}
Then we would require the inequality
\begin{equation}
    \frac{\delta_1^b-\delta_1^a}{\delta_2^a-\delta_2^b} < 1 \, .
\end{equation}
That would imply that the RHS of Eq.~\eqref{eq:lt1} is required to be less than 1,
but that can be ruled out.
First, note that we must have $\delta_2<1-\delta_1$.
That is because $1-\delta_1$ is the probability of the measurement result being \emph{less than} $E_0+\epsilon$, and $\delta_2$ is the probability of the estimate being less than $E_0-\epsilon$ (which of course must be less).
Moreover, we must have $\delta<1$, and so we obtain
\begin{align}
    p\delta + (1-p)\delta_2 &< p+(1-p)(1-\delta_1) \, , \nn
    p\delta/2 + (1-p)\delta_2 &< 1-p\delta/2-(1-p)\delta_1 \, , \nn
    1-[p\delta/2 + (1-p)\delta_2] &> p\delta/2+(1-p)\delta_1 \, , \nn
    \frac{1-[p\delta/2 + (1-p)\delta_2]}{p\delta/2+(1-p)\delta_1} &>1 \, , \nn
\frac{\{1-[p\delta/2 + (1-p)\delta_2]\}^{n-1}}{[p\delta/2+(1-p)\delta_1]^{n-1}} &>1 \, .
\end{align}

Thus we find that it is impossible to obtain a larger error probability by taking a combination of two excited states.
Moreover, this argument shows that it is impossible to obtain a larger error probability with a combination of any number of excited states.
To see that result, we can use $\delta_1^a,\delta_2^a$ and $\delta_1^b,\delta_2^b$ to be the probabilities resulting from disjoint sets of eigenstates.
The reasoning that $\delta_2<1-\delta_1$ holds regardless of the combination of excited states, because the probability of a measurement result below $E_0+\epsilon$ must always be greater than the probability of a result below $E_0-\epsilon$.
Thus for any combination of excited states we can always split the set into two such that one of the subsets gives at least as large a value of $P_{\rm err}$.
Repeating this process gives a single excited state with at least as large an error probability.
Thus we can obtain the maximum error probability with a single excited state, and this result upper bounds the error probability for any spectrum of excited states.

\section{Costing for LKS scheme}
\label{app:LKS}

Here we give a detailed analysis of the cost of unitary synthesis using the method of Low, Kliuchnikov and Schaeffer \cite{low2018trading}.
For the unitary synthesis the approach of Ref.~\cite{low2018trading} is to perform a sequence of $K$ reflections $\openone - 2\ket{v_k}\bra{v_k}$ when $K$ columns of the unitary need be specified (here $K=\bonda$), as well as a diagonal operation (phases in the computational basis).
The reflections are implemented by a sequence of two state preparations (one forward and one reverse).
In the following description we take the dimension to be $\Nunit$, which need not be a power of 2, and the number of qubits to be $n$ so $2^n\ge \Nunit$.
The qubits are ordered such that the most significant qubit is first.
The state preparation is applied by the following procedure.
\begin{enumerate}
    \item Perform a rotation on the first qubit to prepare $\sqrt{p_0}\ket{0}+\sqrt{p_1}\ket{1}$.
    This has Toffoli complexity $b$ for $b$ bits of precision for the rotation, by using a phase gradient state.
    \item Use the state of the first qubit to output $b$ bits for the rotation on the second qubit. 
 Then use that data to perform a controlled rotation on the second qubit, then erase it.
 There is zero Toffoli complexity for the QROM on the single qubit, and again there is complexity $b$ for the rotation.
 \item For qubits $k=3$ to $n-1$, use QROM on qubits 1 to $k-1$ to output a rotation angle.
 This has complexity
 \begin{equation}
     \left\lceil \frac{\Nunit}{2^{n-k+1}\Lambda_k}\right\rceil + (\Lambda_k - 1) b \, ,
 \end{equation}
for the parameter $\Lambda_k$ a power of 2 in the QROM.
Reference~\cite{low2018trading} assumed that $\Lambda$ is taken independent of $k$, but it may be adjusted to minimise complexity.
Then there is complexity $b$ for the rotation on qubit $k$, and erasure of the QROM needs complexity
\begin{equation}
     \left\lceil \frac{\Nunit}{2^{n-k+1}\Lambda'_k}\right\rceil + \Lambda'_k \, ,
\end{equation}
for total complexity
 \begin{equation}
     \left\lceil \frac{\Nunit}{2^{n-k+1}\Lambda_k}\right\rceil + \Lambda_k b 
     +\left\lceil \frac{\Nunit}{2^{n-k+1}\Lambda'_k}\right\rceil + \Lambda'_k \, ,
 \end{equation}
 where $\Lambda_k$ and $\Lambda'_k$ may be chosen independently.
\item At the end the phases are applied.
This requires a QROM on all $n$ qubits, then addition into a phase gradient register, and erasure of the QROM, with complexity
 \begin{equation}
     \left\lceil \frac{\Nunit}{\Lambda_n}\right\rceil + \Lambda_n b 
     +\left\lceil \frac{\Nunit}{\Lambda'_n}\right\rceil + \Lambda'_n \, .
 \end{equation}
\end{enumerate}

Steps 1 to 3 prepare a state with real amplitudes, then the final step applies phases.
This final step applying phases would be applied twice between each of the $K$ reflections.
It is more efficient to combine these phases, so that there are $K$ reflections by states with real coefficients, and $K+1$ diagonal phasing operations.

\section{Threshold analysis for resource estimates}\label{app:threshold_DF_THC}
\subsection{\texorpdfstring{Fe$_{2}$S$_{2}$}{Fe2S2}}
\begin{table}[H]
    \centering
    \begin{tabular}{|c|c|c|c|c|c|c|}
    \hline
    \hline
    $n_{\mathrm{THC}}$ & $\lambda_{\mathrm{THC}}$ & $\|V - V_{\mathrm{THC}}\|$ & $E_{\mathrm{corr}} - E_{\mathrm{corr}}^{*}$ [mHa] & $C_{\mathrm{B.E.}}$ & $\mathrm{QPE}$ & Logical Qubits \\
    \hline
 60  & 104.4673 &  0.1431 &  3.7846                 & 3903 &  4.0030$\times 10^{8}$  &  575 \\ 
 80  & 105.3320 &  0.0400 &  4.9146                 & 4222 &  4.3660$\times 10^{8}$  &  580 \\
100  & 105.4446 &  0.0219 &  6.9844$\times 10^{-1}$ & 4414 &  4.5694$\times 10^{8}$  &  581 \\
120  & 105.5165 &  0.0124 & -1.4958$\times 10^{-1}$ & 4626 &  4.7921$\times 10^{8}$  &  581 \\
140  & 105.5543 &  0.0082 & -4.2948$\times 10^{-1}$ & 4951 &  5.1306$\times 10^{8}$  &  598 \\
160  & 105.5910 &  0.0069 & -3.8850$\times 10^{-1}$ & 5208 &  5.3988$\times 10^{8}$  &  598 \\
180  & 106.0045 &  0.0070 & -3.1358$\times 10^{-1}$ & 5418 &  5.6385$\times 10^{8}$  & 1045 \\
    \hline
    \hline
    \end{tabular}
    \caption{Analysis of THC rank versus accuracy of CCSD(T) for Fe$_{2}$S$_{2}$ with (FeIII, FeIII) oxidation states~\cite{li2017spin}  using a high spin $n_{\uparrow} - n_{\downarrow} = 8$ reference for CCSD(T). CCSD(T) calculations are performed using UCCSD(T) in PySCF.}
    \label{tab:fe2s2_thc_scan1}
\end{table}

\begin{table}[H]
    \centering
    \begin{tabular}{|c|c|c|c|c|c|c|}
    \hline
    \hline
    $n_{\mathrm{THC}}$ & $\lambda_{\mathrm{THC}}$ & $\|V - V_{\mathrm{THC}}\|$ & $E_{\mathrm{corr}} - E_{\mathrm{corr}}^{*}$ [mHa] & $C_{\mathrm{B.E.}}$ & $\mathrm{QPE}$ & Logical Qubits \\
    \hline
 60  & 62.7590  & 0.1431 &  3.7846$\times 10^{0}$   & 3903 &  2.4048$\times 10^{+8}$ &  573 \\ 
 80  & 63.6236  & 0.0400 &  4.9146$\times 10^{0}$   & 4222 &  2.6372$\times 10^{+8}$ &  578 \\
100  & 63.7363  & 0.0219 &  6.9769$\times 10^{-1}$  & 4414 &  2.7620$\times 10^{+8}$ &  579 \\
120  & 63.8082  & 0.0124 & -1.3363$\times 10^{-1}$  & 4626 &  2.8979$\times 10^{+8}$ &  579 \\
140  & 63.8460  & 0.0082 & -4.1996$\times 10^{-1}$  & 4951 &  3.1033$\times 10^{+8}$ &  596 \\
160  & 63.8827  & 0.0069 & -3.8494$\times 10^{-1}$  & 5208 &  3.2663$\times 10^{+8}$ &  596 \\
180  & 64.2961  & 0.0070 & -3.1175$\times 10^{-1}$  & 5418 &  3.4200$\times 10^{+8}$ & 1043 \\
200  & 72.3297  & 0.0069 & -5.5725$\times 10^{-1}$  & 5621 &  3.9915$\times 10^{+8}$ & 1047 \\
    \hline
    \hline
    \end{tabular}
    \caption{Analysis of THC rank versus accuracy of CCSD(T) for Fe$_{2}$S$_{2}$ with (FeIII, FeIII) oxidation states~\cite{li2017spin}  using a high spin $n_{\uparrow} - n_{\downarrow} = 8$ reference for CCSD(T). CCSD(T) calculations are performed using UCCSD(T) in PySCF. The LCU 1-norm $\lambda$ is computed with a number operator symmetry shift computed as the median of $\{f_{i}\}$ where $f_{i}$ are eigenvalues of the one-body operator being block encoded.}
    \label{tab:fe2s2_thc_scan_alpha_shift}
\end{table}
\begin{table}[H]
    \centering
    \begin{tabular}{|c|c|c|c||c|c|c|c|}
    \hline
    \hline
    \multicolumn{4}{|c||}{$\alpha_2 = 0$} & \multicolumn{4}{c|}{$\alpha_2 = 0.1$}\\
    \hline
    DF thresh. &$n_{\mathrm{DF}}$ & $||V - V_{\mathrm{DF}}||$ & $|E_{\mathrm{corr}} - E_{\mathrm{corr}}^{*}|$ [mHa] & DF thresh. &$n_{\mathrm{DF}}$ & $\|V - V_{\mathrm{DF}}\|$ & $|E_{\mathrm{corr}} - E_{\mathrm{corr}}^{*}|$ \\
    \hline
    1.0$\times 10^{-2}$&  62&  0.0473&  7.864&  1.0$\times 10^{-2}$&  62&  0.0471&  8.250\\
    3.0$\times 10^{-3}$&  80&  0.0154&  8.804$\times 10^{-1}$&  1.0$\times 10^{-3}$&  80&  0.0159&  1.606\\
    1.0$\times 10^{-3}$&  93&  0.0044&  2.496$\times 10^{-1}$&  1.0$\times 10^{-3}$&  93&  0.0046&  6.849$\times 10^{-1}$\\
    3.0$\times 10^{-4}$&  111&  0.0014&  1.424$\times 10^{-1}$&  1.0$\times 10^{-4}$&  111&  0.0014&  1.164$\times 10^{-1}$\\
    1.0$\times 10^{-4}$&  131&  0.0005&  6.639$\times 10^{-2}$&  1.0$\times 10^{-4}$&  131&  0.0005&  4.667$\times 10^{-2}$\\
    3.0$\times 10^{-5}$&  152&  0.0001&  6.824$\times 10^{-2}$&  1.0$\times 10^{-5}$&  152&  0.0001&  3.560$\times 10^{-3}$\\
    \hline
    \hline
    \end{tabular}
    \caption{Analysis of the accuracy of double factorization using the CCSD(T) correlation energy (in $mE_{\mathrm{h}}$) for Fe$_{2}$S$_{2}$ with (FeIII, FeIII) oxidation states~\cite{li2017spin} using a high spin $n_{\uparrow} - n_{\downarrow} = 8$ reference for CCSD(T). CCSD(T) calculations are performed using UCCSD(T) in PySCF. Results are shown with no shift ($\alpha_2 = 0$) and with a shift ($\alpha_2 = 0.1$).}
    \label{tab:fe2s2_df_scan}
\end{table}

\subsection{\texorpdfstring{Fe$_{4}$S$_{4}$}{Fe4S4}}
\begin{table}[H]
    \centering
    \begin{tabular}{|c|c|c|c|c|c|c|}
    \hline
    \hline
    $n_{\mathrm{THC}}$ & $\lambda_{\mathrm{THC}}$ & $\|V - V_{\mathrm{THC}}\|$ & $E_{\mathrm{corr}} - E_{\mathrm{corr}}^{*}$ [mHa] & $C_{\mathrm{B.E.}}$ & $\mathrm{QPE}$ & Logical Qubits \\
    \hline
108  & 271.5315  & 1.2684 &  1.9652$\times 10^{1}$  &   6888  & 1.8362$\times 10^{9}$  &  937  \\
144  & 292.3270  & 0.1294 &  2.8155$\times 10^{0}$  &   7384  & 2.1191$\times 10^{9}$  &  942  \\
180  & 292.8878  & 0.0796 &  1.9821$\times 10^{0}$  &   7804  & 2.2440$\times 10^{9}$  &  1081 \\
216  & 293.5319  & 0.0431 &  7.5063$\times 10^{-1}$ &   8175  & 2.3558$\times 10^{9}$  &  1083 \\
252  & 293.7534  & 0.0301 & -1.9193$\times 10^{-1}$ &   8573  & 2.4724$\times 10^{9}$  &  1083 \\
288  & 293.8331  & 0.0273 & -1.7791$\times 10^{-1}$ &   9120  & 2.6308$\times 10^{9}$  &  1151 \\
324  & 293.9603  & 0.0217 &  6.7870$\times 10^{-2}$ &   9590  & 2.7676$\times 10^{9}$  &  1151 \\
360  & 293.9936  & 0.0198 &  2.2911$\times 10^{-1}$ &  10031  & 2.8952$\times 10^{9}$  &  2110 \\ 
    \hline
    \hline
    \end{tabular}
    \caption{Analysis of THC rank versus accuracy of CCSD(T) for Fe$_{4}$S$_{4}$ with (2 FeII, 2 FeIII) oxidation states~\cite{li2017spin} using a high spin $n_{\uparrow} - n_{\downarrow} = 16$ reference for CCSD(T). CCSD(T) calculations are performed using UCCSD(T) in PySCF.}
    \label{tab:fe2s2_thc_scan2}
\end{table}

\begin{table}[H]
    \centering
    \begin{tabular}{|c|c|c|c|c|c|c|}
    \hline
    \hline
    $n_{\mathrm{THC}}$ & $\lambda_{\mathrm{THC}}$ & $\|V - V_{\mathrm{THC}}\|$ & $E_{\mathrm{corr}} - E_{\mathrm{corr}}^{*}$ [mHa] & $C_{\mathrm{B.E.}}$ & $\mathrm{QPE}$ & Logical Qubits \\
    \hline
144  & 166.9151  & 0.1294 &  2.8155$\times 10^{0}$  &   7384  & 1.2100$\times 10^{9}$ &   940 \\ 
180  & 167.4759  & 0.0796 &  1.9821$\times 10^{0}$  &   7804  & 1.2831$\times 10^{9}$ &  1079 \\
216  & 168.1200  & 0.0431 &  7.5063$\times 10^{-1}$ &   8175  & 1.3493$\times 10^{9}$ &  1081 \\
252  & 168.3416  & 0.0301 & -1.9193$\times 10^{-1}$ &   8573  & 1.4169$\times 10^{9}$ &  1081 \\
288  & 168.4212  & 0.0273 & -1.7791$\times 10^{-1}$ &   9120  & 1.5080$\times 10^{9}$ &  1149 \\
324  & 168.5485  & 0.0217 &  6.7870$\times 10^{-2}$ &   9590  & 1.5869$\times 10^{9}$ &  1149 \\
360  & 168.5817  & 0.0198 &  2.2911$\times 10^{-1}$ &  10031  & 1.6602$\times 10^{9}$ &  2108 \\
    \hline
    \hline
    \end{tabular}
    \caption{Analysis of THC rank versus accuracy of CCSD(T) for Fe$_{4}$S$_{4}$ with (2 FeII, 2 FeIII) oxidation states~\cite{li2017spin} using a high spin $n_{\uparrow} - n_{\downarrow} = 16$ reference for CCSD(T). CCSD(T) calculations are performed using UCCSD(T) in PySCF. The LCU 1-norm $\lambda$ is computed with a number operator symmetry shift computed as the median of $\{f_{i}\}$ where $f_{i}$ are eigenvalues of the one-body operator being block encoded.}
    \label{tab:fe2s2_thc_scan3}
\end{table}
\begin{table}[H]
    \centering
    \begin{tabular}{|c|c|c|c||c|c|c|c|}
    \hline
    \hline
    \multicolumn{4}{|c||}{$\alpha_2 = 0$} & \multicolumn{4}{c|}{$\alpha_2 = 0.1$}\\
    \hline
    DF thresh. &$n_{\mathrm{DF}}$ & $\|V - V_{\mathrm{DF}}\|$ & $|E_{\mathrm{corr}} - E_{\mathrm{corr}}^{*}|$ [mHa] & DF thresh. &$n_{\mathrm{DF}}$ & $\|V - V_{\mathrm{DF}}\|$ & $|E_{\mathrm{corr}} - E_{\mathrm{corr}}^{*}|$ \\
    \hline
    3.0$\times 10^{-3}$&  152&  0.0217820&  1.223   &  3.0$\times 10^{-3}$&  152&  0.0221000&  1.444\\
    1.0$\times 10^{-3}$&  181&  0.0070033&  1.579$\times 10^{-1}$&  1.0$\times 10^{-3}$&  181&  0.0070700&  3.610$\times 10^{-1}$\\
    3.0$\times 10^{-4}$&  222&  0.0020602&  1.719$\times 10^{-1}$&  3.0$\times 10^{-4}$&  222&  0.0020400&  2.576$\times 10^{-1}$\\
    1.0$\times 10^{-4}$&  260&  0.0006839&  4.648$\times 10^{-2}$&  1.0$\times 10^{-4}$&  260&  0.0006840&  5.879$\times 10^{-2}$\\
    3.0$\times 10^{-5}$&  312&  0.0002120&  6.671$\times 10^{-3}$&  3.0$\times 10^{-5}$&  312&  0.0002130&  1.180$\times 10^{-2}$\\
    1.0$\times 10^{-5}$&  365&  0.0000736&  9.582$\times 10^{-3}$&  1.0$\times 10^{-5}$&  365&  0.0000731&  1.067$\times 10^{-2}$\\
    \hline
    \hline
    \end{tabular}
    \caption{Analysis of the accuracy of double factorization using the CCSD(T) correlation energy for Fe$_{4}$S$_{4}$  with (2 FeII, 2 FeIII) oxidation states~\cite{li2017spin} using a high spin $n_{\uparrow} - n_{\downarrow} = 16$ reference for CCSD(T). CCSD(T) calculations are performed using UCCSD(T) in PySCF. Results are shown with no shift ($\alpha_2 = 0$) and with a shift ($\alpha_2 = 0.1$).}
    \label{tab:fe4s4_df_scan}
\end{table}

\begin{table}[H]
    \centering
    \begin{tabular}{|c|c|c|c|c|c|c|}
    \hline
    \hline
    $n_{\mathrm{THC}}$ & $\lambda_{\mathrm{THC}}$ & $\|V - V_{\mathrm{THC}}\|$ & $E_{\mathrm{corr}} - E_{\mathrm{corr}}^{*}$ [mHa] & $C_{\mathrm{B.E.}}$ & $\mathrm{QPE}$ & Logical Qubits \\
    \hline 
108  & 261.8652  & 1.1938  &  3.1102$\times 10^{1}$  & 6888  & 1.7708$\times 10^{9}$ & 935 \\
144  & 280.7802  & 0.1477  &  7.2471$\times 10^{0}$  & 7384  & 2.0354$\times 10^{9}$ & 942 \\
180  & 282.0736  & 0.0691  & -6.4487$\times 10^{-1}$ & 7804  & 2.1611$\times 10^{9}$ & 1081 \\ 
216  & 282.4604  & 0.0464  & -6.7771$\times 10^{-1}$ & 8175  & 2.2670$\times 10^{9}$ & 1083 \\
252  & 282.7991  & 0.0280  & -2.0255$\times 10^{-1}$ & 8573  & 2.3802$\times 10^{9}$ & 1083 \\
288  & 282.8973  & 0.0225  & -3.2773$\times 10^{-1}$ & 9120  & 2.5329$\times 10^{9}$ & 1151 \\
324  & 282.9653  & 0.0199  & -7.2980$\times 10^{-1}$ & 9590  & 2.6641$\times 10^{9}$ & 1151 \\
360  & 283.0203  & 0.0175  & -4.2608$\times 10^{-1}$ & 10031 & 2.7872$\times 10^{9}$ & 2110 \\
    \hline
    \hline
    \end{tabular}
    \caption{Analysis of THC rank versus accuracy of CCSD(T) for Fe$_{4}$S$_{4}$ with Fe-4(III) oxidation states~\cite{li2017spin} using a high spin $n_{\uparrow} - n_{\downarrow} = 16$ reference for CCSD(T). CCSD(T) calculations are performed using UCCSD(T) in PySCF.}
    \label{tab:4fe_thc_scan}
\end{table}

\begin{table}[H]
    \centering
    \begin{tabular}{|c|c|c|c|c|c|c|}
    \hline
    \hline
    $n_{\mathrm{THC}}$ & $\lambda_{\mathrm{THC}}$ & $\|V - V_{\mathrm{THC}}\|$ & $E_{\mathrm{corr}} - E_{\mathrm{corr}}^{*}$ [mHa] & $C_{\mathrm{B.E.}}$ & $\mathrm{QPE}$ & Logical Qubits \\
    \hline 
108  & 143.1948  & 1.1938 &  3.1102$\times 10^{1}$  &  6888  & 9.6833$\times 10^{8}$ &  935\\ 
144  & 162.1098  & 0.1477 &  7.2422$\times 10^{0}$  &  7384  & 1.1752$\times 10^{9}$ &  940\\
180  & 163.4031  & 0.0691 & -6.4487$\times 10^{-1}$ &  7804  & 1.2519$\times 10^{9}$ & 1079\\
216  & 163.7900  & 0.0464 & -6.8280$\times 10^{-1}$ &  8175  & 1.3145$\times 10^{9}$ & 1081\\
252  & 164.1287  & 0.0280 & -2.0255$\times 10^{-1}$ &  8573  & 1.3814$\times 10^{9}$ & 1081\\
288  & 164.2269  & 0.0225 & -3.2750$\times 10^{-1}$ &  9120  & 1.4704$\times 10^{9}$ & 1149\\
324  & 164.2949  & 0.0199 & -7.2980$\times 10^{-1}$ &  9590  & 1.5468$\times 10^{9}$ & 1149\\
360  & 164.3499  & 0.0175 & -4.2608$\times 10^{-1}$ & 10031  & 1.6185$\times 10^{9}$ & 2108\\
    \hline
    \hline
    \end{tabular}
    \caption{Analysis of THC rank versus accuracy of CCSD(T) for Fe$_{4}$S$_{4}$ with Fe-4(III) oxidation states~\cite{li2017spin} using a high spin $n_{\uparrow} - n_{\downarrow} = 16$ reference for CCSD(T). CCSD(T) calculations are performed using UCCSD(T) in PySCF. The LCU 1-norm $\lambda$ is computed with a number operator symmetry shift computed as the median of $\{f_{i}\}$ where $f_{i}$ are eigenvalues of the one-body operator being block encoded.}
    \label{tab:4fe_thc_scan_with_symmetry_shift}
\end{table}

\begin{table}[H]
    \centering
    \begin{tabular}{|c|c|c|c||c|c|c|c|}
    \hline
    \hline
    \multicolumn{4}{|c||}{$\alpha_2 = 0$} & \multicolumn{4}{c|}{$\alpha_2 = 0.1$}\\
    \hline
    DF thresh. &$n_{\mathrm{DF}}$ & $\|V - V_{\mathrm{DF}}\|$ & $|E_{\mathrm{corr}} - E_{\mathrm{corr}}^{*}|$ [mHa] & DF thresh. &$n_{\mathrm{DF}}$ & $\|V - V_{\mathrm{DF}}\|$ & $|E_{\mathrm{corr}} - E_{\mathrm{corr}}^{*}|$ \\
    \hline
    3.0$\times 10^{-3}$&  154&  0.0228617&  1.920   &  3.0$\times 10^{-3}$&  154&  0.0228752&  2.908\\
    1.0$\times 10^{-3}$&  185&  0.0075771&  6.028$\times 10^{-1}$&  1.0$\times 10^{-3}$&  185&  0.0075597&  5.635$\times 10^{-1}$\\
    3.0$\times 10^{-4}$&  226&  0.0020788&  7.355$\times 10^{-2}$&  3.0$\times 10^{-4}$&  226&  0.0020692&  1.129$\times 10^{-1}$\\
    1.0$\times 10^{-4}$&  265&  0.0006872&  8.432$\times 10^{-2}$&  1.0$\times 10^{-4}$&  265&  0.0006910&  7.745$\times 10^{-2}$\\
    3.0$\times 10^{-5}$&  316&  0.0002163&  7.598$\times 10^{-3}$&  3.0$\times 10^{-5}$&  316&  0.0002147&  1.807$\times 10^{-2}$\\
    1.0$\times 10^{-5}$&  369&  0.0000738&  1.080$\times 10^{-3}$&  1.0$\times 10^{-5}$&  369&  0.0000740&  3.570$\times 10^{-3}$\\
    \hline
    \hline
    \end{tabular}
    \caption{Analysis of the accuracy of double factorization using the CCSD(T) for Fe$_{4}$S$_{4}$ with Fe-4(III) oxidation states~\cite{li2017spin} using a high spin $n_{\uparrow} - n_{\downarrow} = 16$ reference for CCSD(T). CCSD(T) calculations are performed using UCCSD(T) in PySCF. Results are shown with no shift ($\alpha_2 = 0$) and with a shift ($\alpha_2 = 0.1$).}
    \label{tab:fe4IIIs4_df_scan}
\end{table}

\begin{table}[H]
    \centering
    \begin{tabular}{|c|c|c|c|c|c|c|}
    \hline
    \hline
    $n_{\mathrm{THC}}$ & $\lambda_{\mathrm{THC}}$ & $\|V - V_{\mathrm{THC}}\|$ & $E_{\mathrm{corr}} - E_{\mathrm{corr}}^{*}$ [mHa] & $C_{\mathrm{B.E.}}$ & $\mathrm{QPE}$ & Logical Qubits \\
    \hline 
108  & 280.7099  & 0.7869 &  3.0641$\times 10^{1}$  &  6888   & 1.8982$\times 10^{9}$ &   937 \\ 
144  & 291.3346  & 0.1912 &  5.4497$\times 10^{0}$  &  7384   & 2.1120$\times 10^{9}$ &   942 \\
180  & 293.2292  & 0.0799 & -3.8260$\times 10^{0}$  &  7804   & 2.2466$\times 10^{9}$ &  1081 \\
216  & 293.8136  & 0.0466 &  2.2190$\times 10^{0}$  &  8175   & 2.3581$\times 10^{9}$ &  1083 \\
252  & 294.1204  & 0.0286 &  1.8563$\times 10^{0}$  &  8573   & 2.4755$\times 10^{9}$ &  1083 \\
288  & 294.2313  & 0.0206 &  5.1673$\times 10^{-2}$ &  9120   & 2.6344$\times 10^{9}$ &  1151 \\
324  & 294.2698  & 0.0192 &  3.0606$\times 10^{-1}$ &  9590   & 2.7705$\times 10^{9}$ &  1151 \\
360  & 294.3255  & 0.0168 & -7.6609$\times 10^{-1}$ &  10031  & 2.8985$\times 10^{9}$ &  2110 \\
    \hline
    \hline
    \end{tabular}
    \caption{Analysis of THC rank versus accuracy of CCSD(T) for Fe$_{4}$S$_{4}$ with 2Fe(III)2Fe(II) oxidation states~\cite{li2017spin} using a high spin $n_{\uparrow} - n_{\downarrow} = 16$ reference for CCSD(T). CCSD(T) calculations are performed using UCCSD(T) in PySCF.}
    \label{tab:fe2s2_thc_scan4}
\end{table}

\begin{table}[H]
\centering
    \begin{tabular}{|c|c|c|c|c|c|c|}
    \hline
    \hline
    $n_{\mathrm{THC}}$ & $\lambda_{\mathrm{THC}}$ & $\|V - V_{\mathrm{THC}}\|$ & $E_{\mathrm{corr}} - E_{\mathrm{corr}}^{*}$ [mHa] & $C_{\mathrm{B.E.}}$ & $\mathrm{QPE}$ & Logical Qubits \\
    \hline 
108 &  155.1929 &  0.7869 &  3.0641$\times 10^{1}$ &  6888  & 1.0495$\times 10^{9}$ &  935 \\ 
144 &  165.8176 &  0.1912 &  5.4497$\times 10^{0}$ &  7384  & 1.2021$\times 10^{9}$ &  940 \\
180 &  167.7122 &  0.0799 & -3.8260$\times 10^{0}$ &  7804  & 1.2849$\times 10^{9}$ & 1079 \\
216 &  168.2966 &  0.0466 &  2.2190$\times 10^{0}$ &  8175  & 1.3507$\times 10^{9}$ & 1081 \\
252 &  168.6034 &  0.0286 &  1.8563$\times 10^{0}$ &  8573  & 1.4191$\times 10^{9}$ & 1081 \\
288 &  168.7143 &  0.0206 &  5.1673$\times 10^{-2}$ &  9120  & 1.5106$\times 10^{9}$ & 1149 \\
324 &  168.7528 &  0.0192 &  3.0606$\times 10^{-1}$ &  9590  & 1.5888$\times 10^{9}$ & 1149 \\
360 &  168.8085 &  0.0168 & -7.6609$\times 10^{-1}$ & 10031  & 1.6624$\times 10^{9}$ & 2108 \\
396 &  168.8265 &  0.0174 & -3.3251$\times 10^{-2}$ & 10386  & 1.7214$\times 10^{9}$ & 2110 \\
432 &  168.8812 &  0.0082 & -3.6832$\times 10^{-2}$ & 10757  & 1.7835$\times 10^{9}$ & 2110 \\
468 &  168.9190 &  0.0073 &  3.4664$\times 10^{-1}$ & 11156  & 1.8501$\times 10^{9}$ & 2110 \\
504 &  168.9603 &  0.0078 & -6.8758$\times 10^{-1}$ & 11561  & 1.9177$\times 10^{9}$ & 2110 \\
540 &  168.9765 &  0.0058 & -1.3886$\times 10^{-1}$ & 12167  & 2.0184$\times 10^{9}$ & 2242 \\
    \hline
    \hline
    \end{tabular}
    \caption{Analysis of THC rank versus accuracy of CCSD(T) for Fe$_{4}$S$_{4}$ with 2Fe(III)2Fe(II) oxidation states~\cite{li2017spin} using a high spin $n_{\uparrow} - n_{\downarrow} = 16$ reference for CCSD(T). CCSD(T) calculations are performed using UCCSD(T) in PySCF. The LCU 1-norm $\lambda$ is computed with a number operator symmetry shift computed as the median of $\{f_{i}\}$ where $f_{i}$ are eigenvalues of the one-body operator being block encoded.}
    \label{tab:fe2s2_thc_scan_one_body_shift}
\end{table}

\begin{table}[H]
    \centering
    \begin{tabular}{|c|c|c|c||c|c|c|c|}
    \hline
    \hline
    \multicolumn{4}{|c||}{$\alpha_2 = 0$} & \multicolumn{4}{c|}{$\alpha_2 = 0.1$}\\
    \hline
    DF thresh. &$n_{\mathrm{DF}}$ & $\|V - V_{\mathrm{DF}}\|$ & $|E_{\mathrm{corr}} - E_{\mathrm{corr}}^{*}|$ [mHa] & DF thresh. &$n_{\mathrm{DF}}$ & $\|V - V_{\mathrm{DF}}\|$ & $|E_{\mathrm{corr}} - E_{\mathrm{corr}}^{*}|$ \\
    \hline
    3.0$\times 10^{-3}$&  152&  0.0221358&  7.301$\times 10^{-1}$&  3.0$\times 10^{-3}$&  152&  0.0224483&  2.619$\times 10^{-1}$\\
    1.0$\times 10^{-3}$&  183&  0.0069379&  2.493$\times 10^{-1}$&  1.0$\times 10^{-3}$&  181&  0.0070526&  6.233$\times 10^{-1}$\\
    3.0$\times 10^{-4}$&  222&  0.0020665&  8.187$\times 10^{-2}$&  3.0$\times 10^{-4}$&  222&  0.0020752&  1.715$\times 10^{-2}$\\
    1.0$\times 10^{-4}$&  260&  0.0006887&  4.144$\times 10^{-2}$&  1.0$\times 10^{-4}$&  260&  0.0006901&  1.664$\times 10^{-2}$\\
    3.0$\times 10^{-5}$&  312&  0.0002146&  1.359$\times 10^{-2}$&  3.0$\times 10^{-5}$&  312&  0.0002151&  2.107$\times 10^{-2}$\\
    1.0$\times 10^{-5}$&  365&  0.0000741&  4.842$\times 10^{-3}$&  1.0$\times 10^{-5}$&  365&  0.0000736&  2.165$\times 10^{-2}$\\
    \hline
    \hline
    \end{tabular}
    \caption{Analysis of the accuracy of double factorization using the CCSD(T) for Fe$_{4}$S$_{4}$ with 2Fe(III)2Fe(II) oxidation states~\cite{li2017spin} using a high spin $n_{\uparrow} - n_{\downarrow} = 16$ reference for CCSD(T). CCSD(T) calculations are performed using UCCSD(T) in PySCF. Results are shown with no shift ($\alpha_2 = 0$) and with a shift ($\alpha_2 = 0.1$).}
    \label{tab:feIII2feII2s4_df_scan}
\end{table}

\subsection{FeMoCo}

\begin{table}[H]
    \centering
    \begin{tabular}{|c|c|c|c||c|c|c|c|}
    \hline
    \hline
    \multicolumn{4}{|c||}{$\alpha_2 = 0$} & \multicolumn{4}{c|}{$\alpha_2 = 0.1$}\\
    \hline
    DF thresh. &$n_{\mathrm{DF}}$ & $\|V - V_{\mathrm{DF}}\|$ & $|E_{\mathrm{corr}} - E_{\mathrm{corr}}^{*}|$ [mHa] & DF thresh. &$n_{\mathrm{DF}}$ & $\|V - V_{\mathrm{DF}}\|$ & $|E_{\mathrm{corr}} - E_{\mathrm{corr}}^{*}|$ \\
    \hline
    5.00$\times 10^{-3}$&  312&  0.0578095&   2.414   &  5.00$\times 10^{-3}$&  312&  1.706   &  0.0580414\\
    2.50$\times 10^{-3}$&  344&  0.0289546&   1.470   &  2.50$\times 10^{-3}$&  344&  1.621   &  0.0289732\\
    1.25$\times 10^{-3}$&  394&  0.0144793&   5.330$\times 10^{-2}$&  1.25$\times 10^{-3}$&  394&  3.872$\times 10^{-1}$&  0.0144793\\
    5.00$\times 10^{-4}$&  470&  0.0057589&   4.048$\times 10^{-1}$&  5.00$\times 10^{-4}$&  472&  4.500$\times 10^{-1}$&  0.0057756\\
    2.50$\times 10^{-4}$&  526&  0.0028081&   6.060$\times 10^{-2}$&  2.50$\times 10^{-4}$&  522&  6.970$\times 10^{-2}$&  0.0028139\\
    1.25$\times 10^{-4}$&  589&  0.0013772&   3.989$\times 10^{-2}$&  1.25$\times 10^{-4}$&  585&  7.884$\times 10^{-3}$&  0.0013765\\
    5.00$\times 10^{-5}$&  679&  0.0005509&   1.028$\times 10^{-2}$&  5.00$\times 10^{-5}$&  679&  7.430$\times 10^{-3}$&  0.0005514\\
    \hline
    \hline
    \end{tabular}
    \caption{Analysis of the accuracy of double factorization using the CCSD(T) correlation energy for FeMoco~\cite{li2017spin} using a high spin $n_{\uparrow} - n_{\downarrow} = 35$ reference for CCSD(T). CCSD(T) calculations are performed using UCCSD(T) in PySCF. Results are shown with no shift ($\alpha_2 = 0$) and with a shift ($\alpha_2 = 0.1$).}
    \label{tab:femoco_df_scan}
\end{table}

\section{Validation of overlap extrapolation protocol}
\label{app:valov}

\begin{figure}[tbh]
    \centering
    \includegraphics[width=0.8\linewidth]{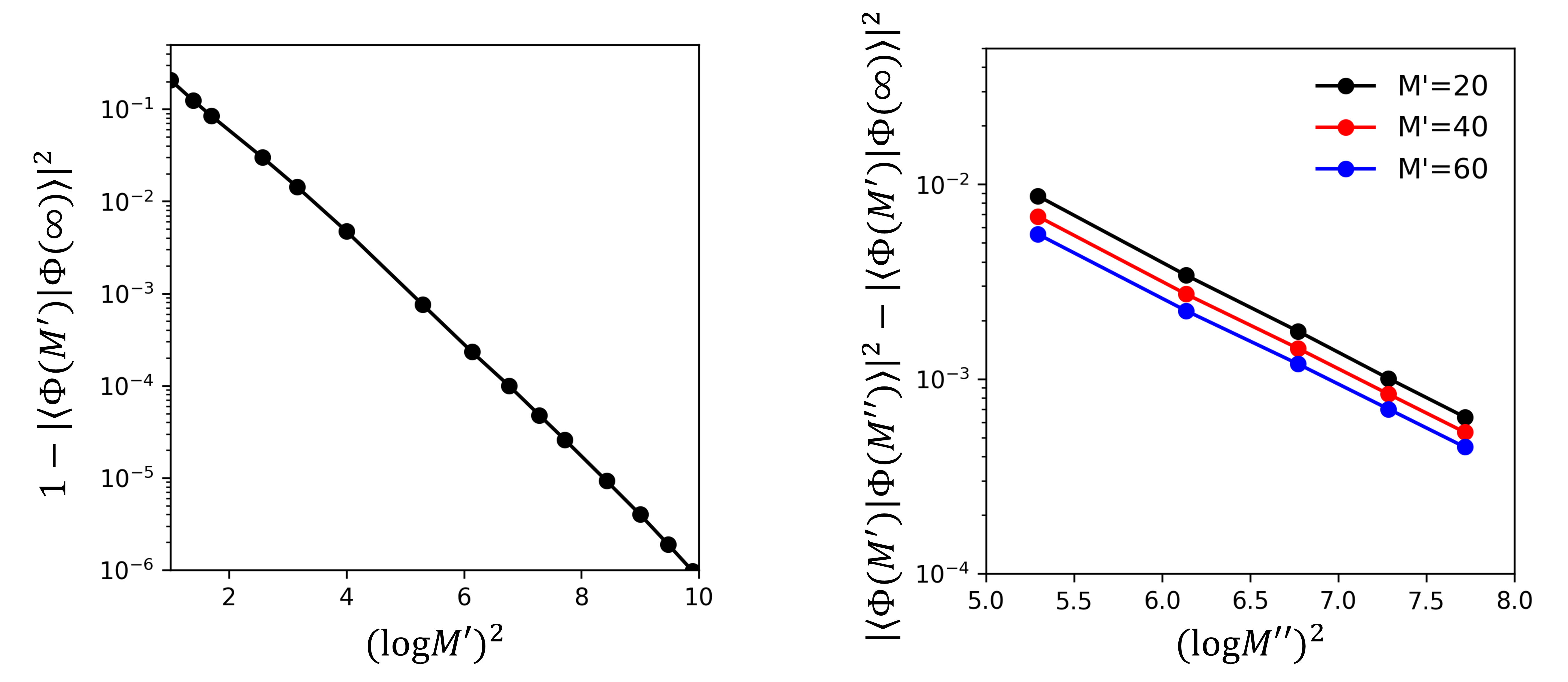}
    \caption{Two empirical linear relations for the \ce{Fe_2S_2} system. We demonstrate the first linear relation using the overlap between MPSs for $\bondb'=[10, 1400]$ and the exact MPS, while the second linear relation is demonstrated using the overlap between MPSs for $\bondb'=[20, 60]$, MPSs for $\bondb''=[200,600]$, and the exact MPS.}
    \label{fig:dimer_verf}
\end{figure}

Here we show several examples that demonstrate the validity of the overlap extrapolation protocol for the \ce{Fe_2S_2} and \ce{Fe_4S_4} systems.
For the \ce{Fe_2S_2} system, the exact wave function ($\Phi(\infty)$) for the active space model of CAS(30e,20o) is accessible. 
We obtained the exact MPS and MPSs with several bond dimensions and calculate the overlap between them.
The left and right panels of Fig.~\ref{fig:dimer_verf} show the plots corresponding to Eqs.~\eqref{eq:lin1} and~\eqref{eq:lin2}, respectively. We can see that both empirical linear relations fit remarkably well.

\begin{figure}[tbh]
    \centering
    \includegraphics[width=0.95\linewidth]{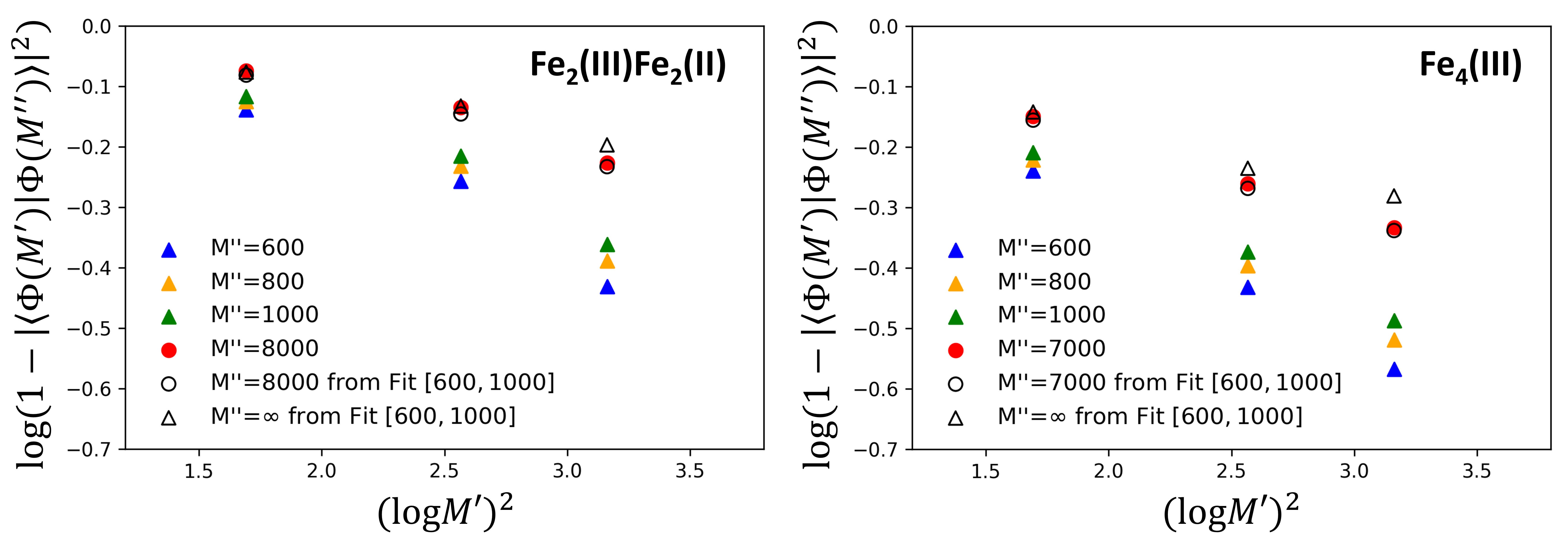}
    \caption{ Validation of the overlap extrapolation for the \ce{Fe_4S_4} systems. The data from the overlaps of the MPSs for $\bondb'=20, 40, 60$ and $\bondb''=600, 800, 1000, 8000$, obtained by DMRG, are represented by the colored dots. While, the data obtained through extrapolation based on the linear relation in Eq.~\eqref{eq:lin2} are represented by the empty dots. The close agreement between the empty circles and red circles demonstrates the robustness of the extrapolation protocol.}
    \label{fig:cubane_verf}
\end{figure}

However, the exact wave functions for the complete active space models of the \ce{Fe_4S_4} systems, namely CAS(54e,36o) and CAS(52e,36o), are not accessible. 
Therefore, while we are unable to directly confirm the empirical linear relations for these systems, Fig.~\ref{fig:cubane_verf} provides strong evidence supporting the validity of the overlap extrapolation.
Each empty triangle corresponds to $|\langle \Phi(\bondb') | \Phi(\infty)\rangle|$ for each value of $\bondb'=20, 40,$ and $60$, and was obtained by fitting the blue, yellow, and green triangles just below it, as discussed in Fig.~\ref{fig:cubane}.
Based on these $|\langle \Phi(\bondb') | \Phi(\infty)\rangle|$ values, we applied the linear relation of Eq.~\eqref{eq:lin2} to predict the values of $|\langle \Phi(\bondb') | \Phi(\bondb''=8000)\rangle|$, corresponding to the empty circles in Fig.~\ref{fig:cubane_verf}.
On the other hand, we obtained the MPS for $\bondb''=8000$ using DMRG, and directly computed the overlaps between this MPS and those for $\bondb'=20, 40,$ and $60$ corresponding to the red circles.
The excellent agreement between the red and empty circles provides indirect validation of $|\langle \Phi(\bondb') | \Phi(\infty)\rangle|$ for $\bondb'=20, 40,$ and $60$, indicating the reliability of the extrapolation.

\end{document}